\documentclass[12pt]{article}
\usepackage{graphicx,color}
\usepackage{XoohmE}
\usepackage{booktabs}
\usepackage{amsmath}
\usepackage{epstopdf}

\def\4{\oplus}
\def\8{\otimes}
\def\B{\mathop{\raisebox{1.5pt}{$\SSS\bullet$}}\nolimits}

\def\bAr{\begin{array}}
\def\eAr{\end{array}}

\def\iso#1{\buildrel{\sss#1}\over\simeq}
\def\mitrix#1{\hbox{\scriptsize$\begin{matrix}#1\end{matrix}$}}
\def\cA{\mathcal{A}}
\def\sB{\mathscr{B}}

\def\Db{{\skew3\bar{D}}}

\def\bcD{\BM{\cal D}}
\def\sD{\mathscr{D}}

\def\sF{\mathscr{F}}

\def\cI{{\cal I}}
\def\cJ{{\cal J}}
\def\cK{{\cal K}}
\def\im{\mathop{\rm im}\nolimits}

\def\sO{\mathscr{O}}

\def\SO{\mathop{\rm SO}}

\def\Span{\mathop{\rm Span}\nolimits}

\def\Spin{\mathop{\rm Spin}}

\def\SU{\mathop{\rm SU}}

\def\sU{\mathscr{U}}
\def\vC#1{\vcenter{\hbox{\hss#1\hss}}}

\def\bz{\mathbb{Z}}
\def\hgt{\mathop{\rm hgt}\nolimits}
\def\dist{\mathop{\rm dist}\nolimits}

\newtheorem{definition}{Definition}[section]
\newtheorem{theorem}{Theorem}[section]
\newtheorem{proposition}[theorem]{Proposition}
\newtheorem{corollary}[theorem]{Corollary}
\newtheorem{conjecture}[theorem]{Conjecture}

\newenvironment{proof}{\noindent{\bf Proof:}~}%
                {$\Box$\llap{\raise2pt\hbox to9pt{\Large$\Chekk$}}}{}{}{}
\long\def\TC#1#2{{\color{#1}{#2}}}
\definecolor{Hey}{rgb}{.9,.05,.4}
\definecolor{orange}{rgb}{.8,.4,0}
\definecolor{plum}{rgb}{.4,0,.6}

\font\cfnt=lcircle10 at 9pt
\def\lplus{\mathop{\kern2pt
            \raise1.275ex\hbox to0pt{\cfnt\char"07\hss}\kern-.6pt+}}

\def\YT#1#2{\vcenter{\hbox{\vbox{\baselineskip0pt\parskip=\medskipamount%
             \def\B{$\sqcap$\llap{$\sqcup$}\kern-1.9pt}
              \def\Bd{\hbox{\kern2.4pt\raise.4pt\hbox{$\cdot$}\kern-5.7pt\B\kern0pt}}
              \def\4{\raise.25pt\hbox to0pt{\hss\kern2pt--\hss}}
              \def\Z{\hfil\vskip-5.9pt}\lineskiplimit0pt\lineskip0pt%
               \setbox0=\hbox{#1}\hsize\wd0\parindent=0pt#2}\,}}}
\def\Ft#1{\,\footnote{#1}}
\newdimen\parshift\parshift=\parindent
\catcode`@=11
 \long\def\@footnotetext#1{\insert\footins{\reset@font\footnotesize\interlinepenalty%
  \interfootnotelinepenalty\splittopskip\footnotesep\splitmaxdepth\dp\strutbox%
   \floatingpenalty\@MM\hsize\columnwidth\addtolength{\hsize}{-2\parindent}
    \@parboxrestore\protected@edef\@currentlabel{\csname p@footnote\endcsname\@thefnmark}
      \color@begingroup
       \@makefntext{\rule\z@\footnotesep\ignorespaces#1\@finalstrut\strutbox}
        \color@endgroup}}
 \long\def\@makefntext#1{\hglue\parshift
                         \vbox{\noindent\hb@xt@0em{\hss\@makefnmark}#1}}
\catcode`@=12
 \def\ba{\left(\begin{array}}
 \def\ea{\end{array}\right)}
 
 \def\vd{\partial}
 \def\brr{\begin{eqnarray}}
 \def\err{\end{eqnarray}}

 \newcommand{\fr}[2]{{\textstyle\frac{#1}{#2}}}
 \HarvTitles 
 \seceq

\begin{document}

\begin{flushright}
\today
\end{flushright}

\begin{center}
{\Large\bf On Graph-Theoretic Identifications of Adinkras,\\[1mm]
           Supersymmetry Representations and Superfields}\\[3mm]
{\bf C.F.\,Doran$^a$, M.G.\,Faux$^b$, S.J.\,Gates, Jr.$^c$,
     T.\,H\"{u}bsch$^d$, K.M.\,Iga$^e$ and G.D.\,Landweber$^f$}\\[1mm]
{\small\it
  $^a$Department of Mathematics,
      University of Washington, Seattle, WA 98105
  \\
  $^b$Department of Physics,
      State University of New York, Oneonta, NY 13825
  \\
  $^c$Department of Physics,
      University of Maryland, College Park, MD 20472
  \\
  $^d$Department of Physics and Astronomy,
      Howard University, Washington, DC 20059
  \\
  $^e$Natural Science Division,
      Pepperdine University, Malibu, CA 90263
  \\
 $^f$Mathematics Department,
     University of Oregon, Eugene, OR 97403-1222
 }\\[3mm]
{\bf ABSTRACT}\\[2mm]
\parbox{5.7in}{\parindent=2pc\noindent
In this paper we discuss off-shell representations of $N$-extended supersymmetry in one dimension, \ie, $N$-extended supersymmetric quantum mechanics, and following earlier work on the subject codify them in terms of
graphs called Adinkras. This framework provides a method of generating
all Adinkras with the same topology, and so also all the corresponding
irreducible supersymmetric multiplets.  We develop some graph theoretic
techniques to understand these diagrams in terms of a relatively small amount of
information, namely, at what heights various vertices of the graph should be ``hung''.

We then show how Adinkras that are the graphs of $N$-dimensional cubes can be
obtained as the Adinkra for superfields satisfying constraints that involve superderivatives.  This dramatically widens the range of supermultiplets that can be described using the superspace formalism and also organizes them.  Other topologies for Adinkras
are possible, and we show that it is reasonable that these are also the result of constraining superfields using superderivatives.

We arrange the family of Adinkras with an $N$-cubical topology, and so also the sequence
of corresponding irreducible supersymmetric multiplets, in a
cyclic sequence, which we call the {\em main sequence}.  We produce the $N{=}1$
and $N{=}2$ main sequences in detail, and indicate some aspects of the situation for higher $N$.
}
\end{center}

\section{Introduction}
Supersymmetry appeals to mathematicians due to its apparent richness and
surprising connection to well established and developed concepts.  Supersymmetry
also appeals to physicists' desire to forge a unified picture of nature, and has
a seemingly miraculous ability to speak to disparate conundrums, offering hope
for their resolution.
 However, from a mathematical standpoint, physical supersymmetry has yet to be fully
 and properly formulated. This is especially so regarding the classification
 of off-shell representations of supersymmetry. The purpose of this paper is to describe some
 recent progress into this problem.

 {}From the point of view of theoretical physics, supersymmetry is a crucial
 ingredient in string theory---the familiar rubric for a large contemporary
 attempt to formulate a quantum theory of nature which includes all known
 matter and all of its known interactions, including gravity.  Indeed, the
 primary reason for introducing and trusting quantum physics as a universal,
 fundamental, scientific framework is the stability of atoms. In a similar
 spirit, supersymmetry provides the only known universal mechanism for
 stabilizing the vacuum, both in quantum field theories, and also in superstring
 theory, including its M- and F-theory extensions.

Phenomenologists have long since wrestled with the hierarchy puzzle, \ie, the
perplexing stability of the disparate scales of elementary force couplings (the
electroweak energy scale being some fifteen orders of magnitude less than the
Planck scale); in the absence of seemingly miraculous fine-tuning, such differences
should be eradicated due to quantum renormalization effects. Supersymmetry offers
an escape from this problem. The particular boson/fermion dichotomy implied by supersymmetry has, as an ancillary benefit, remarkable non-renormalization effects which remove the need for fine tuning, at the expense of introducing into quantum field theory unexpected complexities with yet unresolved puzzles of their own. An intended purpose of our work is to begin to
speak to these through a mathematical reformulation of supersymmetry.

For mathematicians, supersymmetry provides a virtual playground of structures which beg for a rigorous foundation and complete classification.  However, the term supersymmetry has come to mean slightly different things to physicists and mathematicians.  This has caused some unfortunate mis-communication, which has partially hindered the historic synergy between these respective fields (this in spite of the existing pedagogical literature such as
Refs.\cite{r1001,rBK,rDel,rDF}). In our work, we endeavor to speak to both audiences. Consequently, we shall present our ideas, and our approach to the problem at hand, in more detail than is customary in either field.  Nevertheless, we defer the fully rigorous ``mathspeak'' foundation of the work presented here and based on Ref.\cite{rA} to a concurrent but separate effort\cite{r6-1}.

Despite its appeal, the subject of supersymmetry is fraught with more than one conundrum of its own. From the physics standpoint, an obvious one is phenomenological. As of this writing, there has yet to appear any verifiable evidence of fundamental supersymmetry in nature\Ft{This state of affairs stands a hope of improvement in the next few years since CERN's LHC collider, scheduled to be commissioned in 2007, may  provide such evidence (nature willing).}.  There is, however, also a theoretical conundrum associated with supersymmetry.  Taking a more mathematical perspective, this one is more vexing and more pressing than the phenomenological one.
This problem is called the {\em off-shell problem}, and can be understood as follows.

A satisfying aspect of Yang-Mills theories is that the underlying symmetries, described by ordinary Lie algebras, are realized independently from the physics.  The basic fields cleanly represent the Lie algebra without additional, dynamical constraints.  By contrast, this feature holds in known supersymmetric field theories only in a very limited number of cases, and is not valid for the most interesting theories ({\it e.g.}, string theory) involving supersymmetry.  These limited cases usually involve a number of spacetime dimensions less than or equal to four.

For most theories in spacetime dimensions greater than four, supersymmetry has known representations only if the component fields of the representation are subject to particular {\em dynamical} constraints, namely that these fields satisfy Euler-Lagrange equations.
Supersymmetrical representations of this character are said to be ``on-shell".   This state of affairs can be viewed as less than fully satisfactory for a variety of reasons, and puts interesting supersymmetrical theories at variance with most supposedly fundamental descriptions of Nature.

For instance, the separation of the symmetry representations from the physics (\ie, the Lagrangian and its equations of motion) is an ingredient in Yang-Mills theories, including  the standard model of particle physics.  Since Yang-Mills symmetries are realized locally, it is important that the quantum partition function respect these symmetries without anomalies.  Otherwise the quantized theory would not be unitary, and it would therefore have no meaningful predictive power; it would be ill-defined.  From a path-integral point of view, the fields in a quantum field theory are ordinarily not constrained to satisfy the associated Euler-Lagrange equations.
Instead, such solutions merely describe the most probable path---the ``classical" approximation to the quantum theory.

Since higher-dimensional supersymmetric field theories are formulated only on-shell, the program of quantization, in a manner that manifestly realizes the supersymmetry, is seemingly compromised, and it is not entirely clear how or whether a manifestly supersymmetric unitary quantum partition function should exist for these constructions.  The lack of a formulation of these interesting theories without the imposition of Euler-Lagrange equations is called the ``off-shell problem''.

The off-shell problem is fundamentally connected with the representation theory of Lie super\-algebras. Whereas the representation theory of compact or complex reductive Lie algebras is a mature subject, the classification of representations of Lie superalgebras poses a  more difficult and interesting problem which is not yet fully understood. Whereas mathematicians have made significant progress on certain aspects of this problem (see, for example, Ref.\cite{rKac}), the off-shell field content of representations of physical supersymmetry is generally not known.
This problem is,
perhaps, most interesting and most relevant at the level of supergravity theories.  These are field theories which exhibit supersymmetry as a local invariance. Since the elementary supersymmetry algebra contains the Poincar{\'e} algebra as a subalgebra, and since the gauging of the Poincar{\'e} algebra implies General Relativity, it follows that gauged supersymmetry algebras automatically include gravity.

The mathematical challenge of the ``off-shell problem''
has remained unresolved for more than thirty years (see Ref. \cite{rGLPR}).
This suggests the possibility for uncovering fundamental and interesting new mathematical features of supersymmetry by attempting to meet this challenge.  This duration of time also suggests that a new vantage point or language may aid in achieving this goal. In particular, we propose to use the recently introduced tools called ``garden algebras'' \cite{rGLP} and ``Adinkras'' \cite{rA} described below.

The use of ``garden algebras'' is the assertion that
the key to understanding the to-be-completed classification of
supersymmetry representation is to embed supersymmetry
representations within the firmly established structure of
Clifford modules (see Ref.\cite{rABS}).  This strategy was first suggested by the
work of Ref.\cite{rGR2}.  We note that the essential algebraic
features of supersymmetric field theories are present in those
one-dimensional field theories obtained by dimensional
reduction\cite{rGR0,rGR1,rGR2,rGLPR,rGLP,rFS,rFSK}.  In that
context, we make two propositions.  The first is that the
representation theory of supersymmetry in arbitrary dimensions
is encoded in the representation theory of one-dimensional
superalgebras. The second proposition is that a complete
representation theory of one-dimensional superalgebra is
encoded in the tractable representation theory of so-called
${\cal GR}(d,N)$ algebras introduced in \cite{rGLP}.
Ref.\cite{rA} introduced a diagrammatic
method for classifying and generating representations of these
algebras, and in-turn, one-dimensional superalgebras was introduced.
The diagrams used in this method are called Adinkras.

It should be noted that recently there have appeared works, carried
out by independent groups, in which the ``Garden Algebras'' approach
(and associated concepts) have led to new results for constructing,
understanding and classifying one-dimensional supersymmetrical theories.
One such work appears in Ref.\cite{rBKMO}. where it is shown that the
``root super-fields'' introduced in Ref.\cite{rA} imply a web of
interrelationships between non-linear sigma-models and their associated
geometries all related by the `AD' maps discussed in Ref.\cite{rGLPR}.
In Ref.\cite{rKRT} a forceful demonstration of the power of the
``Garden Algebras'' approach was given in the derivation of previously
unknown and interesting features about supersymmetric representation
theory that is totally independent of superspace. This last work
represents a line of research\cite{rKT,rPT,rT04} that began in
2001\cite{rT01,rT01a} and was initiated after a 1997 communication
between S.J.G. and F.~Toppan.

In the language of graph theory, an Adinkra is a directed graph with some extra information associated to each vertex and edge, intended to describe the supersymmetry transformation in terms of component fields.
 In this paper we present evidence that a subset of such graphs is in one-to-one correspondence with superfields, and, therefore, that Adinkras provide an intriguing and totally {\em {independent}} alternative to a superfield-based description of supersymmetry, partly addressing the conjectures of Ref.\cite{rGLP}.  It is our belief that the graph theoretic context might prove useful for forging a deeper understanding of supersymmetry, and might allow for an off-shell representation theory to be developed.  Thus we hope to generate a study of ``adinkramatics'',
that is, an abbreviated fusion of {\em adinkraic} and {\em grammatical}, or possibly {\em diagrammatics} or {\em mathematics}, which pertains to the
graph-theoretic properties of Adinkras.  In this way, off-shell supersymmetric field theories in dimensions greater than four could be developed.  These, in turn, would likely provide valuable food for thought regarding fundamental questions.

This paper is structured as follows.

In Section~\ref{graph} we briefly review the construction of Adinkras, and explain how these are amenable to classification in terms of graph theory, thereby motivating the relevance of such mathematics to the subject of supersymmetry representation theory.
 Section~\ref{Graph} provides a more rigorous set of definitions pertaining to the particular class of {\em engineerable} Adinkras, in which a {\em height} assignment encodes the supersymmetry action and is associated with the physicists' concept of ``engineering dimension.''  Theorem~\ref{tDetA}
and its immediate Corollary~\ref{cDetA}, giving the necessary and sufficient data to specify an Adinkra, are presented in Section~\ref{sHGT}.
 Section~\ref{sVRaising} then presents Theorem~\ref{tHGT} and its Corollaries~\ref{cHGT} and~\ref{cHGT2}, which state that vertex ``lowerings,''  and similarly ``raisings,'' generate the {\em family} of all Adinkras with the same {\em topology} from any one of them given.
 Section~\ref{sSFnVR} explores the superspace formalism and investigates
 how to determine an Adinkra for a superfield, focusing on examples with $N=1$ and $N=2$ superfields.
For instance, for $N=2$ superfields, Proposition~\ref{pDIUinFI} presents how to read off superfield equivalents for the Adinkras in these cases.
Section~\ref{secSderiv} then generalizes these concepts to show how a wide
range of Adinkras can be described in terms of superfields satisfying
constraints involving superderivatives.  First, Theorem~\ref{tsuperimage} shows
how to turn an Adinkra into the image of a superderivative operator.  Then, Subsection~\ref{sidentify} explains how to turn this into a superfield satisfying certain constraints involving superderivatives.  The overall procedure of
taking an Adinkra and returning constraints on superfields is explained
in Theorem~\ref{pA=SF}.  The dependence of this procedure on the topology
of the Adinkra is contemplated in Subsection~\ref{sTop}, prompting
Conjecture~\ref{c?}.
 Section~\ref{mainseq} describes the vertex raises in Section~\ref{sVRaising}
in terms of superderivatives on superfields, and in the process puts the cubical
Adinkras in a sequence called the {\em main sequence}.  This main sequence is illustrated for $N=1$ in Proposition~\ref{pHGTN1} and for $N=2$ in Propositions~\ref{pHGTN2} and~\ref{pA=SFmap}.  The situation for $N=3$ is described in
Proposition~\ref{pA2SFN3}.
 Finally, Section~\ref{sC} offers some concluding remarks.

\section{Review of Adinkra Diagrams}
 \Label{graph}
 We refer to the elementary $N$-extended Poincar\'e superalgebra in $d$-dimensional
 Minkowski space as the $(d|N)$ superalgebra.  The term ``elementary''
 implies a classical Lie superalgebra without central extensions or the addition of other internal bosonic symmetric.
 As explained
 in the Introduction, we are particularly interested in the special
 case of one-dimensional $(1|N)$ superalgebras.
 In this case, we label the single time-like coordinate $\t$.
 The algebra is then defined in terms of translations,
 generated by the single derivative $\vd_\t$, and by
 a set of $N$ real supersymmetry generators $Q_I$, which
 commute with $\vd_\t$ and are also subject to
 the following anti-commutation relation:
 \brr \{\,Q_I~,\,Q_J\,\}=2\,i\,\d_{IJ}\,\vd_\t~,
 \Label{nalg}\err
 where $\d_{IJ}$ is the Kroenecker delta.

 It is common in the physics literature to define a
 parameter-dependent ``transformation" associated with symmetry
 operations. Accordingly, we define
 \begin{equation}
 \d_Q(\e):=-i\e^I\,Q_I,
 \end{equation}
 where $\e^I$ is a set of $N$ anticommuting
 parameters.
 In terms of this operation, the anticommutator~(\ref{nalg})
 is alternatively described by the following commutator,
 \brr [\,\d_Q(\e_1)~,\,\d_Q(\e_2)\,]
      =2\,i\,\e_1^I\,\e_2^I\,\vd_\t~.
 \Label{nalg2}\err
 The notation is such that the parameter superscripts
 enumerate distinct supersymmetries, while the parameter
 subscripts are merely labels, which indicate distinct
 {\it choices} of the parameter.
 We remark that~(\ref{nalg}) and~(\ref{nalg2}) are
 equivalent.

 A diagrammatic paradigm was introduced in Ref.\cite{rA} for
 classifying the representations of~(\ref{nalg}). The diagrams
 used in this method are called ``Adinkra diagrams", or
 ``Adinkras" for short.
 By way of very brief review, every representation of the
 $(1|N)$ superalgebra, for any value of $N$, decomposes as an
 assembly of some number of
 irreducible representations of the $(1|1)$ superalgebra,
 described by $Q^2=i\,\vd_\t$.
 There are two such elemental representations, each of which
 includes one real commuting field, \ie, a boson,
 and one real anticommuting field, \ie, a fermion.
 The distinction between these two $(1|1)$ supermultiplets is merely
 in the transformation relations, and we list them here together with
 the corresponding Adinkras:

One of the irreducible $(1|1)$ supermultiplets, consisting of boson $\f$
and fermion
$\j$, is described by  the following rules\Ft{Note that Eqs.~(\ref{scalar}) preserve the
reality of $\f$ and $\j$.  To see this, note, for instance, that $i\,\e\,\j$
is invariant under Hermitian conjugation because $\e$ and $\j$ are mutually anticommuting,
and because Hermitian conjugation reverses the operator ordering.
Similar considerations can be used to verify the consistency of all similar expressions
used in this section; see also Appendix~\ref{sRSF}.}:
 \beq
  \left.\begin{aligned}
   \d_Q(\e)\,\f &= i\,\e\,\j~,\\[2mm]
   \d_Q(\e)\,\j &= \e\,\vd_\t\,\f~
   \end{aligned}\right\}\quad\Longleftrightarrow\quad
  \begin{picture}(23,0)(2,0)
   \put(0,-4){\includegraphics[width=1in]{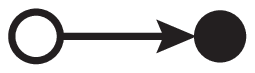}}
   \put(2,5){$\f$}
   \put(21,5){$\j$}
  \end{picture}~,
 \Label{scalar}
 \eeq
where the diagram to the right is the Adinkra corresponding to this multiplet.  This multiplet
is referred to as the elemental scalar multiplet.  The Adinkra codifies these rules
symbolically, by representing bosonic fields using white circles, fermionic fields using black
circles, and by representing the indicated transformations by the direction of the arrow.

The other irreducible $(1|1)$ supermultiplet, dubbed the the elemental spinor multiplet,
consists of a fermion, $\l$, and a boson, $B$, and is analogously described by the Adinkra
and the corresponding transformation rules:
 \beq
  \left.\begin{aligned}
   \d_Q(\e)\,\l &= \e\,B~,\\[2mm]
      \d_Q(\e)\,B &= i\,\e\,\vd_\t\,\l~
   \end{aligned}\right\}\quad\Longleftrightarrow\quad
  \begin{picture}(23,0)(2,0)
   \put(0,-4){\includegraphics[width=1in]{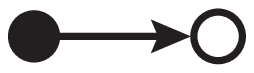}}
   \put(2,5){$\l$}
   \put(21,5){$B$}
  \end{picture}~,
 \Label{spinor}
 \eeq
 Each of the two sets of transformation rules~(\ref{scalar})
 and~(\ref{spinor}) are defined modulo a possible change in the overall
 sign on each of the two rules in the set.  The orientation of the arrow
 is what  identifies~(\ref{scalar}) as the scalar multiplet, in distinction
 to the spinor multiplet~(\ref{spinor}): In the former, the arrow points
 away from the scalar (white) vertex, the {\em source},
 whereas in~(\ref{spinor}) it is the fermion (black) vertex that is the source.
 In either case, this precise correspondence permits us to read off the
 transformation rules from the Adinkra\cite{rA}.

 There actually is an additional specificity involved  in translating an
 Adinkra symbol into transformation  rules. This involves the identification
 of whether or not an additional minus sign should be added to the right-hand
 sides in~(\ref{scalar}) and/or~(\ref{spinor})---a freedom which was mentioned
 above. This choice is encoded in a so-called ``arrow parity", which is
 described more fully in Ref.\cite{rA}.  This issue does pose certain
 restrictions, which are readily resolved, when these elemental $N=1$
 Adinkras are linked together to form more intricate Adinkras associated with
 higher-$N$ supersymmetry.  The results of this paper can be comprehended
 without our explicitly specifying this extra data, however.  As a matter of
 economy, we will largely suppress the issue of arrow parity in this paper.

 Individual $N=1$ Adinkras can be combined to form higher-$N$ Adinkras,
 by using additional arrows to represent additional supersymmetries. In
 this way one can construct Adinkras to represent superalgebras with
 arbitrary $N$. One can keep track of the separate supersymmetries by
 maintaining a partitioning system for the arrows; herein we will use
 colors. In Ref.\cite{rA}, the partitioning was
 arranged by embedding such an Adinkra into an $N$-dimensional Euclidean
 space, such that arrows corresponding to distinct supersymmetry
 generators are directed with mutually orthogonal orientations.
 This orthogonality in depicting Adinkras reflects the `orthogonality'
 of the correspondingly distinct supersymmetry generators:
 $\{Q_I,Q_J\}=0$ for $I\neq J$.

  Consider the following $N=2$ Adinkra:
 \beq
  \vC{\includegraphics[width=1in]{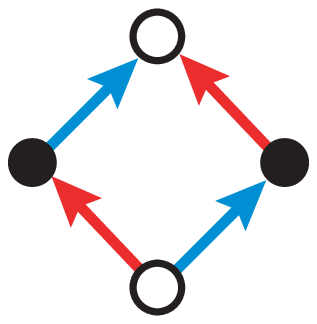}}~. \Label{Adn2scalar}
 \eeq
 The white vertices again represent bosonic fields and the
 black vertices represent fermionic fields. To read off the transformation
 rules associated with this Adinkra, we assign to the lowest and topmost
 bosonic (white) vertices, the names $u$ and $U$, respectively. To the
 left and right fermionic (black) vertices we assign the
 names $\chi_1$ and $\chi_2$, respectively.\Ft{Originally, Adinkras were
 drawn so that arrows point downward\cite{rA}, to mimic the fact that
 in component descriptions of supermultiplets, lowest components are written
 first (and hence, higher on the page).  This led to the unfortunate
 problem that ``higher'' components were lower on the Adinkra, and
 references to the ``lowest'' node could be ambiguous.  Hence, we
 will use the convention that arrows should point upward.  This also
 has the advantage that fields of higher engineering dimension are represented higher on the diagram.}

 Each pair of parallel arrows describes one of two supersymmetries,
 one parameterized by $\e^1$ and the other by $\e^2$.
 Then this diagram translates, using the precise rules described in
 Ref.\cite{rA} or by iterating those given in the displays~(\ref{scalar})
 and~(\ref{spinor}), into the following corresponding transformation rules:
 \begin{equation}\begin{aligned}
  \d_Q\6(\e)\,u &= i\,\e^I\,\chi_I~,\\[2mm]
      \d_Q\6(\e)\,\chi_I &=\ve_{IJ}\,\e^J\,U+\e_I\,\dot{u}~,\\[2mm]
      \d_Q\6(\e)\,U &= -i\,\ve^{IJ}\,\e_I\,\dot{\chi}_J~,
\end{aligned} \Label{n2scalar}
 \end{equation}
 where $I=1,2$, the two-dimensional anti-symmetric Levi-Civita symbol $\ve_{IJ}$ satisfies
 $\ve_{12}=\ve^{12}=1=-\ve_{21}= -\ve^{21}$, and a dot represents a
 time derivative,
 \eg, $\dot{u}:=\vd_\t\,u$. Notice that $\{Q_I,Q_J\}=0$ for $I\neq J$
 implies that the graph~(\ref{Adn2scalar}) must be understood as an
 {\em anti-commutative} diagram: both successive operations, $Q_1Q_2$ and
 $Q_2Q_1$ will transform the field represented by the bottom white circle
 into the field represented by the top one, but there will be a relative
 sign difference in the results. This is equivalent to the observation
 made in Ref.\cite{rA}, that product of all {\em signs} in the
 transformation rules~(\ref{n2scalar}) around square~(\ref{Adn2scalar})
 must be $-1$.

Different $N=2$ multiplets correspond to different choices of arrow directions on square-shaped Adinkras similar to the one shown above.
Various ``duality" maps, inter-connecting the distinct $N=2$ multiplets,
can be described in terms of arrow reversals and global exchanges of white vertices with black vertices; the latter are dubbed ``Klein flips'' (see
Ref.\cite{rGK}). These operations have been explained in a number of previous papers\cite{rGR0,rGR1,rGR2,rA}. It is well-known that similar duality maps can be implemented via differential operations on superfields. In the following section we introduce a graph-theoretic context for the duality operations described above, enabling a more precise correspondence between these Adinkra mutations, to be followed by their superspace analogues.

 Adinkras can be constructed for arbitrary $N$ by iterating the
 above procedure.
 However, in cases where $N$ is larger than 3, more compact
 diagrammatic rules are needed to render the diagrams
 comprehensively in a two-dimensional medium.  There are different
 ways of accomplishing this.  One methodology, espoused in
 Ref.\cite{rA}, was to consider ``folding" operations,
 which combines those vertices whose adjacent edges possess
 the same arrows.
 This system allows one to identify an interesting index
 associated with Adinkras, given by the minimal number of
 dimensions spanned by a ``fully folded" Adinkra. The Adinkras
 which cannot be folded into a linear form Ref.\cite{rA} calls
 ``Escheric", for reasons explained there.
 The fully foldable Adinkras, which are {\em not Escheric}, are the
 subject of our present study.  For reasons explained more fully below,
 these fully foldable, non-Escheric Adinkras are also called ``engineerable".

In a fairly obvious sense, the folding of Ref.\cite{rA} ultimately results in a maximal ``compression'' of each Adinkra. At times, this may not be desirable, as it obscures a possibly useful level of detail, and we briefly digress to describe another, intermediate, option. Recall that the super-multiplets that we are discussing, in 1-dimensional spacetime, may well
have been obtained by dimensional reduction from a $d$-dimensional spacetime.
The various component fields in a $\Spin(1,d{-}1)$ super-multiplet thus become represented by corresponding sets of white or black vertices in the Adinkra. Coalescing each such set of vertices into a single corresponding vertex, we obtain an Adinkra in which the vertices represent
$\Spin(1,d{-}1)$ irreducible representations. Alternatively, one can in the same manner preserve only the massless or the massive little group,
$\SO(d{-}2)$ or $\SO(d{-}1)$, respectively, or indeed any other symmetry group of interest. Indeed, similar graphs have appeared in the literature\cite{rFre}, but have been neither formalized nor used consistently. Such variations of these graphs are called equivariant Adinkras, and will be discussed in a separate effort\cite{r6-1}.

\section{A Graph-Theoretic Description of Adinkras}
 \Label{Graph}
 In the language of graph theory, an Adinkra is, in fact, a
 finite, directed, vertex-bipartite, edge-$N$-partite graph.
 For the benefit of readers less versed in graph theory,
 this terminology can be understood as follows\cite{rG1,rG2}:

 A {\it finite graph} $(V,E,I)$ is a finite set of ``vertices" $V$,
 a finite set of ``edges" $E$, and an incidence function $I$ which maps
 each edge to an unordered pair of vertices, $\{v,w\}$, where $v\in V$
 and $w\in V$.

 A finite graph is {\it directed} if the incidence function
 $I : E \to V \times V$ maps each edge to an {\em ordered} pair of vertices.
 In other words, each edge is endowed
 with a direction, such that the edge {\em points} ``from" one
 incident vertex (the {\em source of that edge}),
 ``to" the other incident vertex (the {\em target of that edge}).
 More specifically, for each edge $e\in E$, the incidence function $I(e)=(v,w)$
 designates that this edge is directed from the vertex $v$ to the vertex $w$.

 A finite graph is {\it bipartite} (or alternatively {\it vertex-bipartite} to agree with our terminology edge-$N$-partite) if its vertices are
 partitioned into two disjoint sets $V_0$ and $V_1$, such that
 every edge is incident with one vertex in $V_0$ and one vertex in
 $V_1$. For our purposes, we call the vertices in $V_{0}$ bosons and the vertices in $V_{1}$ fermions.
 We observe that a
 bipartite graph has the feature that no
 edge can be incident with a given vertex twice.

 \begin{definition}
 A finite graph is {\it edge-$N$-partite} if its edges are partitioned into $N$
 disjoint sets $E_1,...,E_N$, such that each vertex is incident
 with precisely one edge in each $E_i$.
 \end{definition}

 We observe that if we count an edge pointing from a vertex back to itself as being incident to that vertex twice, then this
 property  likewise eliminates graphs with such edges.

 \begin{definition}
 An {\em Adinkra} is a finite, directed, bipartite, and
 edge-$N$-partite graph, that has an {\em edge-parity} assignment
 $\pi: E\rightarrow{\mathbb Z}_2$.
 \end{definition}

If we ignore the edge-parity assignment, the directedness of the edges, the
bipartitioning of the vertices, and the partitioning of the edges, we are
left with an ordinary finite graph.  This graph will be called the
{\em topology} of the Adinkra.  Of course, the fact that there was a
bipartitioning of the vertices and partitioning of the edges means that not
all finite graphs can be a valid Adinkra topology.

The most important topology is the {\em cubical topology} (or more specifically
the $N$-cubical topology), which is the topology obtained by the vertices and
edges of the cube $[0,1]^N$.  This is the main case studied in Ref.~\cite{rA},
though also mentioned there in the case $N=4$ is the dimensional reduction of
the $d=4$, $N=1$ chiral superfield, whose topology is the result of taking the
4-cubical graph and identifying opposite nodes.  Other topologies are possible as well for higher $N$.

 \begin{definition}
 Given two vertices $a$ and $b$,
 a {\em path} from $a$ to $b$ is a finite sequence of edges
 $e_1, \dots, e_m$ and a finite sequence of vertices $v_0, \dots, v_m$
 such that $v_0=a$, $v_m=b$ and, for each $i$, $e_i$ is incident with
 $v_{i-1}$ and also with $v_i$.
 We call the integer $m$ the {\em length} of the path.
 A path connecting two vertices is {\em minimal} if no shorter path,
 \ie, a path having a smaller length, exists.
 \end{definition}
 \Remk
 A path of length $m=0$ is a trivial, {\em empty} path consisting of
 one vertex and no edges. A path in a directed graph need not follow
 the direction of the arrows.

 \begin{definition}
 If two vertices, $v$ and $w$, are connected by a path, the {\em distance}
 between them, $\dist(v,w)$, is the length of a minimal path that
 connects $v$ to $w$; otherwise, $\dist(v,w)=\infty$.
 The relation $dist(v,w)<\infty$ on vertices of the graph is an
 equivalence relation. The equivalence class of vertices, together
 with the edges that connect them in the equivalence class, is
 called a {\em connected component} of the graph.
 \Label{defdist}
 \end{definition}
 \Remk
 The function $\dist(v,w)$ defines a metric on
 the set of vertices of each connected component of the graph.
 If $v$ and $w$ are vertices in the same connected component,
 then a minimal path exists.  The minimal path from any vertex to
 itself is the trivial path, which has no edges, so $\dist(v,v)=0$.

 \begin{definition}
 Given a path having edges $e_1, \dots, e_m$, the {\em net ascent} of
 this path is the
 number of $e_i$ directed along the path minus
 the number of $e_i$ directed against the path.
 If the length of the path is zero, so is the net ascent.
 \end{definition}

 \begin{definition}
 A {\em target} is a vertex such that every edge incident with it is
 directed toward it.  A {\em source} is a vertex such that every edge
 incident with it is directed away from it.
 \end{definition}

 \begin{definition}
 \Label{hgtadd}
 A {\em height assignment} of a directed graph $(V,E,I)$ is a map
 $\hgt:V\to \bz$ so that for every edge going from vertex $a$ to
 vertex $b$, $\hgt(b)=\hgt(a)+1$.
 \end{definition}
 \Remk
 While it is natural to consider integral increments for a height function,
 the corresponding physical concept of {\em engineering dimension} is
 the half-integral $\inv2\hgt$, plus a possible constant. This agrees
 with the unfortunate but well-entrenched discrepancy between
 half-integral  and integral weights used in  physics and mathematics,
 respectively.

 A finite directed graph with a height assignment necessarily has at
 least one vertex of maximal height and at least one vertex of minimal
 height. Such vertices are targets and sources, respectively.
 
 \begin{definition}
 A directed graph is {\em engineerable} if, given vertices $a$
 and $b$, any two paths connecting $a$ and $b$ have the same net
 ascent.
 \end{definition}

 Note that a directed graph is engineerable if and only if every closed
 path (for which $v_0=v_m$) has net ascent zero.

 \begin{proposition}
 If $(V,E,I)$ is a directed graph, then it is engineerable if and
 only if there exists a height assignment for $(V,E,I)$.
 \end{proposition}

 \begin{proof}
First note that these properties are preserved under disjoint
union, and therefore it suffices to prove this for connected
graphs.

Suppose $\hgt$ is a height assignment for $(V,E,I)$.  Let $a$ and
$b$ be vertices, and consider a path involving a sequence of edges
$e_1, \dots, e_m$ and a sequence of vertices $v_0, \dots, v_m$,
with $a=v_0$ and $b=v_m$.  For each $i$, we note that
$\hgt(v_i)-\hgt(v_{i-1})$ is $+1$ if the edge is directed along
the path, and $-1$ if the edge is directed against the path.
Adding these up, we see that $\hgt(b)-\hgt(a)$ is the net ascent
along this path, and thus the net ascent is independent of the path.
Thus, the graph is engineerable.

Conversely, if the graph is engineerable, pick a vertex $v\in V$.
For every vertex $w\in V$, define $\hgt(w)$ to be the net ascent
of a path that connects $v$ to $w$.  This is well-defined because
the graph is engineerable.  If $e$ goes from $a$ to $b$, then take
any path $P_1$ from $v$ to $a$, and append $e$ and $b$ to it to
form the path $P_2$.  Now the net ascent of $P_2$ is one more than
the net ascent of $P_1$, and therefore $\hgt(b)=\hgt(a)+1$.
\end{proof}

 The difference between any two distinct height assignments for a given
 engineerable graph is a function that is constant on each connected
 component. For the case of bipartite graphs we can use this freedom to
 ensure that the height of every boson is even and the height of
 every fermion is odd.
 To do this, it suffices to choose, from each
 connected component, a single vertex $v$, and add a constant to
 that connected component to ensure that $\hgt(v)$ is even if $v$
 is a boson, and that $\hgt(v)$ is odd if $v$ is a fermion.  Then, since
 every edge connects a boson with a fermion, and also connects a
 vertex with even height to a vertex with odd height, by induction
 we can show that $\hgt$ is even on bosons and odd on fermions.

We conclude this section with a proposition which limits the
values of heights on targets, since this condition will be needed in
in Section~\ref{sHGT} below.

\begin{proposition}
Suppose we have an engineerable Adinkra with height function $\hgt$,
and suppose $s_1$ and $s_2$ are either both targets or both sources.  Then
\beq
\dist(s_1,s_2)>|\,\hgt(s_1)-\hgt(s_2)\,|.\Label{hgtcond}
\eeq
\end{proposition}

\begin{proof}
To prove
\beq
     \dist(s_1,s_2)\ge |\,\hgt(s_1)-\hgt(s_2)\,|
\eeq
consider a minimal path from $s_1$ to $s_2$ in the Adinkra, and let
the sequence of vertices in this path be $s_1=v_0, v_1, \dots, v_m=s_2$.
Then $|\,\hgt(v_i)-\hgt(v_{i+1})\,|=1$, and when we take these for all
$i$ from $0$ to $m-1$, and add, the triangle inequality for absolute
values gives the above inequality.

To prove this inequality must be strict, note that if equality holds,
then all $\hgt(v_i)-\hgt(v_{i+1})$ must be the same, either $1$ or $-1$.
For this to be the case, the arrows must either all point along the path,
or all point against the path.  Thus, $s_{1}$ and $s_{2}$ can be neither
both targets nor both sources.
\end{proof}

\section{The `Hanging Gardens' Theorem}
 \Label{sHGT}
 In this section we will prove that a given engineerable Adinkra
 is determined uniquely by specifying its (1)~topology, (2)~bipartition
 into bosons and fermions,
 (3)~which vertices are targets, and
 (4)~the height (the value of $\hgt$) for each of these targets.  This
 theorem suggests a usefully intuitive way to envision engineerable
 Adinkras, whereby the Adinkra is imagined as a collection of weighted
 balls corresponding to the vertices, connected by segments of
 string\Ft{These strings are not to be confused
 with fundamental strings, the putative ultimate essential stuff
 of the universe; instead, they represent the supersymmetry action on
 the component fields represented by the balls within the supermultiplet,
 represented by this macram\'e-like depiction of the Adinkra.}
 corresponding to the edges.
 The theorem can be visualized by suspending those balls which correspond
 to targets from hooks at particular heights, and allowing the rest of
 the balls to
 ``hang" downward under the influence
 of a ``gravitational pull", but are kept in place by the strings.
 Naturally, a number of balls will turn out to be ``locally lowest'' in the
 sense that the
 strings attached to them link only upwards; these balls correspond to
 the sources.
 In this picture, each Adinkra is akin to a unique macram\'e-like
 construction, which, owing to the connection between Adinkras and
 superalgebras, and, in turn, between one-dimensional superalgebras
 and the ${\cal GR}(d,N)$ algebras of Ref.\cite{rGLP}, we call a
 ``hanging ${\cal GR}(d,N)$'' or a ``hanging garden''.

 In Ref.\cite{rA}, Adinkras were represented as cubical
 lattices, with lattice points connected by directed arrows.
 In that paper, a nexus of maps between Adinkras was described
 by certain operations described as AD maps, and pictured in terms
 of Adinkra ``folding" operations.  The macram\'e
 operations described in this section correspond precisely to
 that set of AD maps which preserve the engineerability of the
 Adinkras. A hanging garden is thus identical to a fully unfolded Adinkra,
 as defined in Ref.\cite{rA}.
 However, in step with the physical image of a hanging garden, we depict
 Adinkras so that the value of the $\hgt$ function at each vertex
 corresponds to the physical height of its placement. This permits us to
 dispense with the arrows on the edges; the Reader is welcome to reinsert
 them: they all point upwards. An example is presented in Fig.~\ref{garden1}.
 \begin{figure}[ht]
 \begin{center}
  \includegraphics[width=1.5in]{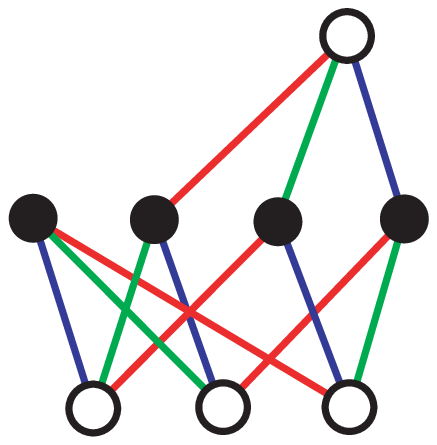}\qquad\quad
  \includegraphics[width=1.5in]{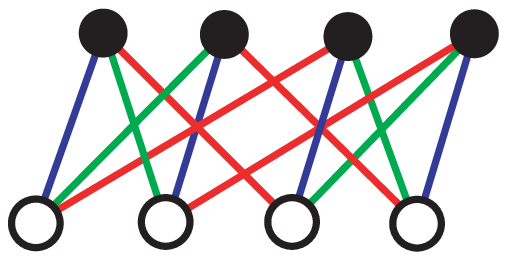}\qquad\quad
  \includegraphics[width=1.5in]{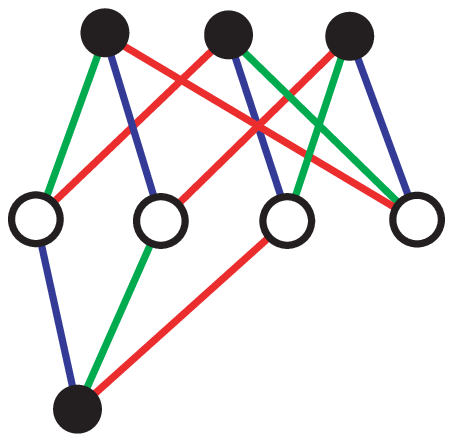}
 \caption{Three examples of $N=3$ hanging gardens.
 The bipartite condition corresponds to the fact that vertices are either
 white (bosons) or black (fermions). The edge-3-partite condition is
 illustrated by coloring the edges.
 The directed condition of engineerable Adinkras is implicitly depicted by
 orienting them so that arrows always point upward. This understood,
 we omit the arrows.}
 \Label{garden1}
 \end{center}
 \end{figure}

 Let us now restate our theorem in a  more formal manner, and then
 prove it:

 \begin{theorem}\Label{tDetA}
Suppose we are given (1)~a topology of an Adinkra (that is, a graph that could be the
 underlying graph of an Adinkra), (2)~a bipartitioning of the vertices into bosons and fermions, (3)~a subset $S$ of the set of vertices of
 this graph, which consists of at least one vertex from each connected
 component of the graph, consisting of what we will call {\em targets}, and
 (4)~a function $h$ from $S$ to $\mathbb{Z}$ (intended to be a height assignment
 restricted to the set of targets) with the following properties:
 \begin{enumerate}
 \item That $h$ applied to bosons is even,
 \item That $h$ applied to fermions is odd, and
 \item That for every pair of distinct elements $s_1$ and $s_2$ of $S$,
 \beq
 \dist(\,s_1~,\,s_2\,)>|\,h(s_1)-h(s_2)\,|,\Label{crith}
 \eeq
 which is the condition given by (\ref{hgtcond}).
 \end{enumerate}
 Then there exists an engineerable Adinkra which has the given topology, whose set
 of targets is $S$, and which has a height assignment $\hgt$ which is an extension
 of $h$ to the set of all vertices.  Furthermore, this is the unique Adinkra with
 this topology, this set of targets, and the given values of $\hgt$ on these targets.
 \end{theorem}

 \begin{proof}
 Let $(V,E,I)$ be a bipartite graph, and let $S\subset V$.
 Suppose we are given a map $h:S\to{\mathbb Z}$ whereby
 $h$ applied to bosons is even and $h$ applied to fermions is odd, and satisfying
 condition~(\ref{crith}) for each possible choice $s_1, s_2\in S$ with $s_1\not=s_2$.

 We may we define $\hgt$ as follows:
 \beq
 \hgt(v)=\max_{s\in S}\bigl(\,h(s)-\dist(v,s)\,\bigr)~.
 \Label{hgtdef}
 \eeq
 We will demonstrate presently that the function $\hgt$, as defined
 in~(\ref{hgtdef}) meets
 the criterion for a height assignment, as given in Definition
 \ref{hgtadd}.

 Between two bosons or two fermions, $\dist$ is even,
 and between a boson and a fermion, $\dist$ is odd.  Thus,
 since $h$ sends bosons to even integers and fermions to
 odd integers, owing to the definition~(\ref{hgtdef}),
 the map $\hgt$ also sends bosons to even integers and
 fermions to odd integers.

 Since $(V,E,I)$ is bipartite, two vertices $v$ and $w$, connected
 by an edge $e$, cannot be of the same type; that is if $v$ is a boson
 then $w$ must be a fermion, and vice-versa.  Thus, based on the
 conclusion in the preceding paragraph, $\hgt(v)$ and
 $\hgt(w)$ cannot be equivalent modulo two, and are therefore
 unequal.  Without loss of generality we choose
 $\hgt(v) < \hgt(w)$.

 Let $s\in S$ be such that $h(s)-\dist(v,s)$ is maximal, and
 let $t\in S$ be such that $h(t)-\dist(w,t)$ is maximal.

 Let $P$ be a minimal path from $t$ to $w$, and let $P'$ be the
 path from $t$ to $v$ obtained by appending $e$ to the terminus of
 $P$.  The length of $P'$, given by $\dist(t,w)+1$, must be
 at least as large as $\dist(t,v)$.  Thus:
\begin{equation}
 \begin{split} \hgt(v) &= h(s)-\dist(v,s)~,\\[2mm]
      &\ge h(t)-\dist(v,t)~,\\[2mm]
      &\ge h(t)-(\,\dist(t,w)+1\,)~,\\[2mm]
      &= \bigl(\,h(t)-\dist(t,w)\,\bigr)-1~,\\[2mm]
      &= \hgt(w)-1~.
 \end{split}
 \Label{hder}
\end{equation}
 Here, the first line follows from~(\ref{hgtdef}) because $s$
 is, by definition, a vertex which maximizes this quantity.
 The second line follows for the same reason. We pass to the
 third line using the aforementioned result
 $\dist(t,w)+1\ge \dist(t,v)$.  We pass to the fourth
 line by rearranging terms, and
 we pass to the final line using~(\ref{hgtdef}).

 Since the function $\hgt$ takes values in ${\mathbb Z}$, and since we have
 determined in~(\ref{hder}) that $\hgt(v)\ge \hgt(w)-1$, it
 follows that the equality holds, {\rm i.e.}, that
 \brr \hgt(v)=\hgt(w)-1~.
 \Label{cone}\err
 If we choose the direction of edge $e$ from $v$ to $w$,
 and if this procedure is applied to all edges, then~(\ref{cone})
 satisfies the criterion, given in Definition
 \ref{hgtadd}, needed to verify that the function $\hgt$, as defined in~(\ref{hgtdef})
 is, in fact, a height assignment.

 Now consider a vertex $s\in S$.  Let $t$ be such that
 $h(t)-\dist(t,s)$ is maximal.  Then
 $h(t)-\dist(t,s)\ge h(s)-\dist(s,s)=h(s)$.
 Thus, $|\,h(t)-h(s)\,|\ge \dist(t,s)$, in
 violation of criterion~(\ref{crith}), unless
 $s=t$.  It follows that $s$ {\it is} the unique element $t$ of
 $S$ that maximizes $h(t)-\dist(t,s)$, and
 therefore that $\hgt(s)=h(s)$, for any $s\in S$.

 Now consider a vertex $s\in S$, and suppose it is not a target.
 Then there exists an edge $e$ directed from $s$ to another vertex $w$,
 and $\hgt(w)=\hgt(s)+1$.  Let $t\in S$ be such that $h(t)-\dist(t,w)$ is
 maximal, and thus equal to $\hgt(w)=\hgt(s)+1$.  The previous paragraph
 proves that $h(t)-\dist(t,s) < \hgt(s)$, and, putting this all together, we
 get $\dist(t,w)+1 < \dist(t,s)$.  On the other hand, since we can take a
 minimal path from $t$ to $w$ and append the edge $e$, we have
 $\dist(t,s)\le \dist(t,w)+1$.  This is a contradiction, and thus,
 every element of $S$ is a target.

Let $v$ be any vertex and $s\in S$ be such that $h(s)-\dist(v,s)$
is maximal.  Let $P$ be a minimal path joining $s$ to $v$.  Let
$e$ be the penultimate edge of $P$ and $w$ the penultimate vertex
of $P$.  Let $P'$ be the path resulting from deleting the last
edge and vertex from $P$.  Now $P'$ must be a minimal path joining
$s$ to $w$, or else $P$ would not be minimal.  Thus,
$\dist(w,s)=\dist(v,s)-1$, and therefore $\hgt(v)=\hgt(w)-1$.
Proceeding likewise, we see that $P$ consists only of edges
directed against the path.

To prove that every target is an element of $S$, consider a vertex
$v\not\in S$.  Let $s\in S$ and $P$ be a path as in the previous
paragraph.  Then $P$ will be a path that ends in an edge directed
away from $v$, so that $v$ is not a target.

To show that all directed engineerable Adinkras arise in this way,
suppose $(V,E,I)$ is directed and engineerable.  Then there exists
a height assignment $\hgt:V\to \bz$, and a set of targets $S$.
Note that every connected component of the graph contains at least
one target.  Define $h:S\to\bz$ to be the restriction of $\hgt$ to
$S$, and let
\begin{equation}
\hgt'(v)=\max_{s\in S} \bigl(\,h(s)-\dist(s,v)\,\bigr).\Label{ehgtfromh}
\end{equation}
First consider two targets $s_1, s_2\in S$, and without loss of
generality assume $\hgt(s_1)\le \hgt(s_2)$.  Consider a minimal
path $P$ from $s_1$ to $s_2$.  Let $u$ be the number of edges in
$P$ directed along the path and $d$ the number of edges of $P$
directed against the path (in the original graph, not the one
constructed with $\hgt'$).  Then $|h(s_2)-h(s_1)|=u-d\le u+d =
\dist(s_1,s_2)$, where equality can only happen if $d=0$.  But
that requires that the first edge in the path $P$ must go away
from $s_1$, and thus that $s_1$ is not a target.  Therefore
\begin{equation}
|h(s_2)-h(s_1)|<\dist(s_1,s_2).
\end{equation}

Now we wish to show that $\hgt=\hgt'$.
For every vertex $v\in V$, if it is a target, then
$\hgt'(v)=h(v)=\hgt(v)$.  If it is not a target, construct a path
$P$ from $v$, along edges directed along the
path, until no such edges are available at the current vertex
(\ie, until a target $s$ is reached).  This process is finite
because the graph is finite and $\hgt$ increases at each step.
Then the length of $P$ is $\hgt(s)-\hgt(v)$.  Since $(V,E,I)$ is
engineerable, all other paths from $v$ to $s$ must have the same
net ascent, and, since $P$ has only edges directed along the path,
all other paths from $v$ to $s$ must be at least as long.  Thus,
$\dist(s,v)$ is the length of $P$, which is
$\hgt(s)-\hgt(v)=h(s)-\hgt(v)$.  Thus,
\begin{equation}
\hgt(v)=h(s)-\dist(s,v)\le \hgt'(v).
\end{equation}
On the other hand, let $t\in S$ be such that $h(t)-\dist(t,v)$ is
maximal, and let $Q$ be a minimal path from $v$ to $t$.  Let $u$
be the number of edges in $Q$ that go along the path and let $d$
be the number that go against it (in the original directed graph,
not the one constructed with $\hgt'$).  Then $\dist(t,v)=u+d$ and
$\hgt(t)-\hgt(v)=u-d$, and
\begin{equation}
\hgt'(v)=\hgt(t)-\dist(t,v)=\hgt(v)-2d\le \hgt(v).
\end{equation}
Therefore $\hgt(v)=\hgt'(v)$.

Now suppose we have two engineerable bipartite directed graphs
that have the same topology,
and suppose they have the same set of targets $S$, and initial height
function $h:S\to \bz$. We can use this procedure to obtain a height
function $\hgt'$ (obviously the same in both cases) which is a
height assignment for both directed graphs, and thus the directed
graphs must be equal.
\end{proof}

A virtually identical proof establishes:
\begin{corollary}\Label{cDetA}
Suppose we are given (1)~a topology of an Adinkra,
(2)~a bipartitioning of the vertices into bosons and fermions, (3)~a subset $S$ of the set of vertices of
 this graph, which consists of at least one vertex from each connected
 component of the graph, and
 (4)~a function $h$ from $S$ to $\mathbb{Z}$ with the following properties:
 \begin{enumerate}
 \item That $h$ applied to bosons is even,
 \item That $h$ applied to fermions is odd, and
 \item That for every pair of distinct elements $s_1$ and $s_2$ of $S$,
 \beq
 \dist(\,s_1,\,s_2\,)>|\,h(s_1)-h(s_2)\,|,\Label{ccrith}
 \eeq
 which is the condition given by (\ref{hgtcond}).
 \end{enumerate}
 Then there exists an engineerable Adinkra which has the given topology, whose set of sources is $S$, and which has a height assignment $\hgt$ which is an extension of $h$ to the set of all vertices.  Furthermore, this is the unique Adinkra with this topology, this set of sources, and the given values of $\hgt$ on these sources.
 \end{corollary}

\section{`Vertex Raising' and `Vertex Lowering' Operations}
 \Label{sVRaising}
 In this section we introduce operations, which we call vertex
 raising and vertex lowering operations, which change the
 height of the placement of some of the vertices in a given
 hanging garden. In the following section we explain how these
 operations generate maps connecting all possible engineerable
 Adinkras of the same topology, starting from any given
 representative.
 This is analogous to the concept of a root-tree espoused in
 Ref.\cite{rA}, which groups supersymmetric multiplets
 {\em via} interconnections generated by transformations
 encoded by Adinkra folding operations.

 Given any bipartite graph $(V,E,I)$, and a vertex $v\in V$,
 we can use the above construction to direct the edges in $E$ so
 that $v$ is the only target.  The result is that the $\hgt$ function
 is determined simply by distance from $v$ (up to a constant
 $\hgt(v)$). Intuitively, this is the result of hanging the graph
 on a single hook at $v$.

 \begin{definition}
 An engineerable directed bipartite graph $(V,E,I)$ is called {\em
 one-hooked} if there is only one target $v\in V$.  In this case,
 the graph is said to be {\em hooked on $v$}.
 \end{definition}

 Given a bipartite graph $(V,E,I)$ and a vertex $v\in V$, the
 graph {\em one-hooked} on $v$ is the unique engineerable directed
 bipartite graph which has as the same topology, but such
 that $v$ is the only target.

 Suppose we have an engineerable directed bipartite graph $(V,E,I)$
 and $v$ is a target.
 We may change the orientation of all the arrows incident with $v$,
 producing a related, engineerable, directed bipartite graph.
 Equivalently, in a hanging garden, we can
 hook all the vertices $w_i$ for which
 $\dist(v,w_i)=1$, and unhook $v$, letting it drop by its weight.
 This operation is called a {\em vertex lowering}.

 This operation affects the height function by reducing the height
 of $v$ by two, turning it into a source,
 and leaving the height of all other vertices unchanged.
 Also, some of the $w_i$'s may become new targets.
 Graphically, in terms of hooks, we are pushing down a local maximum,
 though perhaps creating other local maxima nearby.
 \begin{figure}[ht]
 \begin{center}
 \includegraphics[width=3.5in]{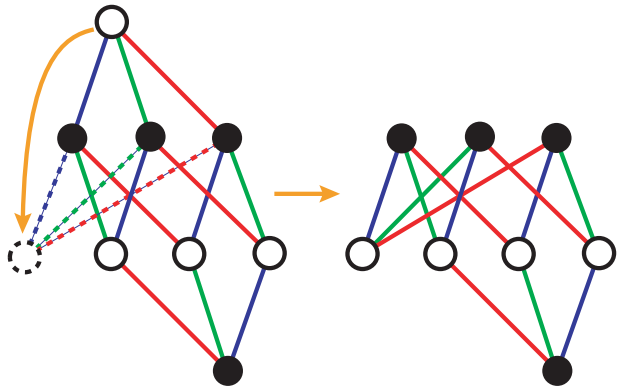}
 \caption{An example of the vertex lowering operation,
 creating a new Adinkra by lowering the single
 white vertex on the top. Notice that all three black vertices that are at
 distance~1 from this vertex have become targets (\ie, are local summits).}
 \Label{garden2}
 \end{center}
 \end{figure}

 There is also the notion of a {\em vertex raising} which can only
 apply to a source $v'$ (\ie, a vertex $v'$ all of whose
incident edges are directed away from $v'$).  The effect of a
vertex raising is to turn $v'$ into a target, to alter $\hgt$
only on $v'$ (increasing it by two), and perhaps to turn some of the $w_i'$
for which $\dist(v',w_i')=1$ into sources; an example of this is shown in Fig.~\ref{garden3}.

 \begin{figure}[ht]
 \begin{center}
 \includegraphics[width=3.5in]{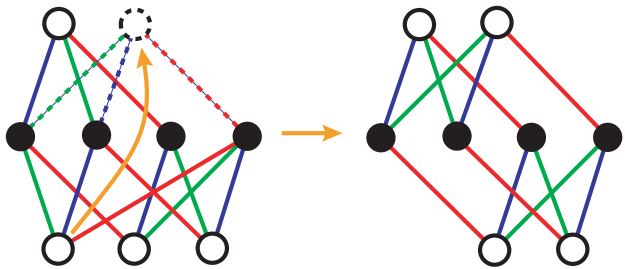}
 \caption{An example of the vertex raising operation, creating another
 Adinkra from one of those shown in Figure~\ref{garden1}, by raising
 one of the lower, white vertices up to the top level. }
 \Label{garden3}
 \end{center}
 \end{figure}

Note that this is a restricted version of one of the two notions
of the so-called {\em automorphic duality} mentioned in Ref.\cite{rA}. We propose a new use of the term automorphic duality,
as follows: Let $(V,E,I)$ be a directed graph, and consider
$(V,E,I')$ the directed graph that results from reversing the
orientations of all the edges.  If $(V,E,I)$ was engineerable before,
with height assignment $\hgt$, then $(V,E,I')$ will be
engineerable, with height assignment $-\hgt$.

We will now see how vertex lowerings and raisings can be used to
relate all engineerable directed bipartite graphs that have
the same topology.

\begin{theorem}\Label{tHGT}
Let $(V,E,I)$ be a finite engineerable directed bipartite graph,
and let $v\in V$ be any vertex.  Then there is a sequence of
vertex lowerings that takes $(V,E,I)$ to a graph of the same
topology, one-hooked on $v$.
\end{theorem}

\begin{proof}
Choose a height assignment $\hgt:V\to \bz$.  Let $S$ be the set of
targets for $(V,E,I)$, and let $S'$ be $S-\{v\}$. (If $v$ is not contained in $S$, then we take $S' = S$.)  If $S'$ is
empty, we are done.  Suppose $S'$ is not empty.  Let $M$ be the
largest value of $\hgt$ restricted to $S'$.  The elements of $S'$ where
$M$ is achieved are targets,
and therefore a vertex lowering is allowed on each.  The result is
a new directed graph with height assignment, but now $M$ is smaller.
Note that $\hgt(v)$ is unchanged via this procedure.

Continue to iterate this procedure.  The process must terminate
since $M$ cannot be less than $\hgt(v)-\max_{w\in V} \dist(v,w)$.
The only way for this to terminate is if $S'$ is empty at some
stage, in which case $v$ is the only target.
\end{proof}

\begin{corollary}\Label{cHGT}
Starting from a one-hooked graph, it is possible to obtain any other
engineerable directed graph of the same topology by a series
of vertex raisings.
\end{corollary}

\begin{proof}
Use the sequence guaranteed in the previous theorem, and operate
them in reverse.
\end{proof}

Consider a one-hooked graph, and reverse the arrows, replacing
$\hgt$ with $-\hgt$.  The result has one source.  There are corresponding
results analogous to the ones above in this situation.

Now for every topology, we consider the following engineerable directed
graph, called the {\em base Adinkra} in Ref.~\cite{rA}, and in the cubical case corresponding to ${\cal GR}(d,N)$ algebras in
Ref.~\cite{rGLP}.  Choose as sources all the bosons, and define
their heights to all be 0.  Equivalently, we can choose as targets  all the fermions, and define their heights to all be 1.  By Theorem~\ref{tDetA} or Corollary~\ref{cDetA}, there is a unique engineerable directed graph with this characterization.

\begin{theorem}
Let $(V,E,I)$ be a finite engineerable directed bipartite graph.  Then there is a sequence of vertex lowerings that takes $(V,E,I)$ to the corresponding base Adinkra of the same topology.  There is also a sequence of vertex raisings that takes $(V,E,I)$ to that base Adinkra.
\end{theorem}

\begin{proof}
Choose a height assignment $\hgt:V\to \bz$.  Let the smallest and largest
values of $\hgt$ on $V$ be denoted $m$ and $M$, respectively.
Since fermions and bosons may not have the same value of $\hgt$, $M\ge m+1$.
Let $S$ be the set where the height $M$ is achieved.

The elements of $S$ are targets, and therefore a vertex lowering is allowed on each.  The result is a new directed graph with height assignment, but now $M$ is smaller because there are no longer any vertices at height $M$.  Note that
if $M\ge m+2$, then the vertices that used to be at level $M$ are at level $M-2$, which is at least $m$.  So in this case $m$ will not change, but $M$
decreases.  Eventually, then, $M=m+1$, and the vertices have two heights: bosons
on one height and fermions on the other.  If the fermions are of height $m$
and bosons of height $M=m+1$, then we iterate this procedure again, and
the fermions will be of height $m+1$ and bosons of height $m$.  This is the base Adinkra.

To find the sequence of vertex raisings, do the above on the automorphic
dual of $(V,E,I)$.  We reverse the automorphic duality on this sequence.
The result describes a sequence of vertex raisings that takes $(V,E,I)$
to the automorphic dual to the base Adinkra, whereby all the fermions
are sources.  If we vertex raise all the fermions once, the result will be
the base Adinkra.
\end{proof}

\begin{corollary}\Label{cHGT2}
Any two engineerable directed bipartite graphs of the same topology
type can be related through a finite sequence of vertex raisings.
\end{corollary}

\begin{proof}
Take the first directed graph and apply the above theorem to find a sequence
of vertex raisings to the base Adinkra.  Take the second directed graph
and apply the above theorem to find a sequence of vertex lowerings to the
base Adinkra, then reverse these operations.  The result is a sequence of
vertex raisings that turn the base Adinkra into the second directed graph.
The composition of the first sequence to this reverse second sequence
will turn the first directed graph into the second.
\end{proof}

\Remk\Label{RAdClass}
Since vertex raising and lowering does not change the topology,
it follows that all the Adinkras which can be obtained one from another
through vertex raising and lowering have the same topology. This
 provides a coarse classification of Adinkras, and
 prompts the following definition:
\begin{definition}\Label{dFam}
 The collection of Adinkras for any given $N$ that have the same
topology, is called a {\em family} of Adinkras; individual Adinkras
within a family are called {\em members} of the family.
 The minimal number of vertex raisings or lowerings that connects two
members in a family is their {\em(kinship) distance}.
 These names extend to the corresponding supermultiplets and superfields.
\end{definition}

\section{Superderivative Superfields and Vertex Raising}
 \Label{sSFnVR}
 The vertex raising and vertex lowering operations described in
 the previous section provide a
 graph-theoretic basis for maps interconnecting
 supermultiplets.  As is well known, there exist established
 superspace methods for accomplishing similar goals.  In this
 section we examine how superderivative operations
 are alternatively described by vertex raises, and how the latter
 can be used to generate a sequence including each irreducible
 supersymmetry representation for a given value of $N$.

\subsection{$(1|N)$ Superspace}
 \Label{s1N}
Superspace generally is the linear space given by $d$ real commuting coordinates $x^0, \dots, x^{d-1}$,
and $N$ real anticommuting coordinates $\q^1, \dots, \q^N$.
As standard\cite{rBK,rDF}, we will call this superspace $\IR^{d|N}$.  In our case, the superspace on which our fields are defined is $\IR^{1|N}$, and we
will sometimes call this $(1|N)$ superspace.  As before, we denote the single time-like coordinate on $\IR^{1}$ by $\tau$.

We will start with two kinds of superfields, called a {\em scalar
superfield} and a {\em spinor superfield}.  A scalar superfield is
a function from superspace to $\IR^{1|0}$, and a spinor superfield is a function from superspace to $\IR^{0|1}$.

More generally, we can consider functions from $\IR^{1|N}$ to $\IR^{M_0|M_1}$,
but these can be thought of as an $(M_0+M_1)$-tuple of superfields: the first $M_0$ of them scalar superfields, and the other $M_1$ of them spinor superfields.

Let $\sF^N_0=C^\infty(\IR^{1|N},\IR^{1|0})$ be the set of scalar superfields
and $\sF^N_1=C^\infty(\IR^{1|N},\IR^{0|1})$ be the set of spinor superfields.

We define the differential operators $Q_I$ on superfields as follows:
 \beq
      Q_I = i\,\vd_I+\,\d_{IK}\q^K\,\vd_\t~,\Label{Qi}
 \eeq
 where $\vd_I:=\vd/\vd\q^I$ are the fermionic derivatives. When acting on
 superfields $\IU(\t;\q^1,\dots,\q^N)\in \sF^N_a$ for $a=0,1$, these superspace differential operators satisfy the
 algebra~(\ref{nalg})--(\ref{nalg2}):
 \begin{align}
  \{Q_I,Q_J\}\,\IU(\t;\q)
  &=\bigl\{\,i\vd_I+\,\d_{IK}\q^K\,\vd_\t\,,
       \,i\vd_J+\,\d_{JL}\q^L\,\vd_\t\,\bigr\}\,\IU(\t;\q)~,\nn\\
  &=+2i\,\d_{IJ}\,\vd_\t\,\IU(\t;\q)~,\Label{eQQonSF}\\[2mm]
  [-i\e_1Q_I\,,\,-i\e_2Q_J]\,\IU(\t;\q)
  &=\e_1\e_2\{Q,Q\}\,\IU(\t;\q)
   = 2i\e_1\e_2\,\d_{IJ}\,\vd_\t\,\IU(\t;\q)~.\Label{eEQEQonSF}
 \end{align}
and therefore $\sF^N_0$ and $\sF^N_1$ are representations of the $(1|N)$ supersymmetry algebra.

 The oppositely twisted\cite{rEWM} superspace derivatives,
 \beq
      D_I = \vd_I+i\,\d_{IK}\,\q^K\,\vd_\t~,\Label{Di}
 \eeq
 anticommute with the $Q_I$, commute with $\e^I Q_I$, and are therefore invariant under supersymmetry.\Ft{Note that the so-defined $D_I$ satisfy
the same algebra~(\ref{eQQonSF}) as the $Q_I$.}

The relevance for us will be that each $D_I$ is a linear operator that maps $\sF^N_0$ to $\sF^N_1$ and vice-versa that is actually a homomorphism of representations of the supersymmetry algebra.

\subsection{The Adinkra of a superfield}
To think of a superfield in Adinkra terms, we need to consider superfields
as a collection of functions of $\t$.  This can be done by examining what are called {\em component fields} of a superfield.

As is well known\cite{r1001,rBK}, a superfield (whether scalar or spinor)
 $\IU(\,\t\,;\,\q^1,...,\q^N\,)$ may be formally expanded over the fermionic coordinates  $\q^I$.
\beq
\IU = \sum_{\stackrel{\{I_1,\dots,I_k\}\subset \{1,\dots,N\}}{\sss
I_1<\dots<I_k}} U_{I_1,\dots,I_k}(\t)\, \q^{I_1}\dots\q^{I_k}.
\Label{eQ*b}
\eeq
Each $U_{I_1,\dots,I_k}(\t)$ is either an $\IR^{1|0}$- or
$\IR^{0|1}$-valued function over $\IR$, and corresponds to a bosonic or fermionic component field, respectively.

Another way to obtain the components of a superfield is to use the {\em invariant projection}\cite{r1001}, and this is how we will define the
components of the superfield:
\begin{definition}[Components]\Label{dComp}
For any subset ${\cI} = \{I_1,\dots,I_k\}$ of $\{1,\dots, N\}$ with
\beq
     I_1 < I_2 < \dots < I_k
\eeq
we define on the space of superfields the superderivative operator
\beq
D_{\cI} := D_{[I_k}\dots D_{I_1]}\Label{eproj}
\eeq
and the {\em projection operator}
\beq
\pi_{\cI} \IU := D_{\cI}\IU \,|
\eeq
where the final $\,|$ means evaluation at $\q^1=\dots=\q^N=0$.

The components of $\IU$ are then
\beq
U_{\cI}:= U_{I_1,\dots,I_k}:= \pi_{\cI}\IU\,|.
\eeq

\end{definition}
 \Remk
 Note that the $D_I$'s in the superderivative operator $D_{\cI}$~(\ref{eproj}) occur in decreasing order in $I$, for convenience in computation, since the $\q^I$'s are in increasing order in $I$ in the component expansion (\ref{eQ*b}). Furthermore, this projection method applies to all expressions and equations involving superfields, and is the only method of obtaining component-level information within this formalism; see also
 Appendix~\ref{sRSF}.

Now it is clear that these are all of the components of the superfield
$\IU$.  To find the corresponding Adinkra, for every subset
${\cI}\subset \{1,\dots,N\}$, we place a node corresponding to $U_{\cI}$\
in $\IR^N$ at $(y_1,\dots,y_N)$, where for all $I$, $y_I=1$ if $I\in {\cI}$, and $y_I=0$ if $I\not\in {\cI}$.  The node is bosonic if the superfield is a scalar superfield and the number of elements of ${\cI}$ is even, or if the superfield is a spinor superfield and the number of elements of ${\cI}$
is odd.  It is fermionic otherwise.

We then examine the effect of $Q_I$. From the perspective of (\ref{eQ*b}) it is clear that $Q_I$ takes components without $\q^I$ and differentiates them while putting these into components with $\q^I$, and takes components with $\q^I$ and sends them to components without $\q^I$.  It thus connects vertices which differ only in the $I$th component, and draws an arrow from $(y_1,\dots,y_{I-1},0,y_{I+1},\dots,y_N)$ to $(y_1,\dots,y_{I-1},1,y_{I+1},\dots,y_N)$.  The sign is taken as plus or minus depending on the parity of $I$.

Therefore, the Adinkra for the superfield representation is an $N$-dimensional cubical Adinkra, with one source (the $U_\emptyset$ component) which is bosonic if and only if the superfield is a scalar superfield, and one target (the $U_{1,\dots,N}$ component), which has the same statistics as the source node if $N$ is even, and the opposite if $N$ is odd.

It will be convenient to define a few standard set-theoretic notational
conventions.  Given a finite set $\cI$, we denote the number of elements
of $\cI$ as $\#{\cI}$.  Given two sets $\cI$ and $\cJ$, the
{\em symmetric difference} ${\cI}\D{\cJ}$ is the set of elements that
are in $\cI$ or $\cJ$ but not both.

We can put together a height assignment and distance function.  It is
particularly convenient to find the height assignment because superfields
have engineering degrees.  Since a height assignment is supposed to be twice
the engineering degree, up to an additive constant, and since each $D_I$ has
engineering degree $1/2$, we can define a height assignment as follows:
\begin{definition}
For every component field $U_{\cI}$ of the unconstrained
superfield $\IU$, define
\beq
\hgt_0(U_{\cI})=\#{\cI}.
\eeq
Given two component fields $U_{\cI}$ and $U_{\cJ}$, define
\beq
\dist_0(U_{\cI},U_{\cJ})
=\#({\cI}\D{\cJ}).
\eeq
\Label{defsfhgt}
\end{definition}
It is straightforward to show that $\hgt_0$ satisfies the definition of a height assignment as given in Definition~\ref{hgtadd} in Section~\ref{Graph}, and that $\dist_0$ coincides with Definition~\ref{defdist}.

\Remk Note that these are the only Adinkras with one source, which is easy to see by the transitive symmetry of the $N$-cube.

In the remainder of this paper, we will show how, given a cubical Adinkra,
we can recreate the supermultiplet by applying constraints on $M$-tuples
of superfields.  Although we do not have an algorithm to deal with
Adinkras that are quotients of cubes, we note that if we have an Adinkra
with such a topology, we should look at all the one-source Adinkras that
can be made from such a topology, and if we can construct these from
constraining superfields, then the remainder of this paper suffices to
explain how the supermultiplet for the given Adinkra could be constructed from
constraining these superfields further.

\subsection{The case $N=1$}
 \Label{sN=1}
For $N=1$, superspace is determined by its coordinates $(\t,\q)$.  The
 supercharge operator and the superspace derivative
 are given, respectively, by
 \begin{align}
 Q &= i\vd_\q+\,\q\,\vd_\t~,\Label{Q1}\\[2mm]
      D &= \vd_\q+i\,\q\,\vd_\t~\Label{D1}.
 \end{align}

\subsubsection{$N=1$ Superfields}
 There are two distinct $(1|1)$ superfields: a scalar superfield $\F$, and a spinor superfield $\L$, with component fields defined by projection\cite{r1001}:
 \begin{align}
   \f&:=\F|~,\qquad i\j~:=~D\F|~;\Label{cPhi}\\[2mm]
   \l&:=\L|~,\qquad \>B~:=~D\L|~.\Label{cLam}
 \end{align}
The judicious factor of $i$ in the definitions~(\ref{cPhi})--(\ref{cLam}) ensure that the component fields $\f,\j,\l,B$ are all real.  One may also reassemble the component fields into the $\q$-expansions:
 \begin{align}
  \F &= \f+i\,\q\,\j~,\Label{n1sSF}\\[2mm]
  \L &= \l+\q\,B~. \Label{n1fSF}
 \end{align}

 The supersymmetry transformation rules on component fields are extracted by
 projecting the component equations of
  $\d_Q\6(\e)\,\F=-i\e\,Q\,\F$ and $\d_Q\6(\e)\,\L=-i\e\,Q\,\L$, and
 are shown here together with the corresponding Adinkras:
 \beq
  \left.\begin{aligned}
  \d_Q\,\f &= i\,\e\,\j\\[2mm]
  \d_Q\,\j &=\e\,\dot{\f}
  \end{aligned}\right\}\qquad\Longleftrightarrow\qquad
  \vC{\includegraphics[width=.25in]{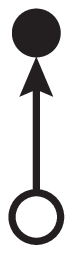}}~,
  \Label{phi}
 \eeq
and
 \beq
  \left.\begin{aligned}
  \d_Q\,\l &= \e\,B\\[2mm]
  \d_Q\, B &= i\,\e\,\dot{\l}
  \end{aligned}\right\}\qquad\Longleftrightarrow\qquad
  \vC{\includegraphics[width=.25in]{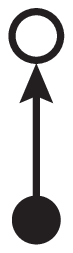}}~.
  \Label{lam}
 \eeq
 These transformation rules are of course identical to~(\ref{scalar})
 and~(\ref{spinor}), respectively.

 Note that the scalar and spinor superfields defined above can be
 defined in superspace modulo an overall multiplicative phase factor.
 For instance, the definitions~(\ref{cPhi})--(\ref{cLam}) and the expansions~(\ref{n1sSF})--(\ref{n1fSF}) may be generalized into:
 \begin{align}
  \F_\a &=
      e^{i\,\a}\,\bigl(\,\f+i\,\q\,\j\,\bigr)~,\Label{n1sSFa}\\[2mm]
  \L_\a &= e^{i\,\a}\,\bigl(\,\l+\q\,B\,\bigr)~.\Label{n1fSFa}
 \end{align}
 The original definitions correspond to the choice $\a=0$.
 The supersymmetry transformation rules for the component fields, extracted
 as above, are independent of the constant $\a$, whence
 the members of such 1-parameter families of superfields are considered
 equivalent. Regardless of the value of $\a$, it is possible to choose $\f$
 and $B$ to be real bosons, and $\j$ and $\l$ to be real fermions, so that
 $\F_\a$ and $\L_\a$ may be regarded as real superfields.
 Thus, in all depictions of the representation of supersymmetry---by component
 fields, by Adinkras, or as superfields---the constant $\a$ is irrelevant.
 Similarly, we can redefine the sign of each component field separately,
 inducing appropriate sign changes in the component transformation rules,
 but without changing their overall structure. Finally,
 we may specify a superfield as an ordered sequence of its component
 fields, listing them by non-decreasing engineering dimensions, with
 groups of equal engineering dimension and statistics separated by
 semicolons, as in $\F=(\f;\j)$ and $\L=(\l;B)$, understanding that the
 component fields are functions of time.

\subsubsection{$N=1$ Superderivative Superfields}
 \Label{sN1SS}
 The superderivative operator $D$ induces maps on the space of
 superfields.  It is instructive to interpret these in terms of
 Adinkra operations.  For instance, consider the following map,
 applied to the scalar multiplet $\F_0$,
 \beq
      D: \F_0 \to (D\,\F_0) ~=~i(\j;\dot{\f})~.
 \Label{dphi}
 \eeq
 The image of this map is a spinor superfield, since its lowest component,
 $i\j$, is a spinor. It is akin to $\L_{\pi/2}$ described in~(\ref{n1fSFa}),
 and we denote this superfield as $\tilde\L_{\p/2}$. The identifications are:
 \beq
      (D\F_0)=:\tilde\L_{\p/2}~,\qquad
      \left\{\begin{aligned}
              \tilde{B}&:=\dot\f~,\\
              \tilde\l &:=\j~.
             \end{aligned}\right.
 \Label{DPhiLam}
 \eeq
 The phase-shift in the $\a$ phase being irrelevant to the component
 fields and the Adinkras, we can represent equation~(\ref{dphi})
 symbolically as follows,
 \beq
  \vC{\includegraphics[width=0.8in]{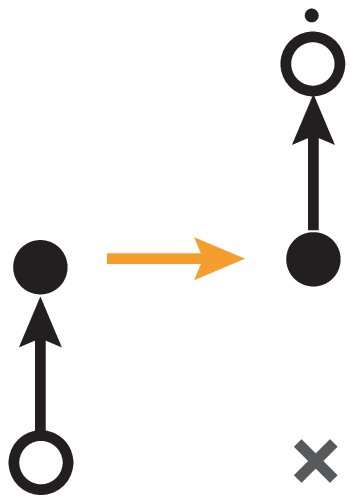}}~,\Label{dphimap}
 \eeq
 and has the obvious effect of raising the bosonic, white vertex.
 The dot on this raised vertex on the right-hand side of the map
 reminds that it corresponds to the component field $\dot{\f}$,
 which plays the r\^ole of the higher component on the right-hand
 side of~(\ref{dphi}). The ``$\times$'' represents the constant
 $\f(0)$, which is annihilated under the $D$ map, and so is
 ``left behind'' in the vertex raising.
 The kernel of this $D$ map is then precisely this constant,
 $\ker(D)=\f(0)$, whereas $(D\F_0)$ may be identified with the
 {\em equivalence class} with respect to such shifts:
 \begin{equation}
   \im(D) = (D\F_0) \iso{D} \{\F_0\equiv\F_0+(c;0)\}~.
 \Label{eDFeq}
 \end{equation}
 Note that this isomorphism, denoted `$\iso{D}$', consists of a
 straightforward identification of the fermionic components,
 $\l(\t)=\j(\t)$, but a derivative identification of the bosonic
 components, $B(\t)=\dot\f(\t)$, corresponding to the {\em vertex raising}.
 In this symbolic Adinkra map, the fermionic vertex remains at the same
 height, as the $D$ map identifies the corresponding component fields.
 We can, therefore, identify this map with the raising of the
 lowest, bosonic vertex two levels up (and then placing a derivative on
 the vertex). This is the simplest example showing how vertex raising
 operation can be interpreted as a superspace derivative, and
 {\em vice versa}.

 The dot on the bosonic vertex on the right-hand side of the Adinkra
 map~(\ref{dphimap}) provides information only in reference to the
 indicated mapping.
 Considering the right-hand-side 2-vertex Adinkra and the corresponding
 superfield all by itself, this dot is meaningless: within a multiplet
 that does not contain $\f(\t)$ itself, we can always {\em rename}
 $\dot{\f}(\t)$ into $B(\t)$.
 Turning this around, the combined operation is tantamount to re-writing
 $\f(\t):=\int^\t B(\t')\,d\t'$, and then expressing the
 superfield on the left, $(\f(\t);\j(\t))$, as
 $(\int^\t B(\t')\,d\t';\j(\t))$---which is defined up to the integration
 constant: this is precisely the equivalence class in~(\ref{eDFeq}).
 In this sense, the constant mode represented by the ``$\times$'' in the
 diagram~(\ref{dphimap}) may then be identified with this integration constant.
 This effectively performs a {\em vertex lowering}:
 it is equivalent to reversing the direction of the vertical arrow in the
 right hand Adinkra, and then rotating the graph so that the arrow points up,
 as per the hanging garden convention. This operation is of course inverse to
 the one represented by the horizontal arrow in the diagram~(\ref{dphimap}).
 The nexus of Adinkra maps generated by arrow reversals is
 discussed in some detail in Ref.\cite{rA};
 here we uncover some more of its structure.

\subsubsection{Supercovariant Mapping of Superfields}
 \Label{sSMaps}
 It will be useful to use the information obtained from the above analysis of the mapping~(\ref{dphi}), and reinterpret this basic transformation it in terms of so-called {\em diagram chasing} technique of homological algebra.

 We note that mapping of superfields $D:\F_0\to\L_{\p/2}$ is covariant with respect to supersymmetry since $D$ is invariant, and both $\F_0$ and $\L_{\p/2}$ are representations of supersymmetry. Without using anything else from the above analysis, we know that this map fits into an {\em exact sequence}:
 \beq
 0 ~\to~ A ~\tooo{~\i~}~ \F_0 ~\tooo{~D~}~ \tilde\L_{\p/2}
 ~\tooo{~\vp~}~\W ~\to~ 0~, \Label{eSeqN1SF0}
 \eeq
where $\i$ is an injection so $A:=\ker(D)$, and $\W:=\cok(D)$ since $\vp$ is a surjection. The auxiliary superfields introduced here, $A$ and $\W$, are another scalar and spinor superfield, akin to $\F$ and $\L$, respectively. This mapping of superfields may be detailed as follows:
 \beq
  \vC{\begin{picture}(125,75)(5,-12)
   \put(-3,-25){\includegraphics[width=5.5in]{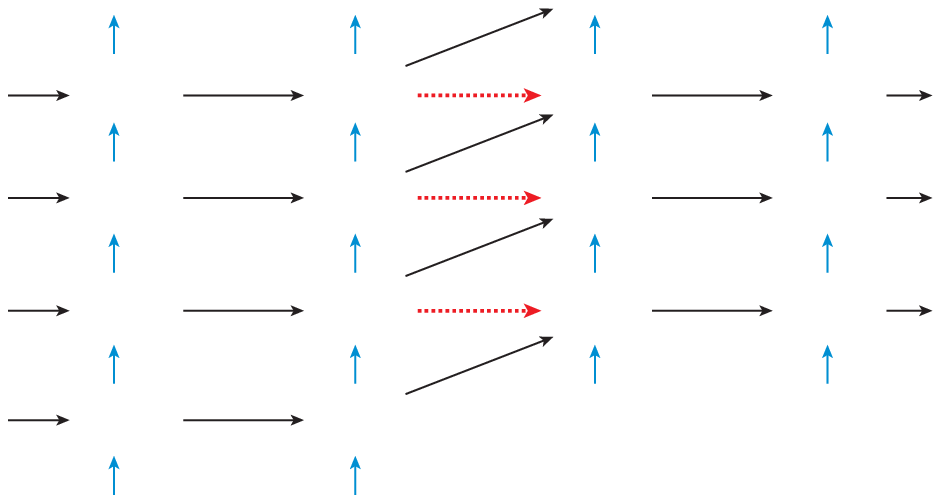}}
   \put(17.2,-12){0}
   \put(49.8,-12){0}
   \put(0,2){0}
   \put(15,2){$a(\t)$}   \put(18,7){\TC{blue}{$\>^Q$}}
   \put(47,2){$\f(\t)$}  \put(50,7){\TC{blue}{$\>^Q$}}
   \put(82,2){0}
   \put(113,2){0}
   \put(0,16){0}
   \put(15,16){$\a(\t)$}  \put(18,22){\TC{blue}{$\>^Q$}}
   \put(47,16){$\j(\t)$}  \put(50.5,22){\TC{blue}{$\>^Q$}}
     \put(60,17.5){\TC{red}{$\sim$}}
   \put(80,16){$\tilde\l(\t)$}  \put(82.5,22){\TC{blue}{$\>^Q$}}
   \put(111.5,16){$\w(\t)$}  \put(113.5,22){\TC{blue}{$\>^Q$}}
   \put(129,16){0~,}
   \put(0,31){0}
   \put(15,31){$\dot{a}(\t)$}  \put(18,37){\TC{blue}{$\>^Q$}}
   \put(47,31){$\dot\f(\t)$}   \put(50.5,37){\TC{blue}{$\>^Q$}}
     \put(60,32.5){\TC{red}{$\sim$}}
   \put(79.5,31){$\tilde{B}(\t)$}        \put(82.5,37){\TC{blue}{$\>^Q$}}
   \put(111,30){$W(\t)$}        \put(113.5,37){\TC{blue}{$\>^Q$}}
   \put(129,30){0~,}
   \put(0,45){0}
   \put(15,45){$\dot\a(\t)$}  \put(18,51.5){\TC{blue}{$\>^Q$}}
   \put(47,45){$\dot\j(\t)$}  \put(50.5,51.5){\TC{blue}{$\>^Q$}}
     \put(60,46){\TC{red}{$\sim$}}
   \put(79.5,45){$\dot{\tilde\l}(\t)$}  \put(82.5,51.5){\TC{blue}{$\>^Q$}}
   \put(111,45){$\dot\w(\t)$}  \put(113.5,51.5){\TC{blue}{$\>^Q$}}
   \put(129,45){0~,}
   \put(17.5,59){$\vdots$}
   \put(48.5,59){$\vdots$}
   \put(82.2,59){$\vdots$}
   \put(113.5,59){$\vdots$}
    \thinlines
   \put(32,4.5){$\SSS\i_{(0)}$}
   \put(67,7){$\SSS D_{(0)}$}
   \put(32,18.5){$\SSS \i_{(1/2)}$}
   \put(67,23){$\SSS D_{({1\over2})}$}
   \put(93,18.5){$\SSS \vp_{(1/2)}$}
   \put(32,33.5){$\SSS \i_{(1)}$}
   \put(67,36.5){$\SSS D_{(1)}$}
   \put(93,33.5){$\SSS \vp_{(1)}$}
   \put(32,47.5){$\SSS \i_{(3/2)}$}
   \put(67,51){$\SSS D_{({3\over2})}$}
   \put(93,47.5){$\SSS \vp_{(3/2)}$}
  \end{picture}}
 \Label{DPhiMap}
 \eeq
where the component fields and their derivatives, obtained by iterative application of $Q$, indicated by vertical arrows in blue, have been stacked in height to represent their engineering dimensions. The superfield maps, $\i,D,\vp$ are accordingly separated into their components, and are here labeled by the engineering dimensions of the fields upon which they act; we have used the above-discussed liberty of setting the engineering dimension of $\f(\t)$, the lowest component of $\F$, to zero, as a convenient reference point. The action of the $D$ map increases the engineering dimension by $\inv2$ so that the engineering dimensions of the component fields of $\tilde\L_{\p/2}$ and $\W$ are by $\inv2$ higher than their counterparts in $A$ and $\F_0$.
 This forces $\tilde\l(\t)$ to have engineering dimension $\inv2$, so $B(\t)$ must have its engineering dimension equal to 1. The actions of $\i$ and $\vp$ being the obvious ones, preserving the engineering dimension, this fixes the engineering dimensions of the component fields of $A$ and $\W$ as indicated by their placement in~(\ref{DPhiMap}).

 The dotted, red  arrows represent isomorphic equivalence maps in the
 superfield formulation, where component fields are defined only up to additive
 time-derivatives of other component fields; the dotted arrows thus represent
 the ``$D-Q$'' difference maps.
 Note that the `lowest-lying' such map, at engineering level $\inv2$, identifies $\tilde\l(\t)\simeq\j(\t)$.
 The next dotted arrow, at the engineering level 1, identifies
 $\tilde{B}(\t)\simeq\dot\f(\t)$, and so on.
 Since all dotted maps represent isomorphisms, it follows that $\a(\t), \dot{a}(\t), \w(\t)$ and $W(\t)$ all vanish. This reduces~(\ref{DPhiMap}) to:
 \beq
  \vC{\begin{picture}(125,75)(5,-12)
   \put(-3,-25){\includegraphics[width=5.5in]{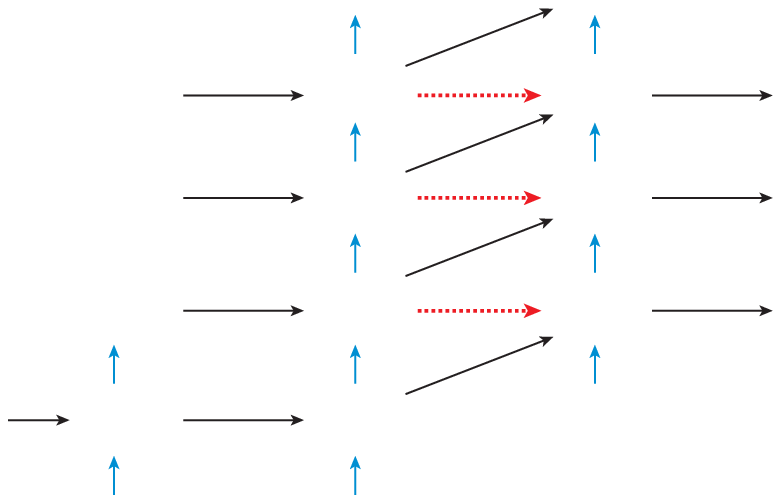}}
   \put(17.2,-12){0}
   \put(49.8,-12){0}
   \put(0,2){0}
   \put(15,2){$\f(0)$}   \put(18,7){\TC{blue}{$\>^Q$}}
   \put(47,2){$\f(\t)$}  \put(50,7){\TC{blue}{$\>^Q$}}
   \put(82,2){0}
   \put(17,16){0}
   \put(47,16){$\j(\t)$}  \put(50.5,22){\TC{blue}{$\>^Q$}}
     \put(60,17.5){\TC{red}{$\sim$}}
   \put(80,16){$\tilde\l(\t)$}  \put(82.5,22){\TC{blue}{$\>^Q$}}
   \put(111,16){0}
   \put(17,31){0}
   \put(47,31){$\dot\f(\t)$}   \put(50.5,37){\TC{blue}{$\>^Q$}}
     \put(60,32.5){\TC{red}{$\sim$}}
   \put(79.5,31){$\tilde{B}(\t)$}        \put(82.5,37){\TC{blue}{$\>^Q$}}
   \put(111,31){0}
   \put(17,45){0}
   \put(47,45){$\dot\j(\t)$}  \put(50.5,51.5){\TC{blue}{$\>^Q$}}
     \put(60,46){\TC{red}{$\sim$}}
   \put(79.5,45){$\dot{\tilde\l}(\t)$}  \put(82.5,51.5){\TC{blue}{$\>^Q$}}
   \put(111,45){0}
   \put(48.5,59){$\vdots$}
   \put(82.2,59){$\vdots$}
    \thinlines
   \put(32,4.5){$\SSS\i_{(0)}$}
   \put(67,7){$\SSS D_{(0)}$}
   \put(32,18.5){$\SSS \i_{(1/2)}$}
   \put(67,23){$\SSS D_{({1\over2})}$}
   \put(93,18.5){$\SSS \vp_{(1/2)}$}
   \put(32,33.5){$\SSS \i_{(1)}$}
   \put(67,36.5){$\SSS D_{(1)}$}
   \put(93,33.5){$\SSS \vp_{(1)}$}
   \put(32,47.5){$\SSS \i_{(3/2)}$}
   \put(67,51){$\SSS D_{({3\over2})}$}
   \put(93,47.5){$\SSS \vp_{(3/2)}$}
  \end{picture}}
 \Label{theDPhiMap}
 \eeq
Finally, the isomorphic equivalence
 $(D_{({1\over2})}-Q_{({1\over2})}):\dot\f(\t)\to\tilde{B}(\t)$
clearly leaves the constant mode $a(0)=\f(0)$ to span the kernel of the superfield mapping $D$. On the other hand, since all components of the superfield $\W$ vanish, $\cok(D)=0$. Therefore, $a(\t)=\f(0)$, as indicated. Note that in fact, the superfield $A$, consisting of a single constant scalar component, $a(0)$, is indeed a representation of supersymmetry, often referred to as a ``zero mode''.
 We combine these facts into the sequence of superfield mappings:
 \beq
 0 ~\to~ (\f(0);0) ~\tooo{~\i~}~ \F_0 ~\tooo{~D~}~ \tilde\L_{\p/2}
 ~\to~ 0~,
 \Label{eDFM}
 \eeq
which contains the two component field mappings, read off from~(\ref{theDPhiMap}) by collapsing all derivatives of all fields:
\begin{eqnarray}
 0~&\tooo{~\i_{(f)}~}&\j(\t)~\tooo{\,D_{(b)}-Q_{(b)}\,}~
 \tilde\l(\t)~\to~0~, \Label{eDFMo}\\[2mm]
 0~\to~\f(0)&\tooo{~\i_{(b)}~}&\f(\t)~\tooo{~~~~``D^2\,"~~~~}
  ~\tilde{B}(\t)~\to~0~,  \Label{eDFMe}\\[3mm]
 \mbox{where}&&\mkern-50mu
  ``D^2\,":=D_{(f)}\circ(D_{(b)}-Q_{(b)})^{-1}\circ D_{(b)}~\propto~\vd_\t~,
\end{eqnarray}
or present these results as the super-constraint equations
\begin{align}
 \tilde\L_{\p/2} &= D\F_0~,        \Label{theD}\\[2mm]
 (\f(0);0) &= \{\F:D\F=0\}~. \Label{kerD}
\end{align}
This last representation, in terms of explicit equations, is of course the standard in physics literature, and we hope that the foregoing discussion provides a clear dictionary between this and the above, so-called ``diagram chasing'' (albeit a very simple one).

 In particular, Eq.~(\ref{kerD}) represents the ``failed'' attempt to define the $(1|1)$-supersymmetry analogue of a {\em chiral superfield\/}\Ft{In $d=4$, this is a superfield annihilated by the complex-conjugate half of the total of four $D$'s.}; the superdifferential constraint $D\F=0$ is too strong, and defines a ``trivial'' superfield, consisting of a single scalar constant, $\f(0)$. While trivial for the purposes of defining interesting superfields, constant modes such as $\f(0)$ may well play a r\^ole in ``topological'' considerations.

The foregoing then defines the simplest not-quite-trivial mapping of superfields. Since it maps not only the vector fields spanned by the component fields of the respective superfields, but also the supersymmetry action upon them, acting vertically in the diagrams~(\ref{DPhiMap}) and~(\ref{theDPhiMap}), it is more properly referred to as a {\em supersymmetry morphism\/}\Ft{There are good reasons for distinguishing {\em super-morphisms\/} from {\em supersymmetry morphisms\/}, much as many a super-algebra is not a supersymmetry algebra.}. Its analogues in the Adinkra realm, {\em adinkramorphisms\/}, are defined in precise analogy; in fact, we only need substitute the corresponding Adinkras in the mapping diagrams~(\ref{DPhiMap}) and~(\ref{theDPhiMap}). This further bolsters our present aim, to provide a close translation between the Adinkra realm and the superspace/superfield framework.

\subsubsection{Multiple Superderivatives}
 Clearly, $D$ may also be applied to the spinor multiplet $\tilde\L_{\pi/2}$,
 \beq
  D: \tilde\L_{\pi/2} \to (D\,\tilde\L_{\pi/2})
  ~=~i(\tilde{B};\dot{\tilde\l})~.
 \Label{dlam}
 \eeq
 The image of this map is a scalar superfield,
 akin to the superfield $\F_{\p/2}$ described in~(\ref{n1sSFa}).
 Thus, we can re-write equation~(\ref{dlam}) symbolically
 as follows,
 \beq
  \vC{\includegraphics[width=.8in]{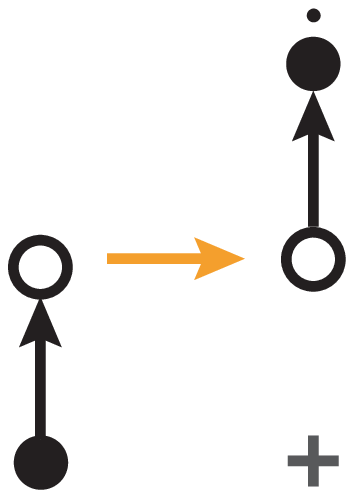}}~. \Label{dlammap}
 \eeq
 We can picture this map as a process of
 raising the lowermost fermionic vertex upward (and then placing a
 derivative on the vertex). Again, the mapping has a kernel, spanned
 by the fermionic constant $\l(0)$, and represented now by the ``$+$''.
 Equivalently, this map is graphically equivalent to
 reversing the arrow direction, and then
 re-arranging the orientation so that the arrow points up,
 as per the hanging garden convention.

 The above discussion illustrates the simplest correlations
 between Adinkra operations and superspace operations.  Similar
 correlations exist for cases with $N>1$, but there are extra
 subtleties which prove rather intriguing.

\subsection{The case $N=2$}
 \Label{sN=2}
 Following the procedure in Section~\ref{graph}, we now extend the previous discussion of $(1|1)$ supersymmetry to the $(1|2)$ case. To simplify the presentation, we represent the superfields by their $\q$-expansion.

\subsubsection{The Scalar and the Spinor Superfields}
We start with an otherwise unconstrained $N=2$ superfield, $\IU$, the
components of which we define by invariant projections
\begin{equation}
 u:=\IU|~,\qquad i\c_I:=D_I\,\IU|~,\quad\hbox{and}\qquad
 iU:=\inv2\ve^{IJ}D_JD_I\,\IU|~.
 \Label{eIU2}
\end{equation}
 As in the $N=1$ case, these can be reassembled into the $\q$-expansion:
 \brr \IU =
      u+i\,\q^I\,\chi_I
      +\fr12\,i\,\ve_{IJ}\,\q^I\,\q^J\,U~,
 \Label{scalarsf}\err
 where $u$ and $U$ are each real bosons, and
 $\chi_I$ is an $\SO(2)$ doublet of real fermions.
 Component transformation rules can be determined
 from~(\ref{scalarsf}) by extracting the components of
 $\d_Q\6(\e)\,\IU=-i\e^I\,Q_I\,\IU$.  In this way, we determine
 \begin{equation}\begin{aligned}
    \d_Q\,u &= i\,\e^I\,\chi_I~,\\[2mm]
      \d_Q\,\chi_I &= \ve_{IJ}\,\e^J\,U
      +\d_{IJ}\,\e^J\,\dot{u}~, \\[2mm]
      \d\,U &= -i\,\ve_{IJ}\,\e^I\,\dot{\chi}^J~.
 \end{aligned}
 \Label{sca2}
 \end{equation}
 These transformation rules readily translate
 into the following Adinkra (edges point upwards):
 \beq
  \vC{\includegraphics[width=1in]{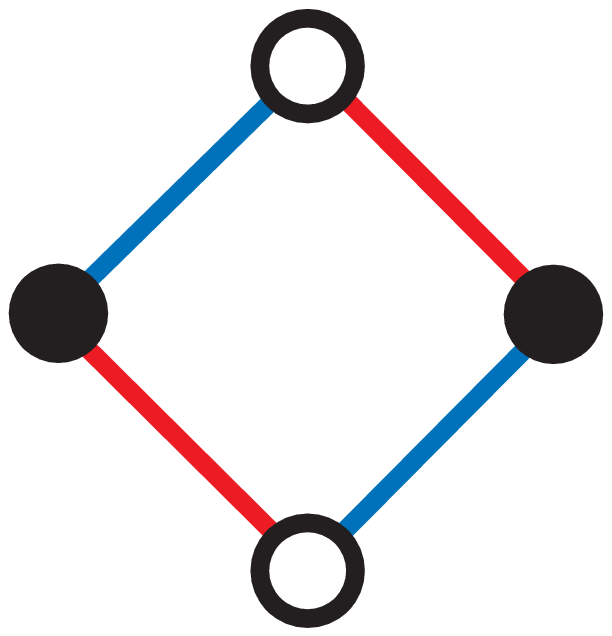}}.\Label{Adsca2}
 \eeq
 Here, the upper-most bosonic vertex corresponds to
 the field $U$, the pair of fermionic vertices correspond to the $\SO(2)$ fermion
 doublet $\chi_I$, and the lower-most vertex corresponds to the field $u$.
 We have used the ``hanging garden" depiction of this Adinkra, whereby
 arrows are suppressed, since all arrows implicitly point upward. We also
 partition the edges: edges of the same color represent the action of the
 same supersymmetry generator, one for $Q_1$ another for $Q_2$. It was also
 possible here to make edges in the same partition parallel to each other.
 Note that the system of transformation rules~(\ref{sca2}) is identical
 to~(\ref{n2scalar}), and the Adinkra~(\ref{Adsca2}) is equivalent to the
 Adinkra~(\ref{Adn2scalar}), although here we have represented this Adinkra
 as a hanging garden.

 A fermionic analog of~(\ref{scalarsf}), known as the
 real $N=2$ spinor superfield, is given by
 \brr \IB =
      \b+\q^I\,B_I
      +\fr12\,i\,\ve_{IJ}\,\q^I\,\q^J\,\varphi
 \Label{bdef}\err
 where $\b$ and $\varphi$ are real fermions and
 $B_I$ is an $\SO(2)$ doublet of real bosons.
 The component transformation rules, determined by computing
 $\d_Q\,\IB$ and extracting the components, are
 \beq
  \left.\begin{aligned}
   \d_Q\,\b &= \e^I\,B_I~,\\[2mm]
      \d_Q\,B_I &= i\,\ve_{IJ}\,\e^J\,\varphi
      +i\,\d_{IJ}\,\e^J\,\dot{\b}~,\\[2mm]
      \d_Q\,\varphi &=
      -\ve_{IJ}\,\e^I\,\dot{B}^J~,
  \end{aligned}\right\}\qquad\Longleftrightarrow\qquad
  \vC{\includegraphics[width=1in]{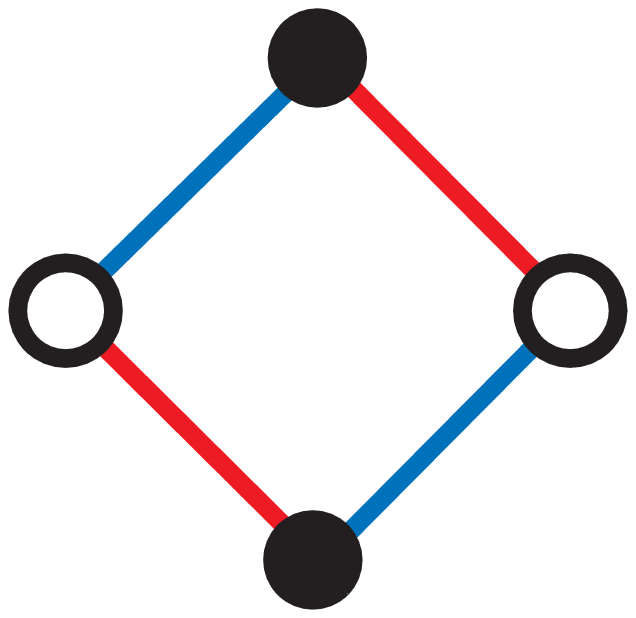}}~.
 \Label{brules}
 \eeq
 Notice that this Adinkra can be obtained from the
 previous one by performing a Klein flip, \ie, replacing all bosonic
 vertices with fermionic vertices, and vice versa.

\subsubsection{$N=2$ Doublet Superfields}
 A reducible supermultiplet is described by the $\SO(2)$
 real doublet superfield defined by
 \brr \IA_I=
      a_I+i\,\q^J\,\a_{JI}
      +\fr12\,i\,\ve_{JK}\,\q^J\,\q^K\,A_I~,
 \Label{supa}\err
 where $a_I$ and $A_I$ each describe $\SO(2)$ doublets of real bosons,
 and $\a_{IJ}$ is an unconstrained real rank-two $\SO(2)$ tensor
 (\ie, a two-by-two matrix) describing
 four real fermion degrees of freedom.  The corresponding
 transformation rules are
 \begin{equation}\begin{aligned}
   \d_Q\,a_I &= i\,\e^J\,\a_{JI}~,\\[2mm]
      \d_Q\,\a_{JI} &=
      \ve_{JK}\,\e^K\,A_I
      +\d_{JK}\,\e^K\,\dot{a}_I~,\\[2mm]
      \d\,A_I &=
      -i\,\ve^{JK}\,\e_J\,\dot{\a}_{KI}~.
\end{aligned}
\Label{Arules}
 \end{equation}
 Since the $\SO(2)$ transformation commutes with supersymmetry,
 it follows that each of the superfields $\IA_1$ and
 $\IA_2$ describe separate $N=2$ multiplets which do
 not mix under supersymmetry.  It is easy to verify this by
 re-writing~(\ref{Arules}) in terms of each
 $\SO(2)$ tensor component.  The Adinkra for this superfield
 is therefore given by
 \beq
  \vC{\includegraphics[width=2.2in]{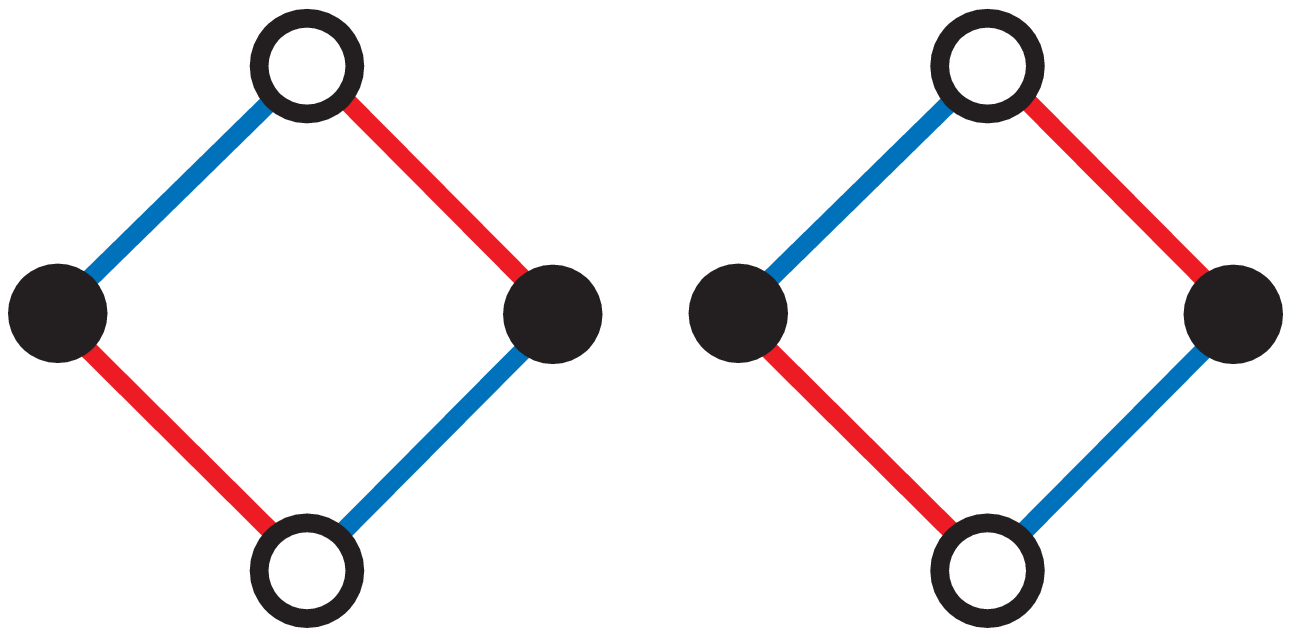}}~. \Label{AdArules}
 \eeq
 In the left connected component, the upper boson vertex
 corresponds to $a_1$, the two fermionic vertices correspond to
 $\a_{11}$ and $\a_{21}$ and the lower bosonic vertex
 corresponds to $A_1$.  In the right connected component
 the upper boson vertex corresponds to $a_2$, the two fermionic
 vertices correspond to $\a_{12}$ and $\a_{22}$, and the
 lower bosonic vertex corresponds to $A_2$.
 Notice that this Adinkra is comprised of two copies of the scalar Adinkra
 described previously. The fact that this multiplet is reducible corresponds
 to the fact that this Adinkra is not connected.
 The $\SO(2)$ transformation however rotates the left connected
 component and the right connected component into each other.

 A fermionic analog of~(\ref{supa}) is given by the
 $\SO(2)$ real spinor doublet superfield
 \beq \IF_I=\omega_I
      +\q^J\,F_{JI}
      +\fr12\,i\,\ve_{JK}\,\q^J\,\q^K\,\Omega_I~,
 \Label{fidef}\eeq
 where $\omega_I$ and $\Omega_I$ each describe $\SO(2)$ doublets
 of real fermions, and $F_{IJ}$ is
 an unconstrained real rank-two $\SO(2)$ tensor
 (\ie, a two-by-two matrix) describing
 four real bosonic degrees of freedom.  The corresponding
 transformation rules are
 \beq
  \left.\begin{aligned}
   \d_Q\,\omega_I &= \e^J\,F_{JI}~,\\[2mm]
      \d_Q\,F_{JI} &=
      i\,\ve_{JK}\,\e^K\,\Omega_I
      +i\,\d_{JK}\,\e^K\,\dot{\omega}_I~,\\[2mm]
      \d\,\Omega_I &=
      -\ve^{JK}\,\e_J\,\dot{\a}_{KI}~,
  \end{aligned}\right\}\quad\Longleftrightarrow~
  \vC{\includegraphics[width=2.2in]{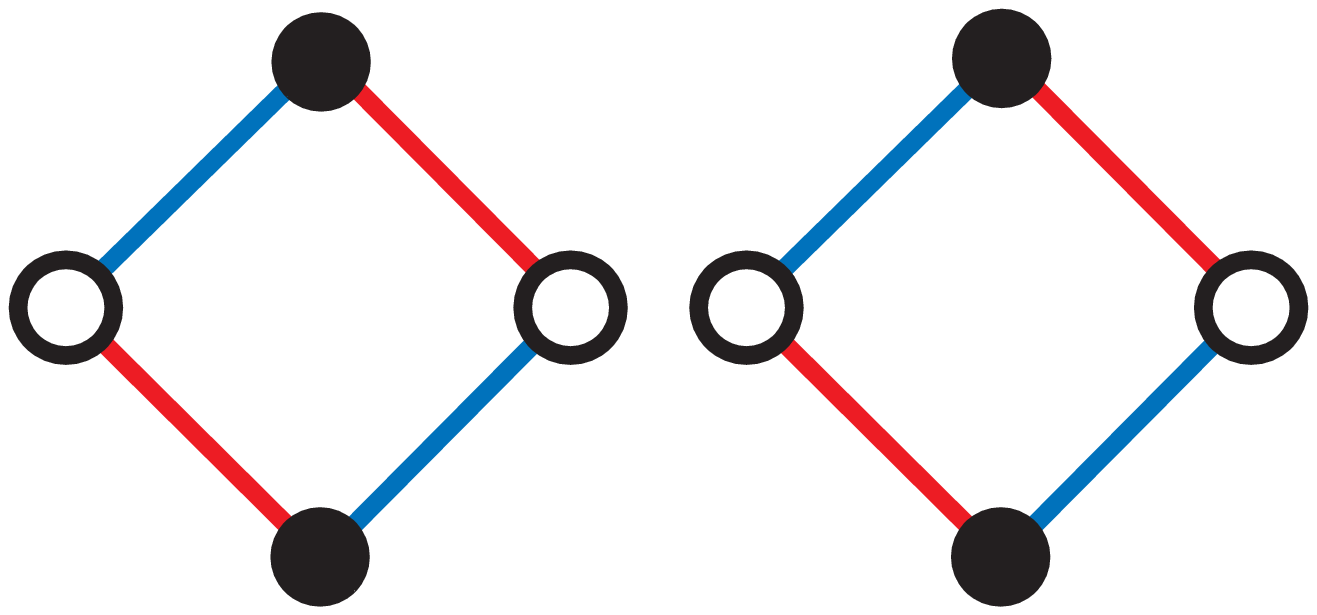}}~.
 \Label{Frules}
 \eeq
 This Adinkra can be obtained from~(\ref{AdArules}) by performing
 a Klein flip, \ie, by replacing all fermionic vertices with
 bosonic vertices, and vice-versa.  The doublet spinor superfield is
 manifestly reducible, as evidenced by the feature that its
 Adinkra is not connected.

\subsubsection{Superderivative superfield pairs}
 It is possible to obtain a real $\SO(2)$ doublet spinor superfield as
 the image of the following map\Ft{In actuality, the superfield $D_I\,\IU$ is
 described by $i$ times a real $\SO(2)$ doublet superfield.  As explained previously,
 the overall phase on superfield ``reality" constraints is irrelevant at the level of
 component transformation rules, and therefore at the level of Adinkras.
 In this sense the matter of these superspace phases is not of direct
 relevance to this paper, and will be suppressed henceforth.},
 \beq
     D_I:\,\IU\rightarrow (D_I\,\IU)~,\Label{DUmap}
 \eeq
 where $\IU$ is an $N=2$ real scalar superfield, \eg, as given in~(\ref{sca2}). We forego a detailed analysis of the maps in the manner of Subsection~\ref{sSMaps}, but should trust the interested Reader to be able to reconstruct the appropriate component-level mappings. We note, however, that the mapping is explicitly covariant with respect to an $\SO(2)$ $R$-symmetry: with respect to this $\SO(2)$ action, the supersymmetry generators, $Q_1,Q_2$, form a doublet, \ie, $\Span(Q_1,Q_2)$ furnishes the 2-dimensional representation
of $\SO(2)$.

 In this case all of the components
 of $D_I\,\IU$ are described by the
 components of the irreducible superfield
 $\IU$.
 It is straightforward to identify the placement of the component degrees
 of freedom into the fermionic $\SO(2)$ doublet Adinkra, vertex-by-vertex.
 This can be done by computing the component expansion for
 $D_I\,\IU$,
 \brr (D_I\,\IU) =
      i\,\bigl(\,\chi_I
      +\q^K\,(\,\ve_{IK}\,U+\d_{IK}\,\dot{u}\,)
      +\fr12\,i\,\ve_{KL}\,\q^K\,\q^L\,
      (\,\ve_{IJ}\,\dot{\chi}^J\,)\,
      \bigr)
 \Label{DIU}\err
 then applying the operator $\d_Q\6(\e)$
 to compute the transformation rules for the resulting components,
 and then translating these transformation rules into an Adinkra
 diagram.  The result is that $(D_I\,\IU)$ is described
 by the following ``decorated'' Adinkra:
 \beq
  \vC{\begin{picture}(75,30)(0,-2)
   \put(0,-2){\includegraphics[width=2.8in]{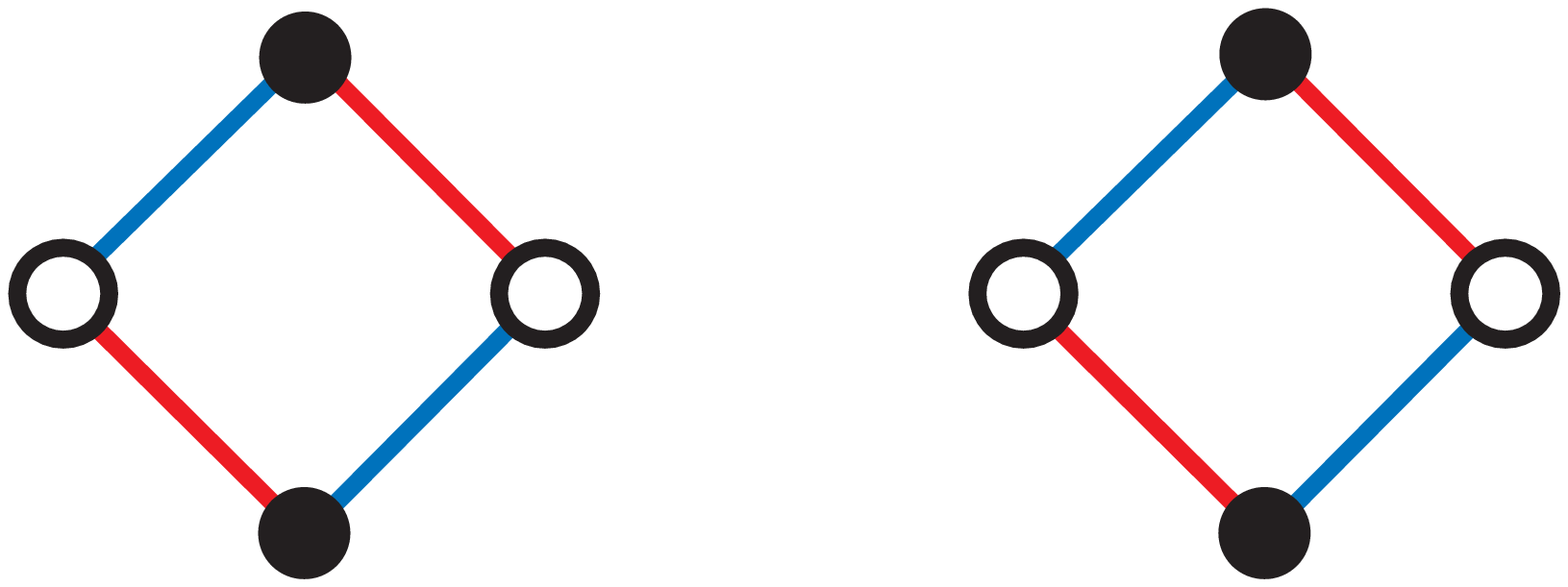}}
   \put(10,0){$\c_1$}
   \put(0,11){$\dot{u}$}
   \put(10,11.5){$(D_1\IU)$}
   \put(30,11){$U$}
   \put(9,23){$\dot\c_2$}
   \put(49,0){$\c_2$}
   \put(39,11){$U$}
   \put(50,11.5){$(D_2\IU)$}
   \put(69,11){$\dot{u}$}
   \put(49,23){$\dot\c_1$}
  \end{picture}}~.
 \Label{double2}
 \eeq
 Here, the left connected component corresponds to
 $(D_1\,\IU)$ and the right connected component
 corresponds to $(D_2\,\IU)$, and we have made explicit
 the identification of the Adinkra vertices with the degrees of
 freedom in the scalar superfield $\IU$.

 It is helpful to exhibit the superderivatives~(\ref{DIU})
 explicitly in terms of their $\SO(2)$ components,
 \begin{align}
   (D_1\,\IU) &=
      i\,\bigl(\,\chi_1+\q^1\,\dot{u}
      +\q^2\,U+i\,\q^1\,\q^2\,\dot{\chi}_2\,\bigr)~,\Label{D1U}\\[2mm]
   (D_2\,\IU) &=
      i\,\bigl(\,\chi_2-\q^1\,U
      +\q^2\,\dot{u}-i\,\q^1\,\q^2\,\dot{\chi}_1\,\bigr)~. \Label{D2U}
 \end{align}
 By comparing the component fields in these expressions with the
 vertices in the Adinkra~(\ref{double2}) a noteworthy correlation
 becomes evident:  The ``lowest component" of the superfield
 $(D_1\IU)$ correlates with the ``lowest vertex" in the $(D_1\,\IU)$ Adinkra.
 The pair of fields appearing at ``first-order" in $\q^I$  in the
 superfield appear at ``height-one" in the Adinkra.
 Finally, the field appearing at highest order in $\q^I$,
 in the superfield, appears as the ``highest vertex" in the Adinkra.
 Indeed, this was the ultimate rationale behind the hanging gardens
 convention of orientating all arrows upward. Furthermore, the partitioning of the edges reveals that the red (NW-rising) edges pertain to $Q_1,D_1,\q^1$, whereas the blue (NE-rising) edges correspond to $Q_2,D_2,\q^2$. We will keep to this color-coding in more complicated cases, as it facilitates reading the Adinkras when the high value of $N$ and the hanging gardens height convention prevents all edges from the same partition to be depicted parallel to each other.

 Since the $\q^I$-expansion structure of the pair of derivative
 superfields~(\ref{DIU}) is identical to that of the
 $\SO(2)$-doublet~(\ref{fidef}), the former clearly maps to the latter,
 and yields the following component field identifications:
\begin{equation}
\begin{aligned}
 \tw\w_1     &= i\c_1 ~,&\qquad     \tw\w_2     &=i\c_2~,     \\[2mm]
 \tw{F}_{11} &= i\dot{u}~,&\qquad  \tw{F}_{12} &=-iU~,       \\[2mm]
 \tw{F}_{21} &= +iU~,&\qquad        \tw{F}_{22} &=i\dot{u}~, \\[2mm]
 \tw\W_1     &= i\dot\c_2~,&\qquad  \tw\W_2     &=-i\dot\c_1~.
\end{aligned}
 \Label{etFi=DiUc}
\end{equation}
 The obvious identities between the component fields of~(\ref{D1U})
 and~(\ref{D2U}) have thus imposed corresponding identities, \ie,
 {\em constraints\/} on the component fields of the latter:
\begin{equation}
\begin{aligned}
 \tw\W_1     &= \dot\w_2 ~,\qquad &\tw\W_2     &=-\dot\w_1~,   \\[2mm]
 \tw{F}_{11} &= \tw{F}_{22}~,\qquad&\tw{F}_{12} &=-\tw{F}_{21}~.
\end{aligned}
 \Label{etFiCon}
\end{equation}
The $\SO(2)$-doublet superfield satisfying these component field constraints
will be denoted $\tw\IF_I$.

 Before we continue examining the action of $D_I$, several remarks are in order:
 It is clear, from the indicated vertex identifications, that this
 {\em pair\/} of superderivative superfields, $\{(D_I\IU),\>I=1,2\}$,
 includes all degrees of freedom in $\IU$, except $u(0)$, which
 here again spans the kernel of the map~(\ref{DUmap}).
 Second, the component fields of $\IU$ are, in fact, {\it almost} completely
 represented in {\em either\/} of the two disconnected diagrams:
 In the left connected component, the zero-mode of $\chi_2$ is
 missing, by virtue of the dot on the corresponding vertex.
 In the right connected component the zero mode of $\chi_1$ is
 similarly excluded.  Next, the pair of component fields corresponding to the white circles in the left-hand graph is to be identified with the corresponding pair in the right-hand graph, thus ``fusing'' the two.
 Once so fused, the presently topmost vertices must be dropped, since each describes a derivative of a degree of freedom already present in the fused Adinkra. This then leaves:
 \beq
 \vC{\begin{picture}(30,20)
  \put(3,0){\includegraphics*[width=.75in]{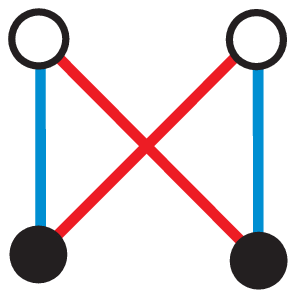}}
  \put(-1,2){$\c_1$}
  \put(22,2){$\c_2$}
  \put(-1,16){$U$}
  \put(22,16){$\dot{u}$}
 \end{picture}}
 \Label{AdDIU}
 \eeq
where we have omitted the disconnected ``$\times$'' representing $u(0)$. We will return to this, below, and provide additional motivation for this identification of the implicitly constrained superderivative superfield pair, $(D_I\IU)$, with the depicted Adinkra.

Finally, it is clear that the pair of superderivative superfields,
$\{(D_I\IU),\>I=1,2\}$, spans a 2-dimensional representation
of the $\SO(2)$ $R$-symmetry. Of course the $\SO(2)$ invariance of the
multiplet does not necessarily imply an $\SO(2)$ invariance of any chosen Lagrangian, or the dynamics obtained by such a Lagrangian.  In fact, even if they were chosen so, boundary
conditions may be selected that break this symmetry. Nevertheless, it
is useful to exhibit the symmetries possible here, and the $\q$-expansion
presenting the component field content of the $\{(D_I\IU),\>I=1,2\}$
pair~(\ref{DIU}) certainly makes use of that.

 As it may not be necessary to maintain $\SO(2)$ equivariance, we also consider the action of the map $D_1$, by itself, upon $\IU$. This can be described in terms of Adinkras as follows,
\beq
 \vC{\begin{picture}(115,40)
  \put(-10,13){$D_1\>:$}
  \put(0,-2){\includegraphics*[width=4.5in]{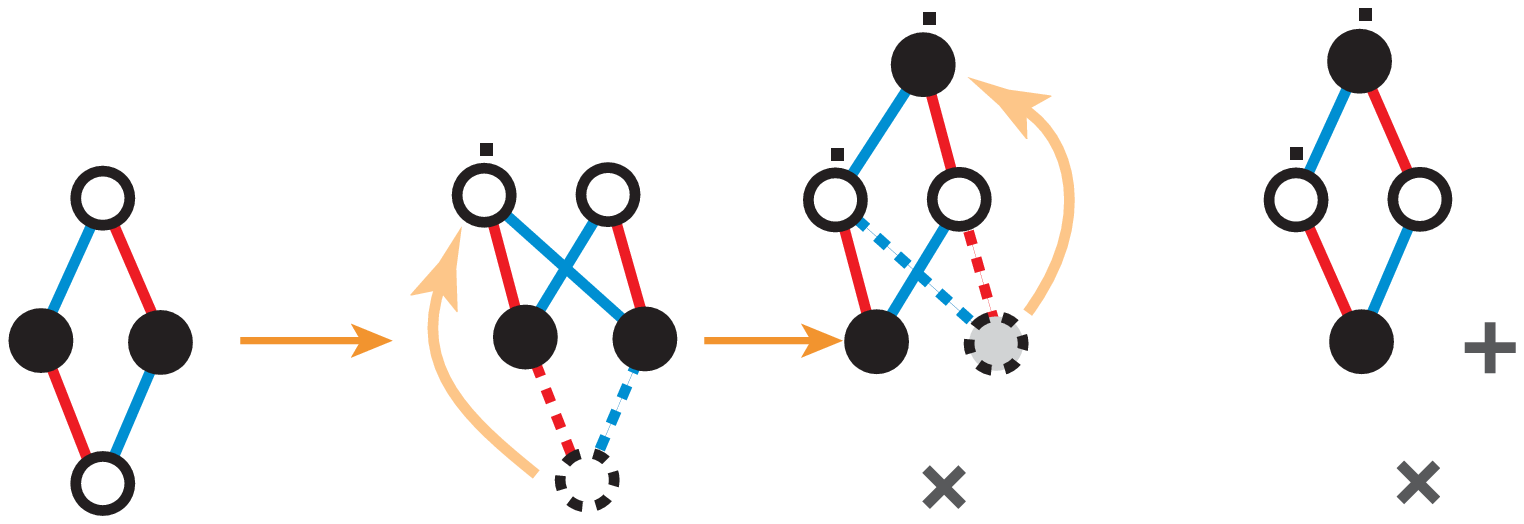}}
  \put(83,12){$\cong$}
   \put(112,12){.}
 \end{picture}}
 \Label{d1map}
\eeq
 In this sequence we see that the action of the $D_1$ map on a scalar
 superfield may be represented as a {\em pair\/} of vertex raises: First
 raise the lowermost, scalar vertex, which is a source; this turns both
 fermionic vertices into sources. Then raise that fermion vertex which
 corresponds to $\c_2$. Note that the fermion component of $\IU$ that
 remains intact is the $D_1\IU|$ one.
 A dot has been placed on the raised vertices to indicate that they
 are time derivatives of the corresponding vertices prior to the
 raising operation.  As described above, the constant mode of
 any field associated with any raised vertex describes the kernel of
 the map, in the sense that the constant mode is not being ``raised",
 as it is annihilated by the derivative action.
 These are indicated in the diagrams above by ``$\times$''
 for bosonic and ``$+$'' for fermionic constants.
 This describes, in an $N=2$ example, the relationship between
 superderivatives vertex raising in Adinkras.

 Similarly, the map $D_2: \IU\to (D_2\,\IU)$ can be described
 in terms of Adinkras as
 \beq
  \vC{\begin{picture}(115,40)
  \put(-10,13){$D_2\>:$}
   \put(0,-2){\includegraphics*[width=4.5in]{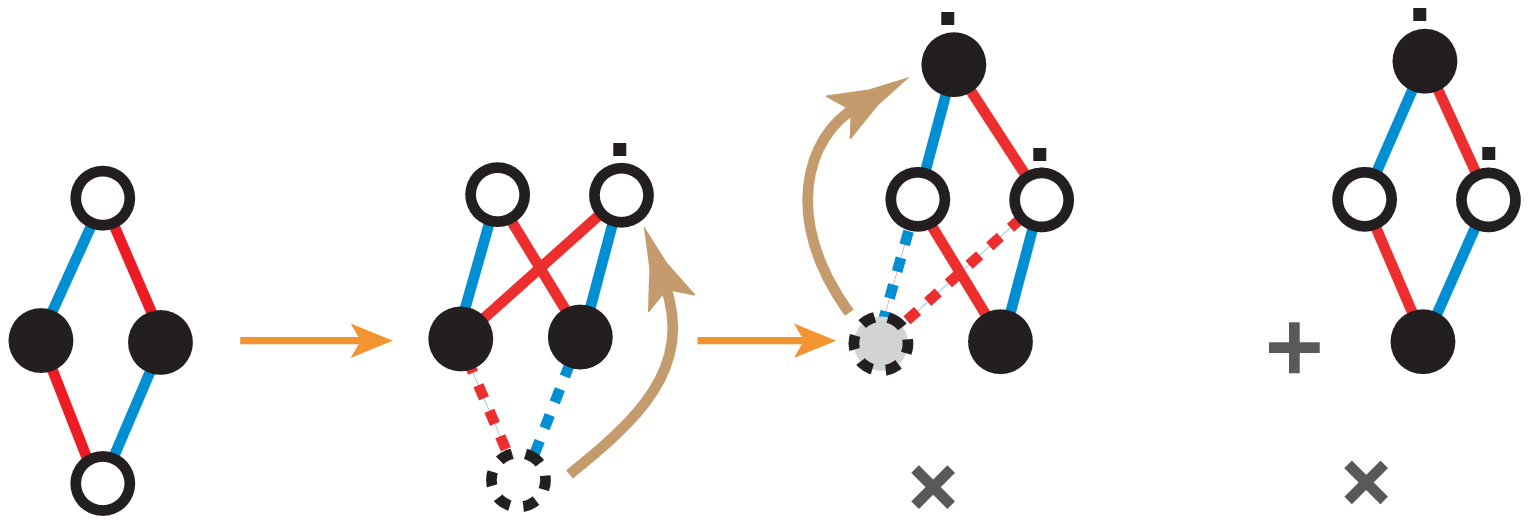}}
   \put(83,12){$\cong$}
   \put(112,12){.}
  \end{picture}}
  \Label{d2map}
 \eeq
 This process is a mirror-image of the $D_1$ map; the only
 difference is in the choice of which of the two fermionic vertices
 is ``raised" in the second step.

 We have seen that a superderivative map $\IU\to(D_I\,\IU)$ is,
 for each of the two values, $I=1,2$, implemented on an Adinkras
 by a two-step sequence involving one vertex raise at each step.
 These vertex raises are manifestations of the term in the
 derivative $D_I$ proportional to $\q^I\,\vd_\t$.
 The operator $\q^I\,\vd_\t$ ``raises" vertices at all but the
 highest component level of any superfield upon which it
 acts\Ft{In a $(1|2)$-superfield, the ``highest" component field
 multiplies $\fr12\,\ve_{IJ}\,\q^I\q^J$, which in turn is annihilated
 by any $\q^I$, and so also by the ``raising'' operator,
 $\q^I\,\vd_\t$.}. Since there are three component
 levels in the superfield $\IU$, this explains why there
 are two steps in the map we have considered.

 Now consider the Adinkra which appears at the intermediate step in the
 two-step process describing the $D_I$ maps,
 \beq
 \vC{\begin{picture}(30,20)
  \put(3,0){\includegraphics*[width=.75in]{adx.eps}}
  \put(-1,2){$\c_1$}
  \put(22,2){$\c_2$}
  \put(-1,16){$\f_1$}
  \put(22,16){$\f_2$}
 \end{picture}}
 \Label{adrx}
 \eeq
 Here we have not included the dots in the diagram since, as explained
 above, these have no intrinsic meaning. Also, we have chosen
 names for the vertices to facilitate translation of the diagram into
 transformation rules, but otherwise this is identical to~(\ref{AdDIU}).
 The corresponding transformation rules are
 given by
 \begin{equation}\begin{aligned}
  \d_Q\,\chi_1 &= \e^I\,\f_I~,\\[2mm]
      \d_Q\,\chi_2 &= -\ve_{IJ}\,\ve^I\,\f^J~,\\[2mm]
      \d_Q\,\f_I &= i\,\ve_{IJ}\,\e^J\,\dot{\chi}_2
      +i\,\e_I\,\dot{\chi}_1~.
 \end{aligned}\Label{xrules}\end{equation}
 It is readily verified that these rules do properly represent~(\ref{nalg2}) when applied to each of the four components
 $\chi_{1,2}$ and $\f_{1,2}$.  Another way to obtain the
 transformation rules~(\ref{xrules}) is to ``lower" the upper
 vertex in the spinor multiplet Adinkra corresponding to
 $D_1\,\IU$.  This can be done by starting with the
 transformation rules~(\ref{brules}), re-defining
 $\b:=\chi_1$, $B_I:=\f_I$, and $\varphi:=\dot{\chi}_2$,
 and then determining the transformation rule for
 $\chi_2$ by removing the time derivative from the rule
 for $\varphi$ shown in~(\ref{brules}).  What results are the
 same transformation rules which one can read off of the Adinkra
 shown in~(\ref{adrx}).
 The degrees of freedom appearing in~(\ref{adrx}) then correspond to
 the $\SO(2)$ spinor doublet superfield $\tw\IF_I:=(D_I\,\IU)$,
 taken as a single, ``fused" superfield.

 The foregoing analysis then proves:
\begin{proposition}\Label{pDIUinFI}
 The superfield $(D_I\IU)$ is the {\em superderivative superfield\/} solution
 to the constraint system~(\ref{etFiCon}), in terms of the otherwise
 unconstrained $N=2$ superfield $\,\IU$.
\end{proposition}
 \Remk It remains to specify the component field constraint
 system~(\ref{etFiCon}) in purely superfield and superderivative terms.  This will be discussed for general $N$ in the next section.

\section{Superderivative Solutions for all $N$}\Label{secSderiv}
The above examples suffice to motivate the main ideas for general $N$, as illustrated in propositions in this section.

\subsection{Superderivative images}
Herein we explore the characteristics of the various linear maps constructed with the aid of the superderivatives $D_I$.

\begin{proposition}
Let ${\cI}$ be a subset of $\{1,\dots,N\}$, and let
$p\equiv \#{\cI} \bmod{2}$.  Let $D_{\cI}$ be the superderivative as
in (\ref{eproj}). Then $D_{\cI}$ maps $\sF^N_0$ surjectively onto $\sF^N_p$.
\end{proposition}

\begin{proof}
Let $k=\#{\cI}$.
Note that ${D_{\cI}}^2$ is $(-1)^{k(k-1)/2}(i\vd_\t)^k$.  Since this is surjective (we can antidifferentiate $k$ times with respect to $\t$, and multiply by the correct power of $i$ to invert), it follows that $D_{\cI}$ is surjective.
\end{proof}

For the next few propositions it will be necessary to recall from Definition~\ref{defsfhgt} that given a component
\begin{equation}
      c := D_{\cI}\IU\,|~, \Label{eDefc}
\end{equation}we have $\hgt_0(c)=\#{\cI}$, and that given two components
\begin{align}
c_1 &:= D_{\cI}\IU\,|~, \Label{eDefc1}\\
c_2 &:= D_{\cJ}\IU\,|~, \Label{eDefc2}
\end{align}
then
\begin{equation}
 \dist_0(c_1,c_2)=\#({\cI}\Delta {\cJ})~. \Label{eDistc1c2}
\end{equation}

\begin{proposition}
Let $\IU$ be a scalar superfield, and let $D_{\cI}$ be a
superderivative.  Let $U_{\cI}:=D_{\cI}\IU\,|$ be the component corresponding to $\cI$.  Then the kernel of $D_{\cI}$ is the set of
superfields $\IU$ so that for all components $c$ of $\IU$,
\beq
\vd_\t^{(\dist_0(U_{\cI},c)-\hgt_0(c)+\hgt_0(U_{\cI}))/2}(c)=0.
\eeq
\end{proposition}

\begin{proof}
For $D_{\cI}\IU$ to be zero would mean every component is zero.  The components are of the form $D_{\cJ}D_{\cI}\IU\,|$ for some subset $\cJ$ of $\{1,\dots,N\}$.

Let ${\cK}={\cI}\Delta{\cJ}$, and let $m=\#({\cI}\cap {\cJ})$.
Then by anticommuting the various $D_I$ past each other, we get
\beq
    D_{\cJ}D_{\cI} = \pm (i\vd_\t)^m D_{\cK}~,
\eeq
so that each requirement that every $D_{\cal J} D_{\cal I}\IU\,|$ be zero turns into a requirement that $\vd_\t^m c=0$ for some component $c$.  Elementary Venn diagram arguments show that $\dist_0(U_{\cI},c)=\#{\cJ}$, $\hgt_0(c)=\#{\cK}$, and $\hgt_0(U_{\cI})=\#{\cI}$, and $m=(\#{\cJ} - \#{\cK}
+ \#{\cI})/2$.
\end{proof}

In the following, we consider a collection of subsets ${\cI}_\a\subset
\{1,\dots,N\}$ with
$\a$ ranging from 1 to $M$.  For every $\a$, we consider
the superderivatives
\beq
     \sD_\a:=D_{{\cI}_\a}~,
\eeq
 and define the corresponding component fields
\beq
     f_\a:=U_{{\cI}_\a}= \sD_\a \IU\,|~.
\eeq
Let $M_0$ be the number of $\a$ for which $\#{\cI}_\a$ is even, and
$M_1$ the number of $\a$ for which $\#{\cI}_\a$ is odd.
Let $\ell_\a$ be non-negative integers.

The primary object of study for much of the remainder of this section will be the superderivative operator
\beq
\bcD :=
 (\vd_\t^{\ell_1}\sD_1,\dots,\vd_\t^{\ell_M}\sD_M)
 :\sF^N_0\rightarrow
\prod_{i=1}^{M_0} \sF^N_0\times
\prod_{i=1}^{M_1} \sF^N_1~.
\Label{emainDmap}
\eeq

We define, for every component $c$ of $\IU$,
\beq
\mu(c):=\min_\a \bigl((\dist_0(f_\a,c)-\hgt_0(c)+\hgt_0(f_\a))/2+\ell_\a\bigr).\Label{edefm}
\eeq
For convenience, if $\cI$ is any subset of $\{1,\dots,N\}$, then we will
also use the notation
\beq
m(\cI) := \mu(U_{\cI})~.
\eeq

\begin{corollary}
In this setting, the kernel of $\bcD$ is the set of superfields $\IU$ so that for all components $c$ of $\IU$,
\beq
\vd_\t^{\mu(c)}(c)=0~.
\eeq
\end{corollary}

\begin{proof}
The kernel of the operator is the intersection of the kernels of each
$\vd_\t^{\ell_\a}\sD_\a$.  Thus the result follows from previous proposition.
\end{proof}

\begin{theorem}\Label{tsuperimage}
Suppose for every $\a$ and $\b$ with $\b\not=\a$, we have
\beq
\hgt_0(f_\a)-\hgt_0(f_\b)+2(\ell_\a-\ell_\b)
< \dist_0(f_\a,f_\b)~.\Label{ehgtcond}
\eeq

The Adinkra for the image of the map
\beq
 \bcD : \sF^N_0 \rightarrow
 \prod_{i=1}^{M_0} \sF^N_0\times\prod_{i=1}^{M_1} \sF^N_1
 \Label{eFatD}
\eeq
will have the same topology as the topology of the Adinkra for $\sF^N_0$, and
will have exactly $M$ sources: $s_1,\dots,s_M$.  For each $\a$ the component corresponding to $s_\a$ and the image of $f_\a$ under this map agree in the $\a$th coordinate.  There is a
height function $\hgt$ on this Adinkra so that for all $1\le \a\le M$,
$\hgt(s_\a)=\hgt_0(f_\a)+2\ell_\a$.
\end{theorem}

This theorem will be used when we have an Adinkra defined by its sources and a height function, as in the Hanging Gardens Theorem (Theorem~\ref{cDetA}).  The condition~(\ref{ehgtcond}), when phrased in terms of $\hgt$, is precisely
the condition~(\ref{ccrith}) necessary to specify an Adinkra.

\begin{proof}
The image of $\bcD$ is isomorphic to $\sF^N_0$ modulo the kernel of $\bcD$.
The kernel of $\bcD$ consists of superfields which satisfy the equations
\beq
\vd_\t^{m({\cI})}(U_{\cI})=0
\eeq
for each component field $U_{\cI}$.

For every $\cI\subset \{1,\dots,N\}$, define
\beq
V_{\cI}:=\vd_\t^{m(\cI)}U_{\cI}.
\eeq
The superfield $\IU$ is specified by determining $V_{\cI}$ uniquely up to
elements in the kernel of $\bcD$.  Therefore, the components of the image
of $\bcD$ are the $V_{\cI}$'s.

We now show that the components corresponding to $f_\a$ are sources.

For $\a\in\{1,\dots,M\}$, consider $f_\a$.  By equation~(\ref{edefm}),
\beq
      \mu(f_\a) = \min_{1\le \b\le M}
      \bigl( [\dist_0(f_\b,f_\a)
             -\hgt_0(f_\a)
              +\hgt_0(f_\b)]/2+\ell_\b\bigr)~.
\eeq
If we use $\b = \a$ in the minimization here, we get
\beq
     \mu(f_\a) \le \ell_\a~,
\eeq
and for $\b \not= \a$, we can use assumption~(\ref{ehgtcond}) to see that
\beq
     \min_{\b\not=\a} \bigl( [\dist_0(f_\b,f_\a)
                             -\hgt_0(f_\a)
                              +\hgt_0(f_\b)]/2+\ell_\b\bigr) > \ell_\a~.
\eeq
 Therefore $\mu(f_\a) = \ell_\a$.

From this, we see that $\vd^{\ell_\a}f_\a$ is one of the components of
$\bcD\,\IU$.  Let $s_\a=\vd^{\ell_\a}f_\a$ be this component.

We now determine the edges of the Adinkra corresponding to $\bcD\,\IU$.  Suppose we have two components $U_{\cI}$, $U_{\cJ}$ of $\IU$ connected by an edge.  Without loss of generality the arrow goes from $U_{\cI}$ to $U_{\cJ}$.  Then $Q_I(U_{\cI})=\pm U_{\cJ}$.  The corresponding components of $\bcD\,\IU$ are
$V_{\cI}=\vd_\t^{m({\cI})}(U_{\cI})$ and
$V_{\cJ}=\vd_\t^{m({\cJ})}(U_{\cJ})$.  We see that
\beq
 \begin{split}
 Q_I(V_{\cI})&=i^{m({\cI})}\vd_\t^{m({\cI})}Q_I(U_{\cI})~,\\[2mm]
 &=\pm i^{m({\cI})}\vd_\t^{m(\cI)} U_{\cJ}~,\\
 &=\pm i^{m(\cI) - m(\cJ)}\vd_\t^{m(\cI) - m(\cJ)} V_{\cJ}~.
 \end{split} \Label{eqtov}
\eeq
where in the last step we are implicitly assuming $m(\cI)-m(\cJ)\ge 0$.  In order to justify this assumption, and more generally discover what $m(\cI)-m(\cJ)$ must be, recall that
\beq
 \begin{split}
 m(\cI)&=\min_\a \bigl( [\dist_0(f_\a,U_{\cI})
                         -\hgt_0(U_{\cI})
                          +\hgt_0(f_\a)]/2+\ell_\a\bigr)~,\\[1mm]
 m(\cJ)&=\min_\b  \bigl( (\dist_0(f_\b,U_{\cJ})
                          -\hgt_0(U_{\cJ})
                           +\hgt_0(f_\b)]/2+\ell_\b\bigr)~.
 \end{split}
\eeq

Let $\a$ and $\b$ be such that
\begin{align}
 m(\cI)&= \bigl(\dist_0(f_\a,U_{\cI})-\hgt_0(U_{\cI})+\hgt_0(f_\a)\bigr)/2
           +\ell_\a~,\Label{emi}\\[1mm]
 m(\cJ)&= \bigl(\dist_0(f_\b,U_{\cJ})-\hgt_0(U_{\cJ})+\hgt_0(f_\b)\bigr)/2
           +\ell_\b~.\Label{emj}
\end{align}
Using the definition of minimum, we see that if we replace $\b$ in
equation~(\ref{emj}) with $\a$, we would get something at least as great as $m(\cJ)$:
\beq
     m(\cJ) \le
     \bigl(\dist_0(f_\a,U_{\cJ})-\hgt_0(U_{\cJ})+\hgt_0(f_\a)\bigr)/2
     +\ell\a~.
\eeq
Now by assumption there is an arrow pointing from the vertex corresponding to $U_{\cI}$ to the vertex corresponding to $U_{\cJ}$ in the $\IU$
Adinkra, so $\hgt_0(U_{\cJ})=\hgt_0(U_{\cI})+1$.  Also note that the
adjacency of $U_{\cI}$ to $U_{\cJ}$ implies that $\dist_0(f_\a,U_{\cJ})=\dist_0(f_\a,U_{\cI})\pm 1$.
Therefore
\beq
 \begin{split}
 m(\cJ)&\le\bigl(\dist_0(f_\a,U_{\cJ})
                 -\hgt_0(U_{\cJ})+\hgt_0(f_\a)\bigr)/2+\ell_\a~,\\[1mm]
 &=\bigl(\dist_0(f_\a,U_{\cI}) \pm 1
                 -(\hgt_0(U_{\cI})+1)+\hgt_0(f_\a)\bigr)/2 +\ell_\a~,\\[1mm]
 &\le\bigl(\dist_0(f_\a,U_{\cI})
                 -\hgt_0(U_{\cI})+\hgt_0(f_\a)\bigr)/2 +\ell_\a\\[1mm]
 &=m(\cI)~.
 \end{split}\Label{emj4}
\eeq
Likewise, plugging in $\a$ into (\ref{emi}) results in
\beq
 \begin{split}
 m(\cI)&\le\bigl(\dist_0(f_\b,U_{\cI})
                 -\hgt_0(U_{\cI})+\hgt_0(f_\b)\bigr)/2 +\ell_\b~,\\[1mm]
 &=\bigl(\dist_0(f_\b,U_{\cJ})\pm 1
                 -(\hgt_0(U_{\cJ})-1)+\hgt_0(f_\b)\bigr)/2+\ell_\b~,\\[1mm]
 &\le\bigl(\dist_0(f_\b,U_{\cJ})+2
                 -\hgt_0(U_{\cJ})+\hgt_0(f_\b)\bigr)/2+\ell_\b~,\\[1mm]
 &=m(\cJ)+1~.
 \end{split}
\eeq
Thus, we have
\beq
     0\le m(\cI)-m(\cJ) \le 1~.
\eeq
Thus we see that equation~(\ref{eqtov}) is justified.

Hence $Q_I(V_{\cI})$ is either $\pm V_{\cJ}$ or $\pm \vd_\t V_{\cJ}$, so that if $U_{\cI}$ and $U_{\cJ}$ are connected by an edge, then $V_{\cI}$ and $V_{\cJ}$ are connected by an edge.  Now if $U_{\cI}$ and $U_{\cJ}$ are not connected by an edge, then there is no $Q_I$ so that $Q_I(U_{\cI})$ is either $\pm U_{\cJ}$ or $\pm \vd_\t U_{\cJ}$.  And since we see that $Q_I(V_{\cI})$ for every $I$ is obtained by an edge in the original Adinkra from $U_{\cI}$, we see that $V_{\cI}$ would not be connected to $V_{\cJ}$ if $U_{\cI}$ were not connected to $U_{\cJ}$.
Therefore the edges are all the same as before, and the Adinkra for $\bcD\,\IU$ has the same topology as the Adinkra for $\IU$.

It immediately follows that the distance function $\dist$ on $\bcD\,\IU$ is the same as the old distance function $\dist_0$ on $\IU$, or more precisely,
\beq
     \dist(V_{\cI},V_{\cJ})=\dist_0(U_{\cI},U_{\cJ}).
\eeq

Now define the following function on the nodes of the image of $\bcD$:
\beq
\hgt(V_{\cI}) :=\hgt_0(U_{\cI}) + 2\, m(\cI).
\eeq

We now verify that $\hgt$ is a height assignment for $\bcD\,\IU$, according
to Definition~\ref{hgtadd}.  As above, suppose $U_{\cI}$ and $U_{\cJ}$ are components
of $\IU$ and $Q_I(U_{\cI})=\pm U_{\cJ}$.  As before, we have the corresponding
components
$V_{\cI}=\vd_\t^{m(\cI)} U_{\cI}$ and
$V_{\cJ}=\vd_\t^{m(\cJ)} U_{\cJ}$
of $\bcD\,\IU$.  Now recall that $V_{\cI}$ and $V_{\cJ}$ are connected
by an edge, where the arrow points from $V_{\cI}$
to $V_{\cJ}$ if $m(\cI)-m(\cJ)=0$ and where it points from $V_{\cJ}$ to $V_{\cI}$ if
$m(\cI)-m(\cJ)=1$.  Now we compute:
\begin{align}
 \hgt(V_{\cI})-\hgt(V_{\cJ})
 &=\hgt_0(U_{\cI}) + 2\, m(\cI) - \hgt_0(U_{\cJ}) - 2 \,m(\cJ)~,\\[1mm]
 &=2 (m(\cI) - m(\cJ)) -1~,\\
 &=\begin{cases}
    1, & \text{if $m(\cI)-m(\cJ)=1$},\\
    -1 &  \text{if $m(\cI)-m(\cJ)=0$}.
   \end{cases}
\end{align}
Thus, $\hgt$ is a height assignment for $\bcD\,\IU$.

For any vertex $V_{\cI}$,
\beq
 \begin{split}
 \hgt(V_{\cI})&=\hgt_0(U_{\cI}) +2 \,m(\cI)~\\
 &=\hgt_0(U_{\cI})+ \min_\a (\dist_0(f_\a,U_{\cI})-\hgt_0(U_{\cI})
                             +\hgt_0(f_\a)+2\ell_\a\bigr)~,\\
 &=\min_\a \bigl(\dist_0(f_\a,U_{\cI})+\hgt_0(f_\a)+2\ell_\a\bigr)~\\
 &=\min_\a \bigl(\dist_0(f_\a,U_{\cI})+\hgt_0(f_\a)+2\,\mu(f_\a)\bigr)~\\
 &=\min_\a \bigl(\dist_0(f_\a,U_{\cI})+\hgt(s_\a)\bigr)~\\
 &=\min_\a \bigl(\dist(s_\a,V_{\cI})+\hgt(s_\a)\bigr)~.
 \end{split}
\eeq
Note that this is the equation for the height function on an Adinkra
that has $M$ sources at $s_1,\dots,s_M$ at heights
 $\hgt(s_\a)=\hgt_0(f_\a)+\ell_\a$ as described in
 Equation~(\ref{ehgtfromh}) in the proof of the Hanging Gardens Theorem,
Theorem~\ref{tDetA} in Section~\ref{sHGT}, though modified to be dealing
with sources instead of targets, as in Corollary~\ref{cDetA}.
This corollary thus guarantees that this must be the Adinkra for the
image of the map $\bcD$~(\ref{eFatD}).
\end{proof}

\subsection{Superderivative constraints}\Label{sidentify}
We would now like to express the image of $\bcD$ by putting superderivative constraints on the range.  That is, instead of saying the multiplet is all $M$-tuples of superfields of the form
\beq
     (\vd_\t^{\ell_1}\sD_1\IU,\dots,\vd_\t^{\ell_M}\sD_M\IU)~,
\eeq
we would rather say the multiplet consists of superfields\Ft{Some of
these are scalar superfields and others are spinor superfields.  The notation
here does not distinguish between them because in this subsection, they are treated
identically.}
\beq
     (\IF_1,\dots,\IF_M)~,
\eeq
satisfying a certain finite set of relations involving $D_I$'s.

To come up with our constraints we will follow the example in
equations~(\ref{etFi=DiUc}) and (\ref{etFiCon}) and identify the components
in each of the $\IF_i$'s that come from the same component of $\IU$.

Every component $U_{\cJ}$ of $\IU$ is determined by a subset $\cJ$ of $\{1,\dots,N\}$.  Recall that the superderivatives $\sD_\a$ were defined using
subsets $\cI_\a$ so that $\sD_\a=D_{{\cI}_\a}$.  For each $\a$,
define the superderivative projection operator $P_{U_{\cJ},\a}(\B) := D_{\cK}\B \,|$,
where $\cK=\cJ\Delta \cI_\a$.  The point is that this
will extract the component of the $\a$th superfield in $\bcD\,\IU$
corresponding to $U_{\cJ}$.  More precisely, we have the following proposition:

\begin{proposition}
Let a number $1\le \a\le M$ and a component $c$ of $\IU$ be given.
If $P_{c,\a}$ is as above, then
\beq
P_{c,\a} \vd_\t^{\ell_\a}\sD_\a \IU = \pm i^m\vd_\t^{m} c
\eeq
for some $m$.  This $m$ will be written $m_\a(c)$.\Label{pcomp}
\end{proposition}

\begin{proof}
Write $c=U_{\cJ}$.
The operator $P_{c,\a}(\B)=D_{\cK}\B\,|$, where $\cK=\cJ\Delta \cI_\a$.
Therefore, the $D_I$ in $P_{c,\a}$ occur whenever $I$ is in $\cJ$ but
not in $\cI_\a$ or vice-versa.  Those that are in $\cI_\a$ but not in $\cJ$
will combine with the $D_I$ in $\sD_\a=D_{\cI_\a}$ to form derivatives.  Those
that are in $\cJ$ but not $\cI_\a$ join with the remaining $D_I$ in $\sD_\a$ to
form $D_{\cI}$.
\end{proof}

As a result, we can identify the components of each $\IF_\a$ that correspond to each component $c$: this is
\beq
     P_{c,\a}\IF_\a=\pm i^{m_\a(c)}\vd_\t^{m_\a(c)} c~.
\eeq
Let $\a$ and $\b$ be distinct integers in $\{1,\dots,M\}$.  Without loss of generality $m_\a(c)\le m_\b(c)$.  Then we should identify the component fields
\beq
P_{c,\a}\IF_\a \, |
 = \pm i^{m_\a(c)-m_\b(c)}\vd_\t^{m_\a(c)-m_\b(c)} P_{c,\b}\IF_\b \,|~.
 \Label{ecompconst}
\eeq
where the choice in $\pm$ should be taken to be compatible with Proposition~\ref{pcomp}.

If we instead write the constraint
\beq
P_{c,\a}\IF_\a
 = \pm i^{m_\a(c)-m_\b(c)}\vd_\t^{m_\a(c)-m_\b(c)} P_{c,\b}\IF_\b~.
 \Label{esfconst}
\eeq
then the result will be more constraints, including (\ref{ecompconst}),
but also the result of applying various superderivatives
$D_{\cI}$.  It is straightforward to see that these will be
the same as the component-wise constraints~(\ref{ecompconst}) that would
be identified anyway through this procedure, or else a certain number
of derivatives applied to such constraints.

Therefore we have proved that the following algorithm works.

\begin{theorem}[Superderivative Identification Algorithm]~%
 \Label{pA=SF}\newline
 Let $\cA$ denote a given engineerable Adinkra, and $\cA_\IU$ an Adinkra of
 the same
 topology for which however a corresponding superfield, $\IU$,
 has been identified. Then $\cA$ has a superderivative superfield
 representation in terms of $\,\IU$ as follows:
  \vspace{-1mm}
\begin{enumerate}\itemsep=-3pt
 \item Let $v_1, \dots, v_M$ be the source vertices in $\cA$.
 \item \Label{i2}
  Transform $\cA$ into $\cA_\IU$ by iteratively lowering vertices, using the
  procedure in Section~\ref{sVRaising} as in Corollary~\ref{cHGT2}.  In the
  process the various $v_\a$ may be lowered at various times.  For each $\a$
  let $\ell_\a$ be the number of times $v_\a$ was lowered in this sequence.
 \item
 Let $\tilde{v}_1, \dots, \tilde{v}_M$ be the vertices in $\cA_\IU$ that
 are the lowered versions of $v_1, \dots, v_M$.
 \item
  For each $\a$, let $\cI_\a$ be the subset of $\{1,\dots,N\}$ so that
  $D_{\cI_\a} \IU\, |$
  is the field corresponding to $\tilde{v}_\a$.  Such is guaranteed
  by~(\ref{eproj}).  Define $\sD_\a=D_{\cI_\a}$.
 \item
 The Adinkra for the image of the map
\beq
\bcD:=(\vd_\t^{\ell_1}\sD_1,\dots,\vd_\t^{\ell_M}\sD_M):\sF^N_0
\rightarrow
\prod_{i=1}^{M_0} \sF^N_0\times\prod_{i=1}^{M_1} \sF^N_1
\eeq
is $\cA$ (Here, $M_0$ is the set of bosonic nodes in Step~1, and $M_1$ is
the set of fermionic nodes).  Henceforth we will use
\beq
     (\IF_1,\dots,\IF_M)
\eeq
for a typical element of the right side, suppressing notationally the
distinction between scalar and spinor superfields.
\item For each component $c$ of $\IU$, and every integer $\a$ in $\{1,\dots,M\}$,
construct $P_{c,\a}$ and determine $m_\a(c)$ as in Proposition~\ref{pcomp}.
\item For each component $c$ and pair of distinct integers $\a$, $\b$ in
$\{1,\dots,M\}$, with $m_\a(c)\ge m_\b(c)$, write down the
superdifferenital constraint
\beq
     P_{c,\a}\IF_\a = \vd_\t^{m_\a(c)-m_\b(c)} P_{c,\b}\IF_j
     \Label{eSCostraints}
\eeq
\item The superfield multiplet $(\IF_1,\dots,\IF_M)$ subject to the above
constraints~(\ref{eSCostraints}) has $\cA$ for its Adinkra.
\end{enumerate}\vspace{-2mm}
\end{theorem}
 \Remk
The system~(\ref{eSCostraints}) is most often redundant: several of the constraints in the system may follow from others, upon an application of
some superderivative $D_{\cI}$.

Since various arrays of superderivatives of $\,\IU$ correspond to each
Adinkra in a family, $\,\IU$ is called the {\em underlying superfield\/}
of this family.

\subsection{Topology}
 \Label{sTop}
Although much of the setup to Theorem~\ref{pA=SF} was through the
unconstrained superfields which have cubical topology,\Ft{Recall from
Section~\ref{Graph} that an $N$-cubical topology is the graph of $[0,1]^N$.} it is
straightforward to see that the above algorithm continues to work as
long as every one-source Adinkra with that topology is the adinkra for
some set of superfields whose specification is in terms of
superderivative constraints.

The question is whether this always happens for an Adinkra that describes
$d=1$ supersymmetry with no central charge.  For instance, for $N=4$, it
turns out that there are two distinct topologies that an Adinkra can have:
\begin{equation}
 \vC{\includegraphics[width=3in]{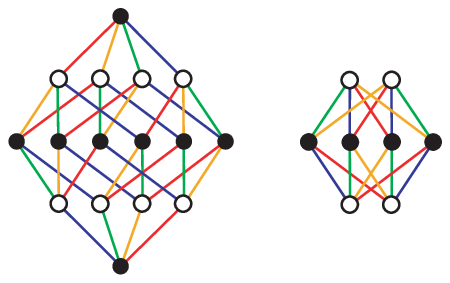}}.
 \Label{eN4}
\end{equation}
In this case, the right-hand Adinkra can be obtained from the
left-hand one via: raising the lowest and lowering the highest
vertex, and then imposing a pairwise, horizontal identification
of vertices.
As it turns out, this Adinkra, though it is not cubical, is the dimensional
reduction of the standard $d=4$, $N=1$ chiral superfield, which is a
superfield under the superderivative constraint
$\bar{D}_{\dot{\a}}\IU=0$.\cite{rA}  This suggests that perhaps
at least some of these cases of non-cubical topology will be describable
in terms of superderivative constraints.

Adinkras that describe $d=1$ supersymmetry with no central charge turn out
to be obtained by quotients of such cubical topologies by imposing a sequence
of certain two-to-one identifications.  The first such $N$ where this appears
is $N=4$.  These ideas will appear in more detail in forthcoming work by the authors.

If the ideas of Subsection~\ref{sidentify} can be made to accomplish these
identifications, it would follow that all such could be described in this fashion.

Considering the `new' topology in the right-hand side Adinkra in~(\ref{eN4}),
we find it fascinating that the {\em Adinkra topology\/} has answered an
old question: ``Why do chiral superfields occur only for $N\geq4$?'' That is, we note that the pair of bosonic, white vertices corresponds to a pair of real component fields which define a {\em complex\/} component field, $A(x)$. At the next level, the two pairs of fermionic, black vertices may be identified with the two, complex components of a Weyl spinor, $\j_\a(x)$ with $\a=1,2$, in 4-dimensional spacetime. Finally, the pair of bosonic, white vertices at the top may be identified with the {\em complex\/} `auxiliary' component field, $F(x)$. Furthermore, the four supersymmetry generators, $Q_I$ with $I=1,2,3,4$, may be combined into two complex generators, $Q_\a$, whereupon this Adinkra has become
\begin{equation}
  \vC{\includegraphics[width=1in]{box2.eps}}~, \Label{AdChiral}
\end{equation}
which appears identical to~(\ref{Adn2scalar}), except that the Adinkra is now understood to be {\em complex\/}: the whole graph, vertices and arrows, represent objects and mappings over the field $\IC$.

 In fact, the reverse of this operation, often called the {\em forgetful functor\/}, can be used to ``double'' {\em any\/} existing real Adinkra. One first complexifies an Adinkra by assigning to each vertex a complex component field and compatible complex supersymmetry transformation to each arrow. Then one forgets the complex structure by splitting the real and imaginary parts of the component fields and of the supersymmetry transformations. This simple operation doubles both the number of vertices and also the ``extendedness'', $N$, of supersymmetry.

Another simple operation consists of the deletion of all edges of a given color, thereby transforming a given $(1|N)$-supersymmetry Adinkra into a disjoint pair of $(1|N{-}1)$-supersymmetry Adinkras. This might be termed fermionic dimensional reduction. The obvious reverse operation, consisting of copying a given Adinkra and then connecting the corresponding vertices by edges of a new color construct a $(1|N{+}1)$-supersymmetry Adinkra from an $(1|N)$-supersymmetry one. This then should be termed fermionic dimensional oxidization.

All these simple operations have a manifest analogue within superfields. Unfortunately, these do not generate (by far) all the possible topologies for larger and larger $N$. Instead, to guarantee the existence of a superfield for every topology---as assumed in Proposition~\ref{pA=SF}---it will be necessary to (1)~devise an iterative algorithm which generates all the two-to-one identification of the cubical Adinkras of higher and higher $N$, and then (2)~determine the superfield analogues of each of those projections. While this work has not been done, we feel the affirmation of this assumption to be sufficiently tempting and suggestive from the foregoing discussion:
\begin{conjecture}\Label{c?}
 For every $N$ and every topology of engineerable Adinkras, there exists
 a corresponding set of superderivative constraints, such that the set of
 superfields satisfying these constraints has an Adinkra of that topology.
\end{conjecture}

\section{The Main sequence of Adinkras}\Label{mainseq}
The process of vertex raising or lowering in Section~\ref{sVRaising} applies
to Adinkras, and are mirrored on superfields by acting by the various $D_I$'s.
In fact, the examples in Sections~\ref{sN=1} and \ref{sN=2} suggest a structure
among Adinkras of this type.

\subsection{The $N=1$ main sequence}
Recall from Subsubsection~\ref{sN1SS} that in $N=1$ there are two kinds of superfields: the scalar superfield $\Phi_\a$ and the fermionic superfield $\Lambda_\a$.  The superderivative operator $D$ maps between them as in mappings~(\ref{dphi}) and (\ref{dlam}).  Recall that these each performed vertex raises.

\begin{proposition}\Label{pHGTN1}
The superderivative maps~(\ref{dphi}) and~(\ref{dlam}) provide the sequence of superfields corresponding to the only $N=1$ {\em main sequence\/} of the Vertex Raising Theorems (Theorem~\ref{tHGT} and Corollaries~\ref{cHGT} and \ref{cHGT2}):
\begin{equation}
 \vC{\includegraphics[width=1.75in]{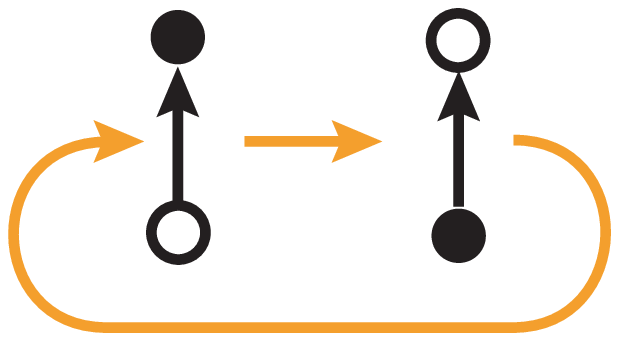}}~.
 \Label{eN1MS}
\end{equation}
\end{proposition}
\begin{proof}
 Concatenating the results of~(\ref{dphi}) and~(\ref{dlam}), and
 using~(\ref{DPhiLam}), we find that
 \beq
      (D\tilde\L_{\p/2})\simeq\dot{\F}_{\p/2}~.
 \Label{eDPhiLam2}
 \eeq
 We effectively apply the $D$ map twice in succession, and an Adinkra
 is raised two levels,
 \brr
  \vC{\includegraphics[width=1.6in]{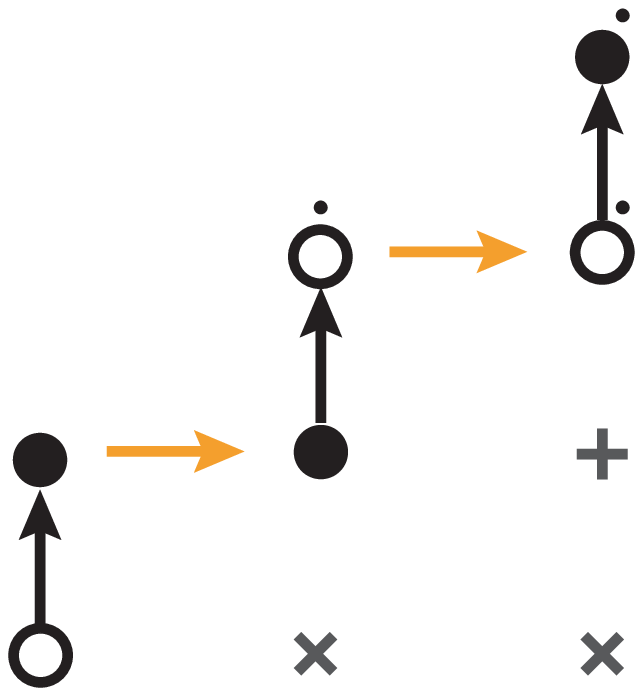}}~, \Label{eDDmap}
 \err
 so that an scalar Adinkra is mapped to another scalar Adinkra. The image
 of this double $D$-map is the time-derivative of the original Adinkra.
 This is a reflection of the algebraic fact that the superspace derivative
 $D$ squares to $i\,\vd_\t$, the phase factor of $i$ is manifest in
 superspace by the physically irrelevant shift in the $\a$ phase, defined
 in~(\ref{n1sSFa})--(\ref{n1fSFa}): $D^2:\F_\a\to(\vd_\t\F_{\a+\p/2})$.
 As discussed above, this overall phase shift is irrelevant when
 considering the superfields, the corresponding super-multiplets and
 their Adinkras as representations of supersymmetry.

 The kernel of the mapping is spanned by the scalar and the spinor constants, represented, respectively, by `$\times$' and `$+$' in the diagram~(\ref{eDDmap}), both comprising the trivial `zero-mode' representations of supersymmetry. As described in Subsection~\ref{sN1SS}, these constants may always be re-absorbed into the component fields as integration constants. The superfields corresponding to the starting and the ending Adinkra in the sequence~(\ref{eDDmap}) may thus be identified.

  The two supersymmetry representations~(\ref{phi}) and~(\ref{lam}) thus comprise the {\em main sequence\/}---and, for $N=1$, the only sequence---of supersymmetry representations, as obtained by vertex raising or, alternatively, by considering either the sequence $\F$--$(D\F)$ or the $\L$--$(D\L)$ one.
\end{proof}

\subsection{The $N=2$ main sequence}
 \Label{sRaiSF}
 We now consider the Adinkras associated to
 $N=2$ superfields described in Subsection~\ref{sN=2}, and the superderivatives associated with them.

 First, recall that, by proposition~\ref{pDIUinFI},
 the mapping $D^{\sss(1)}_I:\IU\to(D_I\IU)$ has its image in $\IF_I$ and
 that $\ker(D^{\sss(1)}_I)=(u\6(0);0;0)$ is a `zero-mode' superfield
 consisting of a single, scalar constant. It remains to determine the
 cokernel of the map $D^{\sss(1)}_I:\IU\to(D_I\IU)$, and to this end we
 now proceed to prove:
\begin{proposition}\Label{pHGTN2}
Let $\,\IU$ denote an unconstrained $N=2$ superfield and $\cA_\IU$ the corresponding Adinkra. Then, there exists a semi-infinite sequence of linear mappings between superfields, generated from $\,\IU$ by the action of superderivative maps constructed from $D_I$~(\ref{Di}), which contains superfields corresponding to all Adinkras of the topology of $\cA_\IU$.
\end{proposition}

\begin{proof}
 Counting dimensions in terms of real-valued {\em functions\/}, we see that\Ft{For a supermultiplet and a superfield, we separate the total number of independent component fields of the same engineering dimension, from those of higher and lower engineering dimension, by the `$|$' divider.}
  $\dim_{\IR}\IU=(1|2|1)$, $\dim_{\IR}\IF_I=(2|4|2)$ and
  $\dim_{\IR}\ker(D^{\sss(1)}_I)=(0|0|0)$. Thus, it must be that
  $\dim_{\IR}\cok(D^{\sss(1)}_I)=(1|2|1)$, so we must identify this cokernel, and a mapping $\m$ that satisfies:
 \begin{equation}
 0\to(u\6(0);0;0)\tooo{~\i~}\IU\tooo{~D^{\sss(1)}_I~}\IF_I
   \tooo{~\m~}\cok(D_I)\to0~. \Label{eCokDi}
\end{equation}
That is, $\m\circ D^{\sss(1)}_I=0$, and moreover $\im(D^{\sss(1)}_I)=\ker(\m)$. A little experimentation provides that
\begin{equation}
 \ID_K{}^JD_J=0~,\quad\hbox{where}\quad
 \ID_K{}^J:=\tw\s_K{}^{IJ}D_I~,\quad
 \tw{\BM{\s}}_K=\begin{cases}\BM{\s}_3 & \text{for $K=1$},\cr
                       \noalign{\vglue1mm}
                       \BM{\s}_1 &\text{for $K=2$},\cr\end{cases}\Label{eIDmap}
\end{equation}
where $\BM{\s}_1,\BM{\s}_3$ are the standard Pauli matrices. Indeed,
\begin{equation}
 \ID_K{}^JD_J=\tw\s_K{}^{IJ}D_ID_J=
  \begin{cases}D_1^2-D_2^2\isBy{(\ref{nalg})}0 &\text{for $K=1$},\cr
         \noalign{\vglue3mm}
         D_1D_2+D_2D_1\isBy{(\ref{nalg})}0 &\text{for $K=2$}.\cr\end{cases}
 \Label{eIDD=0}
\end{equation}
Moreover, $\ID_K{}^J\IF_J$ does not vanish for a general $\SO(2)$-doublet superfield, $\IF_J$, but does vanish precisely when acting upon
$\im(D^{\sss(1)}_J)\subset\IF_J$. Note, however, that the matrix $\ID_K{}^J$ does {\em not\/} transform as a tensor with respect to this $\SO(2)$. Indeed, this should be expected from the Adinkra transformations~(\ref{d1map}) and~(\ref{d2map}) indicated by the separate horizontal orange arrows, the raising of a single vertex. In fact, the two traceless, symmetric matrices $\BM{\s}_1,\BM{\s}_3$ generate the $\SU(2)/\SO(2)$ coset, $\SO(2)\into\SU(2)$ being generated by the imaginary Pauli matrix, $i[\BM{\s}_2]^{IJ}=\ve^{IJ}$. Finally, we note that the matrix $\ID_K{}^J$ satisfies the following properties:
\begin{equation}
 [\ID_K{}^I]
 =\left[\begin{matrix}D_1 & \!-D_2\cr D_2 &~\,D_1\end{matrix}\right]~,\quad\hbox{so}\quad
  \ID_K{}^JD_J=0~,\quad\hbox{and}\quad \ID_L{}^K\ID_K{}^J=0~.\Label{eSqDKJ}
\end{equation}

Of course, the {\em image\/} of the mapping
$\ID^{\sss(2)}_K{}^J:\IF_J\to(\ID_K{}^J\IF_J)$ `lives' in another $\SO(2)$-doublet superfield, akin to $\IA_I$, and we obtain a semi-infinite sequence of superfield mappings:
\begin{equation}
 0\to(u\6(0);0;0)\tooo{~\i~}\IU\tooo{~D^{(1)}_I~}\IF_I
   \tooo{~\ID^{(2)}_J{}^I~}\IA_J\tooo{~\ID^{(3)}_K{}^J~}\IF'_K
    \tooo{~\ID^{(4)}_L{}^K~}\IA'_L\tooo{~\ID^{(5)}_M{}^L~}\cdots
 \Label{eCanon}
\end{equation}
To identify the kernels and cokernels in this sequence of mappings, we recall that any such sequence may be resolved into a zig-zag weave of short exact sequences, of which we show here but the left-most end:
 \begin{equation}
 \vC{\begin{picture}(100,48)(25,-1)
  \put(0,0){\includegraphics[width=5.5in]{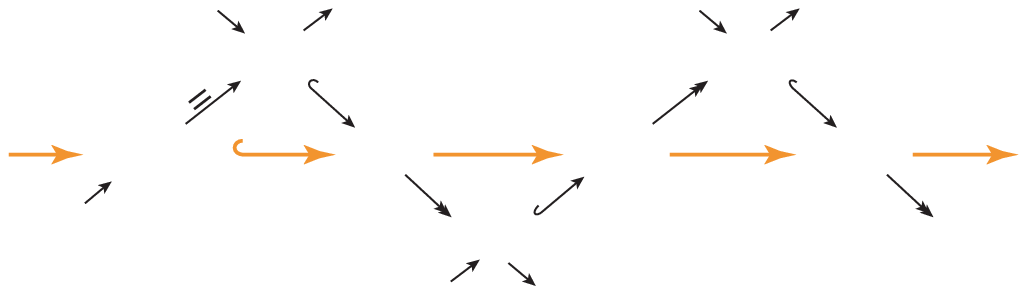}}
  \put(0,20.5){0}
  \put(10,11){0}
  \put(28,42){0}
  \put(50,42){0}
  \put(93,42){0}
  \put(111,42){0}
  \put(60,1){0}
  \put(75,1){0}
  \put(56,9){$\tw\IF_I:=(D_I\IU)$}
  \put(14,20.5){$(u\6(0);0;0)$}
  \put(52,20.5){$\IU$}
  \put(84.5,20.5){$\IF_I$}
  \put(117,20.5){$\IA_J$}
  \put(31,33){$(u\6(0);0;0)$}
  \put(90,33){$\tw\IA_J:=(\ID_J{}^I\IF_I)$}
  \put(39,23){\TC{orange}{$\i_0$}}
  \put(66,23){\TC{orange}{$D^{\sss(1)}_I$}}
  \put(98,23){\TC{orange}{$\ID^{\sss(2)}_J{}^I$}}
  \put(126,23){\TC{orange}{$\ID^{\sss(3)}_K{}^J$}}
  \put(47.5,29){$\SSS\i_0$}
  \put(61,16){$\SSS D_I$}
  \put(75,16){$\SSS\i_1$}
  \put(88.5,28){$\SSS\ID_J{}^I$}
  \put(112,29){$\SSS\i_2$}
  \put(126,16){$\SSS\ID_K{}^J$}
  \put(140,20.5){$\cdots$}
 \end{picture}}
 \Label{eZigZag}
\end{equation}
As indicated in the sequence~(\ref{eCanon}), this continues indefinitely to the right; all inclusion injections are indicated by $\i$. The horizontal, orange maps are the ones appearing in the sequence~(\ref{eCanon}), and are factored by the diagonal sequences of mappings so as to exhibit various kernels and cokernels. In particular, the `fused' pair $\tw\IF_I:=(D_I\IU)$ spans
$\im(D^{\sss(1)}_I)=\ker(\ID^{\sss(2)}_J{}^I)$, and an analogously fused pair
$\tw\IA_J:=(\ID_J{}^I\IF_I)$ spans
$\im(\ID^{\sss(2)}_J{}^I)=\ker(\ID^{\sss(3)}_K{}^J)$.

The simplest is, of course, the beginning at the left, where the left-most exact SE-sequence identifies the `zero-mode' representation of supersymmetry,
$(u\6(0);0;0)$ as $\ker(D^{\sss(1)}_I)$. Conversely, the same sequence
identifies $\tw\IF_I$ as the cokernel of $\i_0$, so that
\begin{equation}
 \tw\IF_I = \cok(\i_0) \iso{D} \IU/\i_0(u\6(0);0;0)
  = \{\IU \equiv \IU + \i_0(u\6(0);0;0)\}~.
 \Label{etFiEq}
\end{equation}
That is, $\tw\IF_I$ may be regarded as the `super-gauge' equivalence class, very much as in~(\ref{eDFeq}). Just as there, here too the isomorphism, denoted `$\iso{D}$' consists of the straightforward identification of both fermionic components, $\w_I(\t)=i\c_I(\t)$, and one bosonic component,
$\inv2(F_{21}-F_{12})=iU(\t)$, but the {\em derivative identification\/} of the other bosonic component, $\inv2(F_{11}+F_{22})=i\dot{u}(\t)$. This, of course, corresponds to the vertex raising in the Adinkra presentation.

The action of the diagonal $\i_i$'s, for $i>0$, is far from trivial, however, as can be gleaned from studying the relationships among the Adinkras in the same zig-zag diagram:
\begin{equation}
 \vC{\begin{picture}(100,48)(25,-1)
  \put(0,0){\includegraphics[width=5.5in]{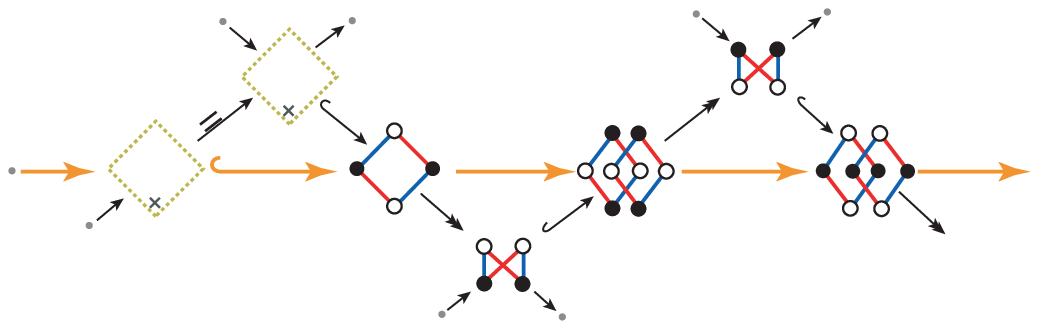}}
  \put(52.5,20.25){$\IU$}
  \put(59,8){$\tw\IF_I$}
  \put(89.5,15.5){$\IF_I$}
  \put(107,34){$\tw\IA_I$}
  \put(109,15.5){$\IA_I$}
  \put(37,23){\TC{orange}{$\i_0$}}
  \put(66,23){\TC{orange}{$D^{\sss(1)}_I$}}
  \put(98,23){\TC{orange}{$\ID^{\sss(2)}_J{}^I$}}
  \put(126,23){\TC{orange}{$\ID^{\sss(3)}_K{}^J$}}
  \put(47.5,29){$\SSS\i_0$}
  \put(61,16){$\SSS D_I$}
  \put(74.5,16){$\SSS\i_1$}
  \put(89.5,29){$\SSS\ID_J{}^I$}
  \put(111,30){$\SSS\i_2$}
  \put(126,16){$\SSS\ID_K{}^J$}
  \put(140,20.5){$\cdots$}
 \end{picture}}
 \Label{eZigZagA}
\end{equation}
First, we note that the two {\em separate\/} components of the horizontal, orange $D_I$ are indeed mapping into the two correspondingly separate components of $\IF_I$: $\{\IF_I\}=\IF_1\oplus\IF_2$, and both components, $\IF_1$ and $\IF_2$ are separately isomorphic to the superfield $\IB$, defined in~(\ref{bdef}).

Next, we note that $\tw\IF_I:=(D_I\IU)$ is included in $\IF_I$ in a non-trivial fashion: Owing to the exactness of the diagonal sequences, the NE-sequence with $\IF_I$ in the middle ensures that $\tw\IF_I=\ker(\ID^{\sss(2)}_J{}^I)$, and this gives an alternate description: Given an independent $\SO(2)$-doublet of superfields $\IF_I$, defined as in~(\ref{fidef}), we obtain:
\begin{equation}
 \tw\IF_I = \{\IF_I\,:\>\ID_J{}^I\IF_I=0\,\}~.
 \Label{etFi}
\end{equation}
which defines $\tw\IF_I$ as a super-constrained superfield.
The fact that the previous, SE-sequence provides that
\begin{equation}
 \tw\IF_I=(D_I\IU)~,
 \Label{etFi=DiU}
\end{equation}
is the {\em superderivative superfield\/} solution\Ft{This nomenclature is perfectly analogous to the standard one in $d=4$, where, \eg, the constrained {\em chiral\/} superfield is defined so as to satisfy the super-constraint $\Db_{\dot\a}\F=0$, and which is solved by $\F=\Db_{\dot\a}\J^{\dot\a}$, in terms of an otherwise unconstrained, 2-component, Weyl fermion superfield, $\J^{\dot\a}$\cite{r1001}. The strange superspace fact that superderivative equations are solved by superderivatives of superfields rather than antiderivatives is a reflection of the fact that fermionic integration is equivalent to superderivatives, and the use of invariant superderivatives ensures invariance with respect to supersymmetry.} of the super-constraint $\ID_J{}^I\IF_I=0$, the supersymmetry-invariant rendition of the component field constraint system~(\ref{etFiCon}). Indeed, applying the invariant projection operators~(\ref{eproj}) on $\ID_J{}^I\IF_I=0$ reproduces the system of component constraints~(\ref{etFiCon}).

In turn, this same short exact NE-sequence also represents:
\begin{equation}
 \tw\IA_J = \cok(\i_1) \iso{\ID} \{\IF_I\}/\i_1(\tw\IF_I)
  = \{\IF_I \equiv \IF_I + \i_1(\tw\IF_I)\}~,
 \Label{etAiEq}
\end{equation}
where, in turn, $\tw\IF_I$ is defined by the equivalence relation~(\ref{etFiEq}). This isomorphism, `$\iso{\ID}$' again includes a vertex raising time-derivative action. The situation in which the generator of an equivalence relation is itself an equivalence class, and when such a concatenation of relationships continues indefinitely, is not unfamiliar in physics, and reminds of the `ghost-for-ghost' phenomenon with BRST symmetry. The sequences~(\ref{eCanon}), (\ref{eZigZag}) and~(\ref{eZigZagA}) then represent a corresponding $(1|1)$-supersymmetry construction.

The next, second SE-sequence then includes $\tw\IA_J$ in the $\SO(2)$-doublet $\IA_J$ as its constrained subset
\begin{equation}
 \tw\IA_j = \ker(\ID^{\sss(3)}_K{}^J) = \{\IA_J\,:\>\ID_K{}^J\IA_J=0\,\},
 \Label{etAi}
\end{equation}
which in turn is solved by the assignment
\begin{equation}
 \tw\IA_J=(\ID_J{}^I\IF_I)~,
 \Label{etAj=DjiFi}
\end{equation}
in terms of otherwise unconstrained $\SO(2)$-doublet, $\IF_I$. Noting that the components of the $\SO(2)$-doublet $\IA_J$, both $\IA_1$ and $\IA_2$ are separately isomorphic to the superfield $\IU$ from which we started, the sequence in a sense comes full circle; the rest of it replicates the foregoing.

Finally, we remind that the superfields defined in the sequence~(\ref{eCanon}) and~(\ref{eZigZag}) have an ever rising engineering dimension. That is, the lowest component field of each one superfield defined in this sequence, $\IU,\IF_I,\IA_J,\ldots$, has an engineering dimension $\inv2$ larger than that of the lowest component of the immediately preceding one. The constrained superfields, $\tw\IF_I,\tw\IA_J,\ldots$ are, of course, included in the $\SO(2)$-doublet superfields $\IF_I,\IA_J,\ldots$, appearing in~(\ref{eCanon}) and the horizontal sequence in~(\ref{eZigZag}), so that the same observation also covers these constrained superfields, as well as their alternate, equivalence-class definitions~(\ref{etFiEq}), (\ref{etAiEq}) and so on.

Finally, the depiction~(\ref{eZigZagA}) of the same sequence shows that all Adinkras of the topology of $\cA_\IU$ appear in the sequence.
\end{proof}

\Remk
The entire sequence~(\ref{eZigZag}) is generated, from the unconstrained superfield $\IU$, by the initial action of the pair $D_I$ and thereafter by the iterated action of the {\em same\/} $2\times2$ matrix of superderivatives $\ID_J{}^I$.

\subsection{The `Main Sequence' of $N=2$ Supermultiplets}
 \Label{sMainS}
 In this section we identify the cyclicality of the sequence~(\ref{eZigZagA}), thus defining the main sequence of $N=2$ superfields, corresponding to the main sequence of Adinkras.

 Consider once again the sequence of vertex raises shown in~(\ref{d1map})
 and~(\ref{d2map}), as repeated in the zig-zag sequences~(\ref{eZigZag}) and~(\ref{eZigZagA}).  If we focus only on the supermultiplet structures which
 appear in these, suppressing any reference to zero-modes lost, to the
 particular ``heights" of the lowest vertices in the Adinkras, and omit
 the dots from vertices since these are without intrinsic meaning, we can
 reproduce the essence of these maps as follows,
 \beq
 \vC{\includegraphics[width=5in]{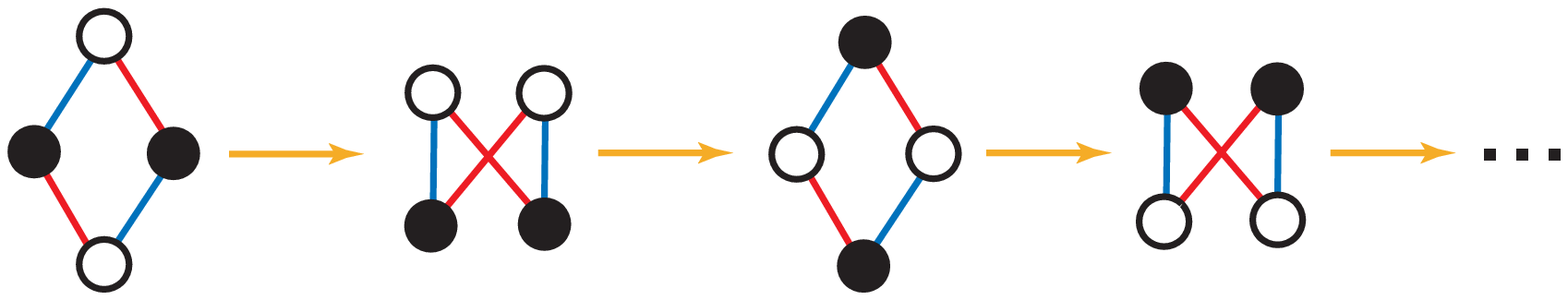}}
 \Label{short}
 \eeq
 The interpretation of this sequence is as follows. We start with a
 particular Adinkra, in this case the Adinkra corresponding to the
 real scalar superfield $\IU$ given in in~(\ref{scalarsf}), then we apply vertex raises to obtain
 new Adinkras.  In each case, one of the ``source" vertices,
 (vertices to which arrows only point away from), is ``raised"
 to generate a new Adinkra.  The second Adinkra in the sequence~(\ref{short})
 corresponds to the multiplet described by~(\ref{xrules}), which,
 in turn, represents a constrained (irreducible) spinor
 doublet superfield $\tw\IF_I$. The third multiplet in~(\ref{short}) corresponds to the
 spinor superfield $\IB$ given in~(\ref{bdef}).  In this
 way we map supermultiplets into one another by a well-defined
 process.  As explained so far, this sequence
 proceeds as $\IU\dto\tw\IF_I\dto\IB\dto\cdots$.

 The ellipses in~(\ref{short}) indicates that we can
 continue the process of raising vertices to generate
 further engineerable Adinkras.  To do this, we take the only
 source vertex in the third Adinkra in~(\ref{short}), and raise
 this to obtain a new Adinkra.  We then raise one of the source
 vertices in the Adinkra which results.  Interestingly,
 this process returns the initial Adinkra in the sequence,
 which therefore becomes cyclic,
 \beq
  \vC{\includegraphics[width=5in]{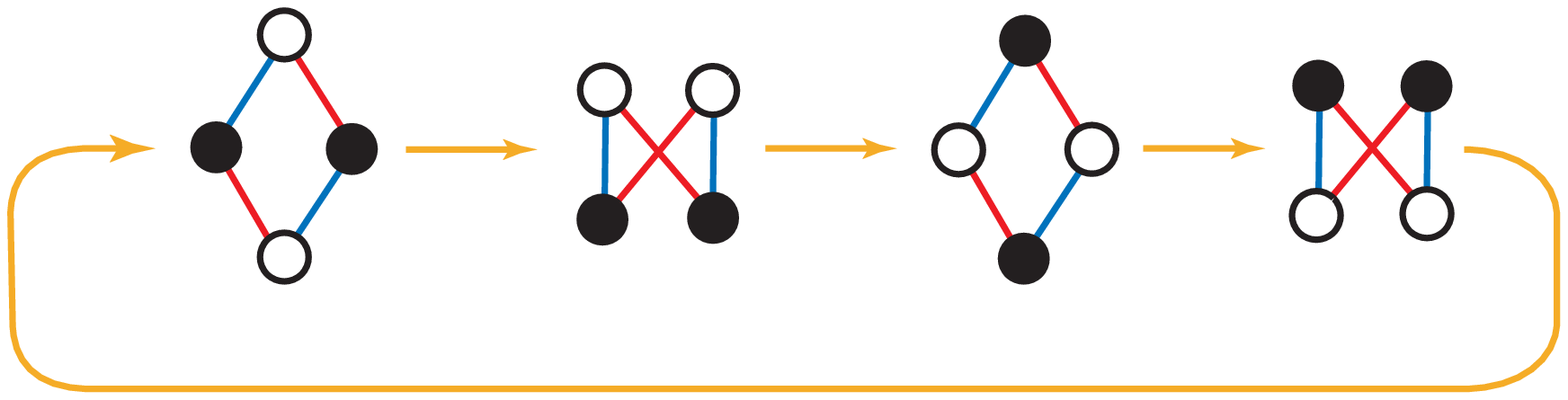}}~.
 \Label{long}
 \eeq
 The fourth Adinkra in this sequence corresponds to an irreducible constrained
 version, $\tw\IA_I$, of the doublet scalar superfield $\IA_I$, described
 in~(\ref{supa}).  Thus, this $N=2$ looping sequence correlates with
 superfields according to
 \beq
  \IU~\dto~\tw\IF_I~\dto~\IB~\dto~\tw\IA_I~\dto~\IU~,
 \Label{eCanon2}
 \eeq
 although, in fact, the superfield obtained on the far right would not
 be $\IU$ identically, but its overall derivative, $\vd_\t\IU$.

 Consider now performing the same process, starting with the same Adinkra, corresponding to the superfield $\IU$, as on the far left of the sequence~(\ref{long}), but now distinguish all four vertices and keep the dots indicating derivatives for relative reference. We obtain:
 \beq
  \vC{\includegraphics[width=5in]{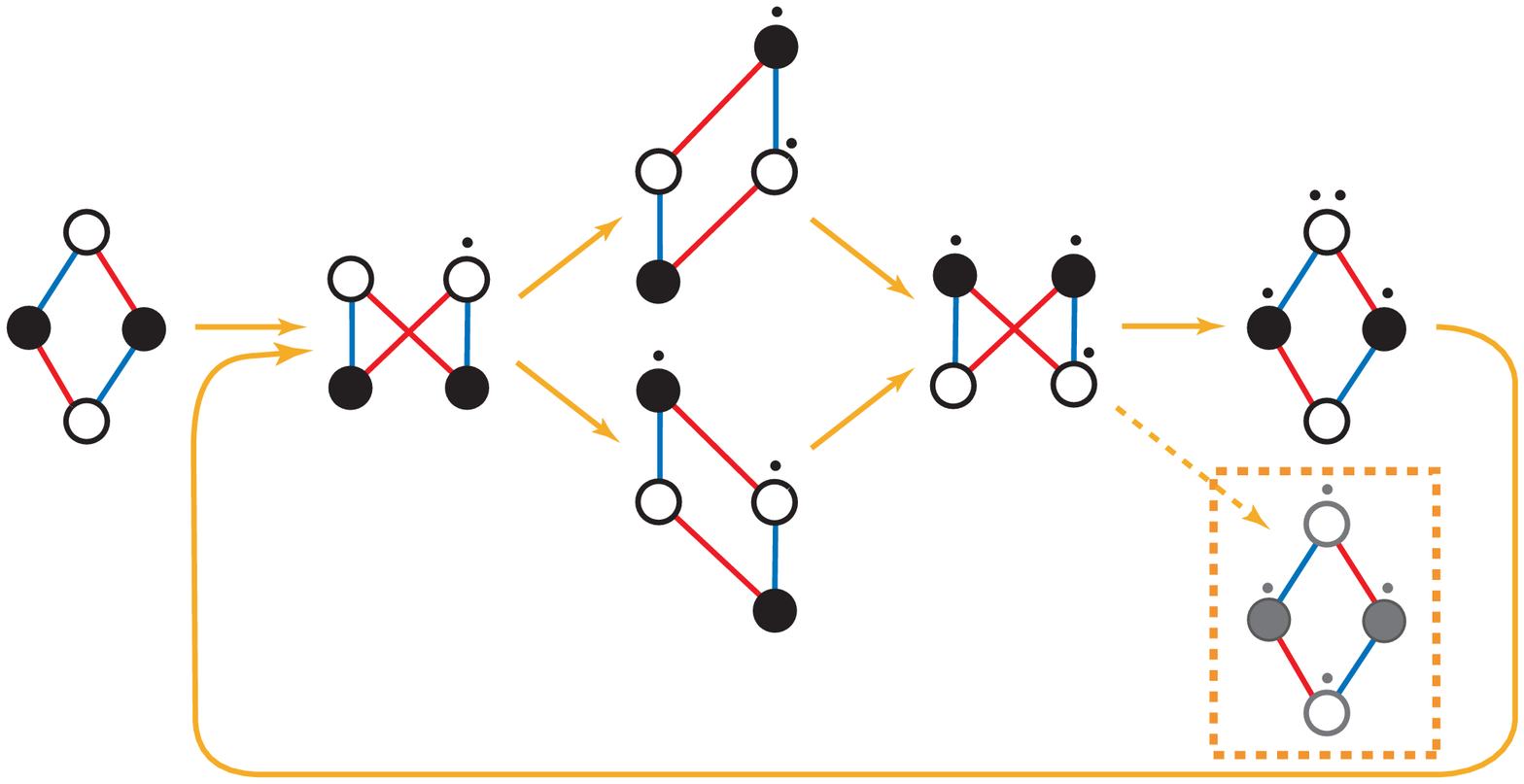}}~.
 \Label{eSeqN2}
 \eeq
The framed, gray Adinkra in the lower right-hand corner represents a supermultiplet that is a ``trivial,'' overall time-derivative of the initial one. The one above it is identical in structure, but corresponds to a supermultiplet in which the top component field of $\IU$ is now the lowest; this happens for all even $N$. There are only four distinct Adinkras in this {\em main sequence\/} since the time-derivative distinctions are irrelevant for considering the supersymmetry action on these supermultiplets; these distinctions however trace the way these supermultiplets are generated in this sequence. Furthermore, every Adinkra generated further to the right is the overall time-derivative of another that is already in the main sequence.

On the other hand, note that enforcing strict $\SO(2)$ equivariance means that the middle pair would have to be skipped, since the two fermionic, black vertices in the second Adinkra must be raised jointly in an $\SO(2)$-equivariant setting, thus producing the penultimate Adinkra directly. The so generated 3-term $\SO(2)$-equivariant main sequence then {\em does not include\/} the omitted Adinkra of the pair in the middle of~(\ref{eSeqN2}). Starting with the superfield $\IB$ in place of $\IU$, one can construct an analogous 3-term {\em complementary\/} $\SO(2)$-equivariant main sequence, which however then would omit the Adinkra corresponding to $\IU$.

 Finally, it is the sequence~(\ref{eSeqN2}) to which the foregoing discussion of superderivative superfields corresponds, and without further ado, we just replace the corresponding Adinkras with the {\em superderivative superfields\/} in terms of $\IU$:
\beq
\vC{\begin{picture}(120,42)(-5,11)
 \put(-20,0){\includegraphics[width=5.2in]{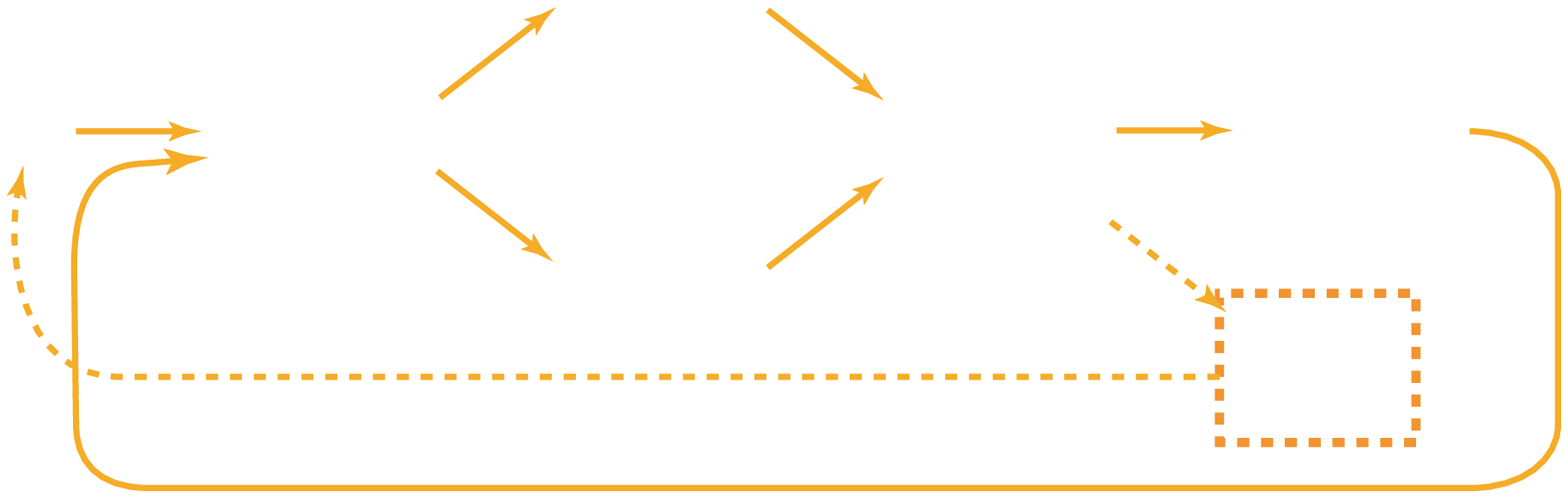}}
 \put(-8,38){$\IU$}
 \put(10,40){$\tw\IF_I=$}
 \put(10,34){$(D_I\IU)$}
 \put(36,28){$(D_1\IU)$}
 \put(36,48){$(D_2\IU)$}
 \put(62,47){$\tw\IA_I=$}
 \put(60,41){$(D_ID_1\IU)$}
 \put(66,37){or}
 \put(60,32){$(D_ID_2\IU)$}
 \put(87,38){$(D_1D_2\IU)$}
 \put(87,20){$(\vd_\t\IU)$}
\end{picture}}
\Label{eSeqN2SF}
\eeq

A comparison of the sequences~(\ref{eSeqN2}) and~(\ref{eSeqN2SF}) then proves:
\begin{proposition}\Label{pA=SFmap}
The cyclic sequence~(\ref{eSeqN2SF}) is the superderivative-superfield rendition, in the $N=2$ case, of the Vertex Raising Theorems (Theorem~\ref{tHGT} and
corollaries~\ref{cHGT} and \ref{cHGT2}); the implied identifications are listed in Table~\ref{tN2A=SF}.
\end{proposition}

\begin{table}[ht]
  \centering
  \begin{tabular}{@{} rl|rl @{}}
    \toprule
 {\bf Adinkra} & {\bf Superderivatives} &
 {\bf Adinkra} & {\bf Superderivatives} \\
    \midrule
 $\vC{\includegraphics[width=15mm]{sca2.eps}}$ & $\IU$
 &$\vC{\includegraphics[width=12mm]{adx.eps}}$ & $(D_I\IU)$ \\[2mm]
 $\vC{\includegraphics[width=15mm]{ferm1.eps}}$
  & either $(D_1\IU)$ or $(D_1\IU)$
 &$\vC{\includegraphics[width=12mm]{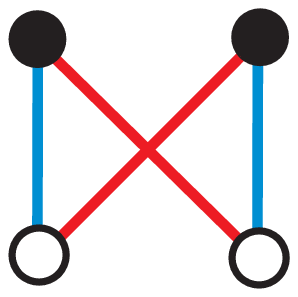}}$
  & either $(D_ID_1\IU)$ or $(D_ID_1\IU)$ \\
    \bottomrule
  \end{tabular}
  \caption{A listing of the Adinkras that appear in the general $N=2$ main sequence and their corresponding {\em superderivative superfields\/}, in terms of an unconstrained, scalar superfields, $\IU$. Here, $I=1,2$, so that $(D_I\IU)$ denotes the pair of superderivative superfields.}
  \Label{tN2A=SF}
\end{table}

 The set of four Adinkras appearing in this sequence describe,
 in fact, the complete set of irreducible and engineerable
 representations of the $(1|2)$ superalgebra.  We refer to this
 sequence of multiplets as the ``main sequence" of $N=2$
 supermultiplets.  The same four Adinkras appearing in~(\ref{long})
 were tabulated, in a completely analogous presentation,
 in Ref.\cite{rA}.  In that reference the main sequence was referred
 to as the ``root tree".  Also in Ref.\cite{rA} a set of operations
 generating all
 elements of this set was described in terms of arrow reversals.
 In this paper, however, we have used the language of vertex
 raising
 as a vehicle for describing the correlation between
 Adinkra operations and superspace derivatives.  We have therefore
 demonstrated more comprehensively the connection between
 Adinkras and superspace.

\subsection{The cases $N>2$}
  \Label{sN>2}
 The constructions of Subsections~\ref{sN=1}--\ref{sMainS}, do not of course stop at $N=2$. There is a straightforward generalization, which we now describe, without going into as many details, however.

\subsubsection{The Main Sequences of Adinkras}
 Owing to Vertex Raising Theorems (Theorem~\ref{tHGT} and Corollaries~\ref{cHGT} and \ref{cHGT2}),
 it is possible to start with an arbitrary engineerable Adinkra, for
 any value of $N$, and to follow through a sequence of vertex raising
 operations. This iteratively cycles through each engineerable
 supermultiplet for that value of $N$, and with the topology
 of the originally chosen Adinkra.

 We refer to the so obtained sequence of Adinkras with a $N$-cubical
 topology as the ``main sequence" of multiplets for each value of $N$;
 see also \SS\,\ref{sTop}.
 The $N=3$ case is presented here:
\begin{equation}
 \vC{\includegraphics[width=160mm]{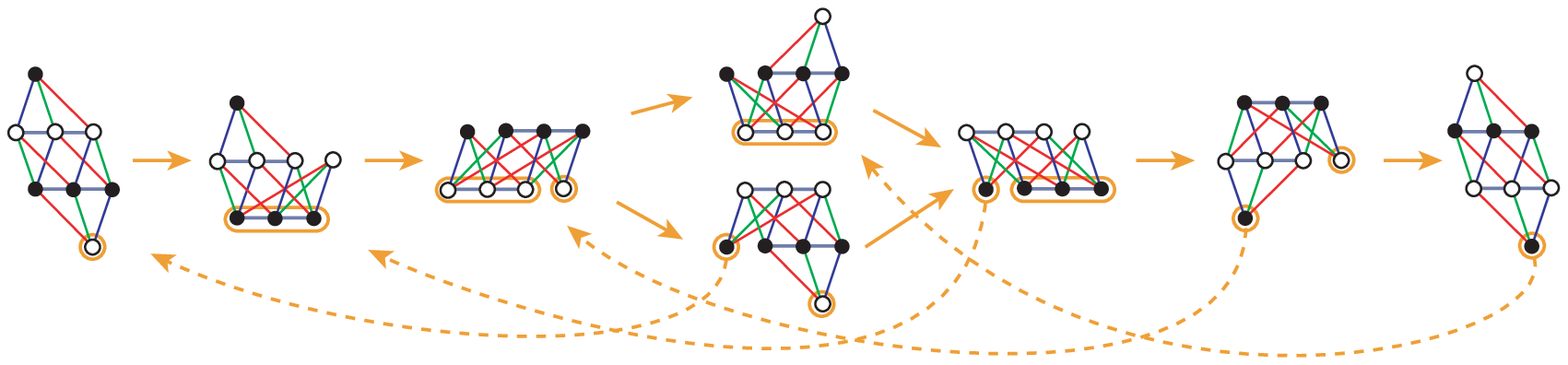}}
 \Label{eSeqN3}
\end{equation}
The grey `rungs' linking the vertices which are at the same level in the left-most Adinkra depict the additional restriction imposed herein: $\SO(3)$-equivariance. That is, any three linked vertices correspond to component fields that jointly span a 3-dimensional representation of $\SO(3)$; the solitary vertices correspond to component fields that span the invariant, 1-dimensional $\SO(3)$-representation. To preserve this $\SO(3)$-action, linked vertices may only be raised or lowered together. In each Adinkra in~(\ref{eSeqN3}), the source vertices are circled, being about to be raised, and the Adinkra(s) to the immediate right represent the result.
 This rule has been followed in the sequence~(\ref{eSeqN3}) and has considerably simplified its construction. The complete cycle, with all eight vertices distinguishable is depicted in Fig.~\ref{fN3All}, however, without the dots indicating the vertex raising that generates the hanging gardens' main sequence.
\begin{figure}[ht]
\begin{center}
\includegraphics[width=6.6in]{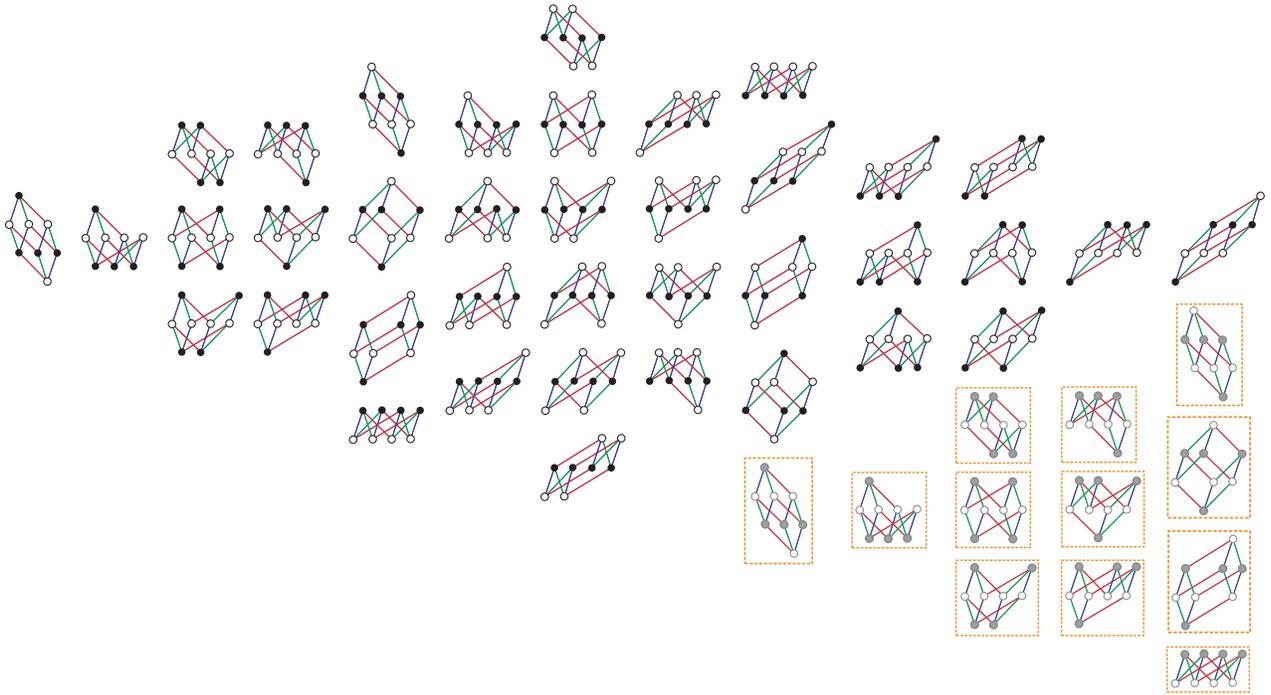}
\caption{The {\em main sequence\/} of $N=3$ Adinkras, generated by the vertex raising, starting from left. All eight vertices were treated as distinguishable, but no labeling was added, to prevent clutter. The gray, boxed Adinkras in the right-hand lower part begin repeating the sequence.}
\Label{fN3All}
\end{center}
\end{figure}
  The $3^{rd}$, $5^{th}$, $6^{th}$ and $7^{th}$ provide two separate ways of raising vertices. One of the two options in the $5^{th}$, $6^{th}$ and $7^{th}$ Adinkra, and the only option in the $8^{th}$ one produce Adinkras already present in the sequence, and the so replicated Adinkra corresponds to a supermultiplet that is the time-derivative of the original. This is indicated by the dashed back-arrows in~(\ref{eSeqN3}); in Fig.~\ref{fN3All}, the similarly repeating Adinkras are depicted grey and framed.
 This provides, in both the $\SO(3)$-equivariant main sequence~(\ref{eSeqN3}) and the general one in Fig.~\ref{fN3All}, for the same cyclicality as in~(\ref{eSeqN2}).

 Furthermore, just as in the $N=2$ main sequence~(\ref{eSeqN2}), here too we see that both the $\SO(3)$-equivariant and the general main sequence have a spindly shape, and both are invariant with respect to a mirror-reflection of sorts, where the $k^{th}$ Adinkra from the left has a counterpart in a corresponding $k^{th}$ Adinkra from the right. Such two Adinkras are each other's up-down reflection, up to the slant in the diagrams in~(\ref{eSeqN3}) and Fig.~\ref{fN3All}, provided so as to suggest a fake perspective to the diagrams. This mirror-reflection is reminiscent of Hodge duality; in particular, the pair of Adinkras in the middle of the sequence~(\ref{eSeqN3}) is such a mirror-pair.

In general, insisting on $\SO(N)$ equivariance precludes vertex raising and lowering of individual vertices from within an $n$-tuplet of vertices corresponding to $n$ component fields that span an irreducible $n$-dimensional representation of $\SO(N)$. Thus, for example, the Adinkras with $(2|4|2)$ vertices appearing in the 3rd column from the left in Fig.~\ref{fN3All} cannot occur in the $\SO(3)$-equivariant main sequence~(\ref{eSeqN3}). This divides {\em families\/} of a topology into equivariant {\em genera\/}. Reducing the equivariance group to $\Ione$ brings us back to the whole family: ``$\Ione$-equivariant genera'' of Adinkras are families of Adinkras.

\subsubsection{Superderivative Solutions for $N=3$}
Unlike the $N=2$ case, an explicit construction for the proof of the $N>2$ analogue of Proposition~\ref{pHGTN2} turns out not to be generated by a finite list of superderivative matrices, and we discuss this briefly, below. However, we do provide a generalization of Proposition~\ref{pA=SFmap}, for each $N$ and each family of Adinkras with the same topology, if at least one Adinkra has an identified, corresponding superfield.

Straightforward iteration of the foregoing constructions in this section proves, as a direct generalization of proposition~\ref{pA=SFmap}:
\begin{proposition}\Label{pA2SFN3}
To the main $\SO(3)$-equivariant main sequence of Adinkras~(\ref{eSeqN3}),
there corresponds an analogous sequence of superderivative superfields,
listed in table~\ref{tN3A=SF}.
\end{proposition}
\Remk
 The diligent Reader should have no difficulty ascertaining the same pairings for the much larger family depicted in Fig.~\ref{fN3All}.
\begin{table}[ht]
  \centering
  \begin{tabular}{@{} rl|rl @{}}
    \toprule
 {\bf Adinkra} & {\bf Superderivatives} &
 {\bf Adinkra} & {\bf Superderivatives} \\
    \midrule
 $\vC{\includegraphics[width=13mm]{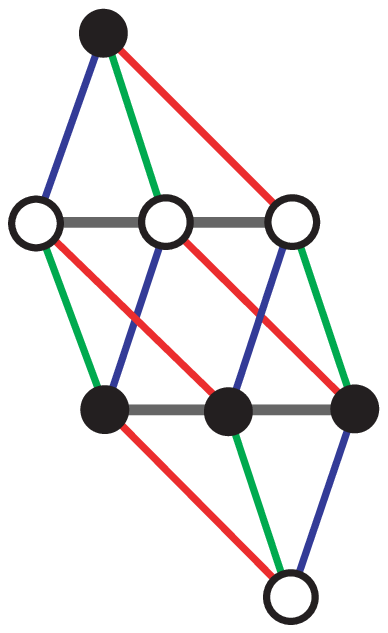}}$ & $\IU$
 &$\vC{\includegraphics[width=15mm]{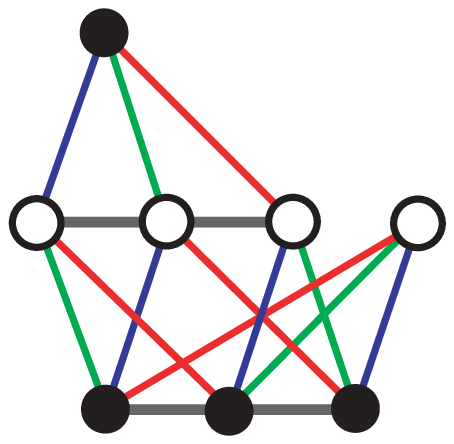}}$ & $(D_I\IU)$ \\
 $\vC{\includegraphics[width=15mm]{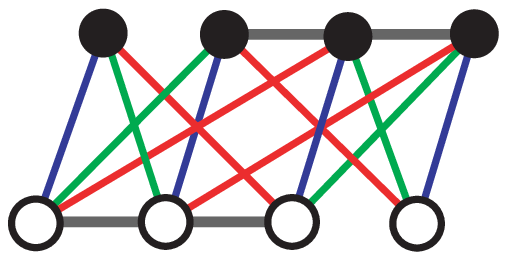}}$
 & $(D_{[I}D_{J]}\IU,\vd_\t\IU)$
 &$\vC{\includegraphics[width=15mm]{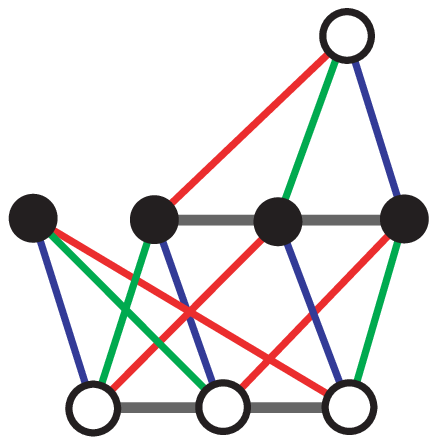}}$
  & $(D_{[I}D_{J]}\IU)$ \\
 $\vC{\includegraphics[width=15mm]{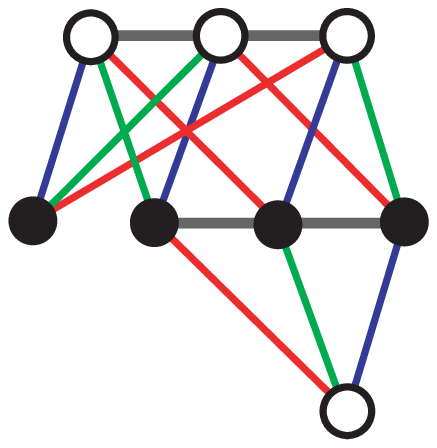}}$
 & $(D_{[1}D_2D_{3]}\IU,\vd_\t\IU)$
 &$\vC{\includegraphics[width=15mm]{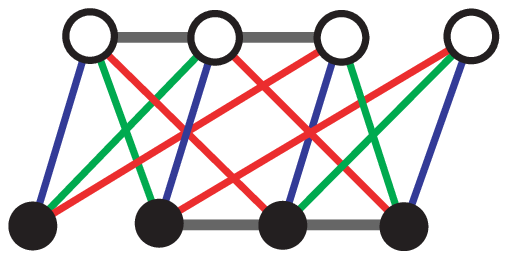}}$
  & $(D_{[1}D_2D_{3]}\IU,\vd_\t D_I\IU)$ \\
 $\vC{\includegraphics[width=15mm]{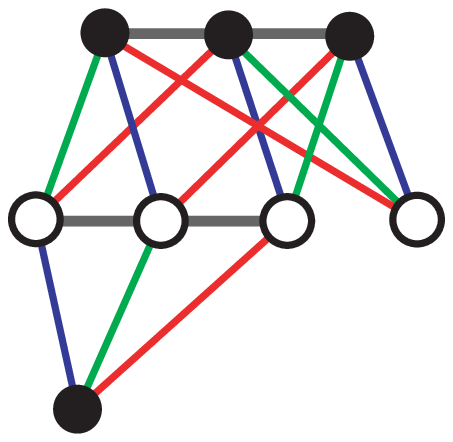}}$
 & $(D_{[1}D_2D_{3]}\IU,\vd_\t^2\IU)$
 &$\vC{\includegraphics[width=15mm]{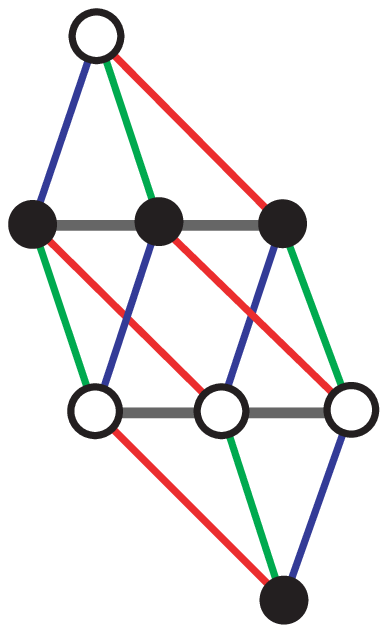}}$
 & $(D_{[1}D_2D_{3]}\IU)$ \\
    \bottomrule
  \end{tabular}
  \caption{A listing of the Adinkras that appear in the $\SO(3)$-equivariant main sequence, their corresponding {\em superderivative superfields\/}, in terms of an unconstrained, scalar superfields, $\IU$. Here, $I,J,K=1,2,3$, so $(D_I\IU)$ denotes the triple of superderivative superfields.}
  \Label{tN3A=SF}
\end{table}
The $\SO(3)$-equivariant sequence~(\ref{eSeqN3}) is, of course, embedded in the general main sequence in Fig.~\ref{fN3All}, and a list of superfield constructions analogous to those in Table~\ref{tN3A=SF} is easy to reconstruct for all the Adinkras therein along the lines described in Theorem~\ref{pA=SF}.

\subsubsection{The Semi-infinite Sequence of Superfields}
The construction of the superfield sequence corresponding to~(\ref{eDDmap}) and the sequence~(\ref{eZigZag}) was fairly straightforward. For $N=2$, the sequence~(\ref{eZigZag}) encodes all the relationships between the:
 ({\it a})~superderivative superfields,
 ({\it b})~the super-constraints that they satisfy, and
 ({\it c})~the dual super-equivalence classes.
The construction of the $N>2$ analogue of the sequence~(\ref{eZigZag}), and so the determination of this data, becomes increasingly more complex with $N$ growing.

 For example, the right-hand side entry in the first row of Table.~\ref{tN3A=SF} gives a superderivative superfield, $(D_I\IU)$, in terms of a single, unconstrained superfield $\IU$, which solves a system of {\em five\/} constraints imposed on an $\SO(3)$-equivariant triple of superfields:
 \begin{eqnarray}
 \ID^{\sss(2)}_{\cal A}{}^I\IF_I&=&0~,\qquad
 \ID^{\sss(2)}_{\cal A}{}^I:=\tw\l_{\cal A}{}^{JI}D_J~,\quad
  I,J=1,2,3,~~{\cal A}=1,\dots,5~,\Label{eN3ker1}\\[2mm]
 \tw\l_{\cal A}&=&\left\{\left(\mitrix{0&1&0\cr1&0&0\cr0&0&0}\right)~,~
                         \left(\mitrix{0&0&1\cr0&0&0\cr1&0&0}\right)~,~
                         \left(\mitrix{0&0&0\cr0&0&1\cr0&1&0}\right)~,~
                         \left(\mitrix{1&\phantom{-}0&0\cr0&-1&0\cr0&\phantom{-}0&0}\right)~,~
                         \left(\mitrix{0&0&\phantom{-}0\cr0&1&\phantom{-}0\cr0&0&-1}\right)
                          \right\}~.
\end{eqnarray}
Note that the five traceless, symmetric matrices, $\tw{\BM{\l}}_{\cal A}$, generate the $\SU(3)/\SO(3)$ coset; the remaining three imaginary Gell-Mann matrices generate $\SO(3)\into\SU(3)$. Akin to Eq.~(\ref{eSqDKJ}), we can display:
\begin{equation}
 [\ID^{\sss(2)}_{\cal A}{}^I]
  =\left[\begin{matrix}D_2&D_1&0\cr D_3&0&D_1\cr 0&D_3&D_2\cr
                 D_1&-D_2&0\cr 0&D_2&-D_3\end{matrix}\right]~,\qquad
 \ID_{\cal A}{}^I\,D_I=0~. \Label{eIDAI}
\end{equation}
Like $\ID_J{}^I$ in~(\ref{eSqDKJ}), this is the maximal-rank first order superderivative linear mapping that annihilates $D_I$. However, unlike~(\ref{eSqDKJ}), this is not even a square matrix, much less nilpotent. The
maximal-rank matrix of first order superderivatives that annihilates
$\ID^{\sss(2)}_{\cal A}{}^I$ may be found by direct computation:
\begin{equation}
 [\ID^{\sss(3)}_{\sB}{}^{\cal A}]=
 \left[\begin{matrix} D_1 &  0  &  0  & D_2 &  0 \cr
                D_2 &  0  &  0  &-D_1 &  0 \cr
                D_3 & D_2 & D_1 &  0  &  0 \cr
                0   & D_1 &  0  & D_3 & D_3\cr
                0   & D_3 &  0  & D_1 & D_1\cr
                0   &  0  & D_2 &  0  & D_3\cr
                0   &  0  & D_3 &  0  & D_2\cr\end{matrix}\right]~,\qquad
 \ID^{\sss(3)}_{\sB}{}^{\cal A}\,\ID^{\sss(2)}_{\cal A}{}^I=0~.
 \Label{eIDBA}
\end{equation}

Like for $N=2$, we have that $D_I\IU$ annihilates $(u(0);0;0)\in\IU$, and so
\begin{equation}
 0~\to~(u(0);0;0) ~\tooo{~\i~}~\IU~\tooo{~D^{(1)}_I~}~\IF_I
   ~\tooo{~\ID^{(2)}_{\cal A}{}^I~}~\IB_{\cal A}
   ~\tooo{~\ID^{(3)}_{\sB}{}^{\cal A}~} \Label{eBegN3}
\end{equation}
is exact. It behooves to make a quick dimension-count, recalling that the discussion leading to the sequences~(\ref{eZigZag}) and~(\ref{eZigZagA}) implies the following:
\begin{enumerate}\vspace{-2mm}
 \item The superderivative maps $D^{\sss(1)}_I$, $\ID^{\sss(2)}_J{}^I$, \etc,
  send a
  $(d_0|d_1|d_2|d_3|\cdots)$-dimensional representation\newline into a
  $(0|d_1|(d_2{+}d_0)|d_3|\cdots)$-dimensional one;\vspace{-2mm}
 \item The inclusion maps $\i_i$, for $i>0$, send a
  $(0|d_1|d_2|d_3|\cdots)$-dimensional representation\newline into a
  $(d_1|d_2|d_3|\cdots)$-dimensional one. Note that the lowest components
  in the resulting superfield have the same engineering dimension as the
  next-to-lowest ones in the initial one.\vspace{-2mm}
\end{enumerate}
This then easily provides the $N=3$ dimension-count version of the sequence~(\ref{eZigZag}):
 \begin{equation}
 \vC{\begin{picture}(100,48)(30,-1)
  \put(0,0){\includegraphics[width=5.5in]{ZigZag.eps}}
  \put(0,20.5){0}
  \put(10,11){0}
  \put(28,42){0}
  \put(50,42){0}
  \put(93,42){0}
  \put(111,42){0}
  \put(60,1){0}
  \put(75,1){0}
  \put(62.5,9){$\SSS(0|3|4|1)$}
  \put(17,21){$\SSS(0|0|0|0)$}
  \put(49,21){$\SSS(1|3|3|1)$}
  \put(80,21){$\SSS(3|9|9|3)$}
  \put(110,21){$\SSS(5|15|15|5)$}
  \put(34,33.5){$\SSS(0|0|0|0)$}
  \put(97,33.5){$\SSS(0|5|8|3)$}
  \put(39,23){\TC{orange}{$\i_0$}}
  \put(66,23){\TC{orange}{$D^{\sss(1)}_I$}}
  \put(98,23){\TC{orange}{$\ID^{\sss(2)}_J{}^I$}}
  \put(126,23){\TC{orange}{$\ID^{\sss(3)}_K{}^J$}}
  \put(47.5,29){$\SSS\i_0$}
  \put(61,16){$\SSS D_I$}
  \put(75,16){$\SSS\i_1$}
  \put(88.5,28){$\SSS\ID_J{}^I$}
  \put(112,29){$\SSS\i_2$}
  \put(126,16){$\SSS\ID_K{}^J$}
  \put(140,20.5){$\cdots$}
 \end{picture}}
 \Label{eZigZag3}
\end{equation}
Now, while the dimension count suffices to identify
\begin{equation}
 \im(D_I)=\vC{\includegraphics[width=15mm]{AdinkraF341.eps}}
 \subset\IF_I
  =\vC{\includegraphics[width=15mm]{AdinkraF1331.eps}}\mkern-20mu
   \vC{\includegraphics[width=15mm]{AdinkraF1331.eps}}\mkern-20mu
   \vC{\includegraphics[width=15mm]{AdinkraF1331.eps}}\mkern-10mu~,
 \Label{eZZ1}
\end{equation}
it does not suffice for an unambiguous identification of the remaining representations. In particular, from the dimension-count alone, it is not clear whether
\begin{equation}
 (0|5|8|3)~=~(0|1|4|3)\4(0|4|4|0)~,\quad\hbox{or}\quad
          ~=~(0|2|4|2)\4(0|3|4|1)~.
 \Label{eZZ2}
\end{equation}
 Thus, the $N=3$ analogue of the sequence~(\ref{eZigZag}) is not only semi-infinite, but the dimensions of the representations appearing grow unboundedly and its structure becomes rather swiftly, rather more and more complex. Necessarily then, the same applies for all $N>2$.

 The emerging structures are strongly reminiscent of a Verma module. The sought-after {\em main sequence\/}, perforce finite, is obtained as a quotient of this module by equivalences, which are easiest defined as the manifest identity of the corresponding Adinkras. A detailed study of this structure is however well beyond the scope of this article and will be addressed separately.

The construction of the semi-infinite superfield sequence in the manner of~(\ref{eZigZag}) becomes considerably more complex with $N$ growing, so the vertex raising technique with Adinkras is superior in constructing the corresponding main sequences with any degree of equivariance and for any $N>3$.

\section{Conclusions}
 \Label{sC}
For $(1|N)$-supersymmetry algebras with $N\leq3$, we have explicitly codified in terms of graph theory the subset of the representation theory that corresponds to engineerable {\em Adinkras\/}. For $N\geq4$, the classification becomes more complex owing to the emergence of more than one topology of Adinkras. The generating process discussed herein applies to all Adinkras of the same topology, which thus form a {\em family\/}. In this way the Adinkra topology provides a coarse classification into families, the one with the $N$-cubical topology called the {\em main sequence\/}. Within each family, our main Theorems~\ref{tDetA}, \ref{tHGT} and the Corollaries~\ref{cHGT} and \ref{cHGT2} of the latter identify and generate all members.

The analogous process has been explicitly reproduced for superfields of $(1|N)$-supersymmetry with $N\leq2$, in terms of superderivatives of superfields, viewed as solutions of superderivative constraints. For $N\geq3$, the generating of the analogous sequence of superderivative superfields and the superderivative constraints that they satisfy becomes rather more arduous. Instead, we have identified an algorithm for corresponding a superderivative superfield to every Adinkra---{\em assuming\/} that at least one Adinkra from each family has a known superfield rendition. This then produces a conditional classification of superderivative superfields directly paralleling the classification of Adinkras.
This condition, that for every Adinkra topology at least one Adinkra has a superfield rendition, remains our conjecture for now.

\begin{flushright}
 \parbox{4in}{\leavevmode\llap{``}%
  \sl There is something fascinating about science:\\
      One gets such a wholesale returns of conjecture\\
      out of such a trifling investment of fact.''\\
      \hglue30mm---~Samuel Clemens}\\[6mm]
 \parbox{4in}{\leavevmode\llap{``}%
  \sl There is no branch of mathematics, however abstract,\\
      which may not some day be applied to phenomena\\
      of the real world.''\\
      \hglue30mm---~Nikolai Lobachevsky}
\end{flushright}

\paragraph{\bf Acknowledgments:}
 The research of S.J.G.\ is supported in part by the National Science
Foundation Grant PHY-0354401.
 T.H.\ is indebted to the generous support of the Department of
Energy through the grant DE-FG02-94ER-40854.

\clearpage
\appendix
\section{Superfields and Component Fields}
 \Label{sRSF}
 The developments described in the main part of
 this paper are applicable largely within the context of classical field
 theory.  However, the quantized versions of these classical
 theories are of particular interest.  Accordingly, there are formal
 considerations which we should not ignore, lest quantization
 be rendered needlessly awkward.  In particular,
 since, in a quantum theory, symmetry transformations are generated on fields
 by operators which are necessarily unitary, it is important that
 the operator corresponding to a supersymmetry transformation, namely
 $\sU_\e:=\exp\{\,\d_Q\6(\e)\,\}$, must be unitary.  This implies
 that the generating operator $\d_Q\6(\e)$ should be anti-Hermitian,
 \begin{eqnarray}
 \d_Q\6(\e)^\dagger~=~-\d_Q\6(\e) \,,
 \Label{edQe}
\end{eqnarray}
with respect to the inner product of superfields in superspace. In
the case of $(1|1)$ supersymmetry, for example, such an inner
product is defined by
\begin{equation}
 \vev{\IA|\IB}:=\int\rd t\,\rd\q~\IA^\dagger\,\IB~,
 \Label{e:<A|B>}
\end{equation}
where $\IA$ and $\IB$ are $(1|1)$ superfields.

In the context of a strictly classical theory, it is mildly puzzling
how to realize an antihermitian supersymmetry operation $\d_Q(\e)$
which maps real fields into real fields.  However, this puzzle is
resolved by defining a supersymmetry transformation to act on any
object $\sO$ according to $\sO\to\sU\sO\sU^{-1}$. In this way even
classical fields are treated as operators, which, in fact, they are!
For unitary $\sU$, this implies $\sO\to\sU\sO\sU^\dagger$. This, in
turn, properly preserves Hermiticity:
\begin{eqnarray}
 0&=&[\sO'-(\sO')^\dagger]=[\sU\sO\sU^\dagger-(\sU\sO\sU^\dagger)^\dagger]
=\sU[\sO-\sO^\dagger]\sU^\dagger~. \Label{eUOUd}
\end{eqnarray}
So, supersymmetry will be understood to act by:
\begin{equation}
 \F\to e^{\d_Q\6(\e)}\,\F\,e^{-\d_Q\6(\e)}\quad\To\quad
 \d_Q\6(\e)\,[\F] =[\,-i\e\, Q\,,\,\F\,]=-i\e\,[\,Q\,,\,\F\,\}
                  =:-i\e\,(Q\,\F)~, \Label{eOpSuSy}
\end{equation}
where, formally,
\begin{equation}
 [\,Q\,,\,\F\,\}:= Q\,\F - (-1)^{|\F|}\,\F\,Q~, \Label{e:[Q,F)}
\end{equation}
and where $|\F|$ is $0$ or $1$ depending on whether $\F$ is
respectively a boson or a fermion. This formal calculation is to be
understood as valid when each term is acting on any test `function'
from a suitable class. So, for $\F=\F^\dagger$ and $\L=\L^\dagger$ a
real scalar and a real spinor superfield, alike:
\begin{align}
 (\d_Q\6(\e)\,\F)^\dagger&=[\,\d_Q\6(\e)\,,\,\F\,]^\dagger
  =[\,\F^\dagger\,,\,\d_Q\6(\e)^\dagger\,]
=\d_Q\6(\e)\,\F~;\\[2mm]
 (\d_Q\6(\e)\,\L)^\dagger&=[\,\d_Q\6(\e)\,,\,\L\,]^\dagger
  =[\,\L^\dagger\,,\,\d_Q\6(\e)^\dagger\,]
=\d_Q\6(\e)\,\L~.\Label{e:e[Q,L]}
\end{align}
For a real, anticommuting parameter $\e$ and the anti-Hermitian anticommuting operator $Q$, we have
that\Ft{\Label{iManifest}We follow the three decades standard convention whereby $(AB)^\dagger=B^\dagger A^\dagger$,
regardless of the parity of $A,B$. Note that Refs.\cite{rDel,rDF} declare $(AB)^\dagger=(-1)^{|A|\cdot|B|}A^\dagger B^\dagger$
instead, thus introducing  extra `$-$' signs in the Hermitian conjugation of anticommuting objects.}:
\begin{equation}
 \d_Q\6(\e)~:=~-i\e\,Q~,\quad\hbox{since}\quad
 (\d_Q\6(\e))^\dagger = i\,Q^\dagger \e^\dagger
 = -i\,Q\,\e = i\e\,Q \Label{eQe}
\end{equation}
is indeed anti-Hermitian~(\ref{edQe}).

With a little forethought and~(\ref{eOpSuSy}) in mind, we {\em define\/}:
\begin{align}
  \f&:=\F|~,\Label{e:F0}\\
 i\,\j&:=[\,D\,,\,\F\,]|~,\Label{e:F1}\\[2mm]
  \l&:=\L|~,\Label{e:L0}\\
   B&:=\{\,D\,,\,\L\,\}|~,\Label{e:L1}
\end{align}
so that, if $\F^\dagger=\F$ and $\L^\dagger=\L$, the component field operators are all
Hermitian\Ft{In turn, only Hermitian operators may be identified with {\em observables\/},
which are the subject of classical theory. This should make the projection of $\f,\j,\l,B$
into {\em classical fields\/} straightforward. Finally, for classical fields, an `operatorial'
expression such as $[\vd_\t,f(\t)]$ is simply to be interpreted as being the result of $\vd_\t$
applied on $f(\t)$, since that is what $[\vd_\t,f(\t)]$ becomes when this is applied on any suitable test function.}:
\begin{align}
 \f^\dagger&=\F^\dagger|=\F|=\f~, \Label{eFOcF0}\\
 \j^\dagger&=(-i\,[\,D\,,\,\F\,])^\dagger|
 =-i\,[\,D\,,\,\F\,]|=\j~; \Label{eFOcF1}\\[2mm]
 \l^\dagger&=\L^\dagger|=\L|=\l~, \Label{eFOcL0}\\
 B^\dagger&=\{\,D\,,\,\L\,\}^\dagger|=\{\,D\,,\,\L\,\}|=B~. \Label{eFOcL1}
\end{align}
Note that we {\em define\/} the components of a superfield expression, or superfield statement, using the projection operator basis,
$\{\,\Ione\,\B\,|~,\,[D,\,\B\,\}|\,\}$, which is {\em dual\/} to the
$\q$-expansion basis, $\{1,\q\}$.

The supersymmetry transformations of the component fields are obtained by applying our basis of projection operators
on the superfield transformation equation,
\begin{align}
 \d_Q\6(\e)\,\F = [\,-i\e\,Q\,,\,\F\,]~,\qquad\hbox{and}\qquad
 \d_Q\6(\e)\,\L = [\,-i\e\,Q\,,\,\L\,]~, \Label{eQonSF}
\end{align}
which produces:
\begin{align}
 \d_Q\6(\e)\,\f&:= \d_Q\6(\e)\,\F|
   = -i\e\,(Q\,\F)|
   = -i\e\,\big((iD+2\q\vd_\t)\F\big)\big|
   = i\,\e\,\j~, \Label{e:dQF0}\\[2mm]
 \d_Q\6(\e)\,\j&:= (-iD\,\d_Q\6(\e)\,\F)|
   =\e\,\dot\f~. \Label{e:dQF1}\\
\noalign{\vglue-2mm\noindent and similarly,}
 \d_Q\6(\e)\,\l&:= \d_Q\6(\e)\,\L|
   =\e\,B~, \Label{e:dQL0}\\[2mm]
 \d_Q\6(\e)\,B&:= (D\,\d_Q\6(\e)\,\L)|
   =i\,\e\,\dot\l~. \Label{e:dQL1}
\end{align}

Furthermore, the projections of $[\d_Q\6(\e_1),\d_Q\6(\e_2)]\F$ yield:
\begin{align}
 [\,\d_Q\6(\e_1)\,,\,\d_Q\6(\e_2)\,]\,\f
 &:=[\,\d_Q\6(\e_1)\,,\,\d_Q\6(\e_2)\,]\,\F|
   =\d_Q\6(\e_1)\big(\d_Q\6(\e_2)\F\big)|-``1\leftrightarrow2"
   =2i\e_1\e_2\,\dot\f~, \Label{e:dQdQF0}\\[2mm]
 [\,\d_Q\6(\e_1)\,,\,\d_Q\6(\e_2)\,]\,\j
 &:=\big(-iD\,[\,\d_Q\6(\e_1)\,,\,\d_Q\6(\e_2)\,]\,\F\big)\big|
   =-iD\,\d_Q\6(\e_1)\big(\d_Q\6(\e_2)\F\big)\big|-``1\leftrightarrow2"\nn\\
 &=2i\,\e_1\e_2\,\dot\j~. \Label{e:dQdQF1}
\end{align}
This is in perfect agreement with the ``operatorial'' equation
\begin{align}
 [\,\d_Q\6(\e_1)\,,\,\d_Q\6(\e_2)\,]
 &=[\,(-i\e_1\,Q)\,,\,(-i\e_2\,Q)\,]
=+2i\e_1\e_2\,\vd_\t~,
 \Label{e:OpdQdQ}
\end{align}
and the general results~(\ref{eQQonSF})--(\ref{eEQEQonSF}). Since operatorial equations are meant to hold when
applied on {\em any suitable function\/}, this proves that the operators $Q$ and $D$, as defined in~(\ref{Qi})
and~(\ref{Di}), respectively, are applicable on superfields. The corresponding component field equations are
obtained by invariant projection, applied upon these superfield equations, consistently with definition~\ref{dComp}.

On the other hand, iterating~(\ref{e:dQF0})--(\ref{e:dQL1}) to obtain the action of $[\d_Q\6(\e_1),\d_Q\6(\e_1)]$
directly upon the pairs of component fields produces:
\begin{align}
 [\,\d_Q\6(\e_1)\,,\,\d_Q\6(\e_2)\,]\,\Big({\f\atop\j}\Big)
 &=\d_Q\6(\e_1)\Big({i\e_2\,\j\atop\e_2\,\dot\f}\Big)
    -``1\leftrightarrow2"
  =-2i\,\e_1\e_2\,\vd_\t\,\Big({\f\atop\j}\Big)~, \Label{eFpair}\\[2mm]
 [\,\d_Q\6(\e_1)\,,\,\d_Q\6(\e_2)\,]\,\Big({\l\atop B}\Big)
 &=\d_Q\6(\e_1)\Big({\e_2\,B\atop i\e_2\,\dot\l}\Big)-``1\leftrightarrow2"
  =-2i\,\e_1\e_2\,\vd_\t\,\Big({\l\atop B}\Big)~. \Label{eLpair}
\end{align}
Being different from the operatorial equation~(\ref{e:OpdQdQ}), these results prove that the component fields
themselves do not belong to the class of suitable functions upon which the operators $Q$ and $D$, as defined
in~(\ref{Qi}) and~(\ref{Di}), respectively, are defined to act.

\newpage
\Refs{References}{[00]}

\Bib{rABS} M.~F. Atiyah, R.~Bott, and A.~Shapiro: {\em Clifford modules\/},
 {\em Topology}, \textbf{3}\,(suppl. 1)\,(1964)\,3--38.

\Bib{rBKMO} S.~Bellucci, S.~Krivonos, A.~Marrani and E.~Orazi:
 {\em ``Root'' Actions for $N=4$ Supersymmetric Mechanics Theories\/},
 {\tt hep-th/0511249}.

\Bib{rG1} B.~Bollob\'as: {\em Graph Theory: An Introductory Course\/},
 (Springer--Verlag, New York, 1979).

\Bib{rBK} I.~L. Buchbinder and S.~M. Kuzenko: {\em Ideas and methods of
 supersymmetry and supergravity\/}, Studies in High Energy Physics Cosmology
 and Gravitation, (IOP Publishing Ltd., Bristol, 1995).

\Bib{rCRT} H.L.~Carrion, M.~Rojas and F.~Toppan: {\em Octonionic
Realizations of One-Dimensional Extended Supersymmetries: a
Classification\/}, \MPL{A18}(2003)\,787--798, {\tt hep-th/0212030}.

\Bib{rG2} G.~Chartrand and L.~Lesniak: {\em Graphs \&\ Digraphs\/},
 2nd ed., (Wadsworth, Belmont, CA, 1986).

\Bib{rDel} P.~Deligne and D.~S. Freed: {\em Supersolutions\/}, in
 {\em Quantum fields and strings: a course for mathematicians, Vol.~1~\&~2,
 Princeton, NJ, 1996/1997}, p.227--355, (Amer. Math. Soc., Providence, RI,
 1999).

\Bib{r6-1} C.~Doran, M.~Faux, S.~J. Gates, Jr., T.~H\"ubsch, K.~Iga, and
 G.~D. Landweber: {\em Off-shell supersymmetry, Clifford modules, and
 Adinkras\/}, forthcoming.

\Bib{rA} M.~Faux and S.J.~Gates, Jr.: {\em Adinkras: A Graphical Technology
 for Supersymmetric Representation Theory\/}, \PR{D71}(2005)\,065002,
 {\tt hep-th/0408004}.

\Bib{rFS} M.~Faux and D.~Spector: {\em Duality and Central Charges in Supersymmetric Quantum Mechanics\/}, \PR{D70}(2004)\,085014.

\Bib{rFSK} M.~Faux, D.~Kagan and D.~Spector: {\em Central Charges and Extra Dimensions in Supersymmetric Quantum Mechanics\/}, {\tt hep-th/0406152}.

\Bib{rFre} P.~Fr{\'e}: {\em Introduction to harmonic expansions on coset
 manifolds and in particular on coset manifolds with Killing spinors\/},
 in {\em Supersymmetry and supergravity 1984 (Trieste, 1984)},
  p.~324--367, (World Sci.\ Pub., Singapore, 1984).

\Bib{rDF} D.~S. Freed: {\em Five lectures on supersymmetry\/},
 (Amer. Math. Soc., Providence, RI, 1999).

\Bib{r1001} S.~J.~Gates,~Jr., M.~T. Grisaru, M.~Ro\v{c}ek, and W.~Siegel:
 {\em Superspace}, (Benjamin/Cummings Pub. Co., Reading, MA, 1983).

\Bib{rGK}S.~J.~Gates,~Jr. and S.~ÊV. Ketov: {\em 2D (4,4) hypermultiplets.
 II. Field theory origins of dualities\/},
 \PL{B418}(1998)\,119--124.

\Bib{rGLPR} S.~J.~Gates,~Jr., W.~Linch, J.~Phillips, and L.~Rana:
 {\em The fundamental supersymmetry challenge remains\/},
 {\em Gravit.\,Cosmol.\,\bf8}\,(2002)\,96--100, {\tt hep-th/0109109}.

\Bib{rGLP} S.J.~Gates, Jr., W.D.~Linch, III and J.~Phillips:
 {\em When Superspace Is Not Enough}, {\tt hep-th/0211034}.

\Bib{rGR0} S.~J. Gates, Jr. and L.~Rana: {\em Ultra-Multiplets: A New
 Representation of Rigid $2D$, $N=8$ Supersymmetry\/},
 \PL{B342}(1995)\,132--137, {\tt hep-th/9410150}.

\Bib{rGR1} S.~J. Gates, Jr. and L.~Rana: {\em A theory of spinning
 particles for large {$N$}-extended supersymmetry\/},
 \PL{B352}(1995)\,50--58, {\tt hep-th/9504025}.

\Bib{rGR2} S.~J. Gates, Jr. and L.~Rana: {\em A theory of spinning
 particles for large {$N$}-extended supersymmetry. II\/},
 \PL{B369}(1996)\,262--268, {\tt hep-th/9510151}.

\Bib{rHSS} T.~H\"ubsch: {\em Haploid (2,2)-Superfields in
 2-Dimensional Spacetime}, \NP{B555}(1999)\,567--628, {\tt hep-th/9901038}.

\Bib{rKac} V.~G. Kac: {\em Lie superalgebras\/},
 {\em Advances in Math.\,\bf26}\,(1)\,(1997)\,8--96.

\Bib{rKRT} Z.~Kuznetsova, M.~Rojas and F.~Toppan: {\em Classification of
 irreps and invariants of the $N$-extended Supersymmetric Quantum
 Mechanics\/}, {\tt hep-th/0511274}.

\Bib{rKT} Z.~Kuznetsova, F.~Toppan: {\em Constrained Generalized
 Supersymmetries and Superparticles With Tensorial Central Charges:
 A Classification\/}, \JHEP{0505}(2005)\,060, {\tt hep-th/0502178}.

\Bib{rPT} A.~Pashnev, F.~Toppan: {\em On the Classification of
 $N$-Extended Supersymmetric Quantum Mechanical Systems\/},
 {\it J.\,Math.\,Phys.\,\bf42}\,(2001)\,5257--5271, {\tt hep-th/0010135}.

\Bib{rT04} F.~Toppan: {\em Central Extensions, Classical Nonequivariant
 Maps and Residual Symmetries\/}, \NP{B127}(Proc.\,Suppl.~2004)\,201--206,
 {\tt hep-th/0307118}.

\Bib{rT01a} F.~Toppan: {\em Division Algebras, Extended Supersymmetries
 and Applications\/}, \NP{B102}(Proc.\,Suppl.~2001)\,270--277,
 {\tt hep-th/0109073}.

\Bib{rT01} F.~Toppan: {\em Classifying $\cal N$-Extended 1-Dimensional
 Supersymmetric Systems\/}, in {\em Kiev 2000, Noncommutative structures
 in mathematics and physics\/}, p.195--201, {\tt hep-th/0109047}.

\Bib{rEWM} E.~Witten: {\em Supersymmetry and Morse Theory\/},
 {\it J.\,Diff.\,Geom.\,\bf17}\,(1982)\,661--692.

\endRefs
\end{document}